\documentclass[journal]{IEEEtran}

\ifCLASSINFOpdf

\else

\fi

\usepackage{rotating}

\usepackage{etex}
\usepackage{tikz}

\usetikzlibrary{shapes,arrows}

\usetikzlibrary{shapes,arrows,trees,calc}

\usetikzlibrary{arrows,shapes,positioning,shadows,trees}

\tikzset{
  basic/.style  = {draw, text width=2cm, drop shadow, font=\sffamily, rectangle,},
  root/.style   = {basic, rounded corners=2pt, thin, align=center,
                   fill=green!30},
  level 2/.style = {basic, rounded corners=6pt, thin,align=center, fill=green!60,
                   text width=12em},
  level 3/.style = {basic, thin, align=left, fill=pink!60, text width=8.5em}
}

\usepackage[T1]{fontenc}
\usepackage{tikz-qtree}
\usepackage{epigraph}
\usepackage{adjustbox}
\usepackage{setspace}
\usepackage{hyperref}
\usepackage{cite}

\usepackage{url}
\hypersetup{
  hidelinks 
}
\usepackage{epsf}\epsfverbosetrue
\usepackage{graphics,epsfig,color}
\usepackage{graphicx,epsfig}

\usepackage{epsfig}
\usepackage{multirow}
\usepackage{slashbox}
\usepackage{alltt}
\usepackage{subfig}
\usepackage{color}
 \usepackage{float} 
\usepackage{graphicx}
\usepackage{verbatim} 
\usepackage{footnote} 
\usepackage{latexsym} 
\usepackage{amsmath}
\usepackage{sidecap} 
\usepackage{wrapfig}
\usepackage{algorithm}
\usepackage{eso-pic}
\usepackage{fix-cm}
\usepackage{mcite}

\usepackage[normalem]{ulem}
\usepackage{wrapfig}

\usepackage{algorithm}
\usepackage{layout}
\usepackage{amsmath}
\usepackage{textcomp}
\usepackage{multirow}
\usepackage{colortbl}
\usepackage{soul,xcolor}
\usepackage{amsmath}
\usepackage{verbatim}
\usepackage{smartdiagram}
\usepackage{metalogo}
\usepackage{amsfonts}
\usepackage{enumitem}

\usepackage{pifont}

\usepackage{tikz}
\usetikzlibrary{shapes}

\usepackage{amsmath}
\usepackage{xspace}

\newcommand{\stkout}[1]{\ifmmode\text{\sout{\ensuremath{#1}}}\else\sout{#1}\fi}

\usetikzlibrary{trees}
\tikzstyle{every node}=[draw=black, thin, minimum height=3em]

\usepackage{tikz}
\makeatletter 
\@namedef{color@1}{red!40}
\@namedef{color@2}{green!40}   
\@namedef{color@3}{blue!40} 
\@namedef{color@4}{cyan!40}  
\@namedef{color@5}{magenta!40} 
\@namedef{color@6}{yellow!40}    

\newcommand{\graphitemize}[2]{%
\begin{tikzpicture}[every node/.style={align=center}]  
  \node[minimum size=5cm,circle,fill=gray!40,font=\Large,outer sep=1cm,inner sep=.5cm](ce){#1};  
\foreach \gritem [count=\xi] in {#2}
{\global\let\maxgritem\xi}  
\foreach \gritem [count=\xi] in {#2}
{%
\pgfmathtruncatemacro{\angle}{360/\maxgritem*\xi}
\edef\col{\@nameuse{color@\xi}}
\node[circle,
     ultra thick,
     draw=white,
     fill opacity=.5,
     fill=\col,        
     minimum size=3cm] at (ce.\angle) {\gritem };}%
\end{tikzpicture}  
}%

\usepackage{booktabs}
\newcommand{\saim}[1]{\textcolor{black}{#1}}
\newcommand{\mage}[1]{\textcolor{black}{#1}}

\newcommand{\blue}[1]{\textcolor{black}{#1}}

\newcommand*{\rom}[1]{\expandafter\@slowromancap\romannumeral #1@}
\newcommand{\Rmnum}[1]{\expandafter\@slowromancap\romannumeral #1@}
\makeatother

\usepackage{bbding}
\usepackage{pifont}
\usepackage{wasysym}
\usepackage{amssymb}
\usepackage{multirow}
\usepackage{comment}
\usepackage{graphics}

\usepackage{smartdiagram}
\usepackage{metalogo}
\usepackage{dtklogos}
\usetikzlibrary{arrows,shapes,positioning,shadows,trees}

\usepackage[utf8]{inputenc}
\usepackage[nil,showlanguages]{babel}
\usepackage{array,longtable,booktabs,makecell}

\begin{document}

\title{MAC Protocols for Terahertz Communication: A Comprehensive Survey}

\author{Saim Ghafoor, Noureddine Boujnah, Mubashir Husain Rehmani and Alan Davy
\thanks{Saim Ghafoor, Noureddine Boujnah, and Alan Davy are with Emerging Network Laboratory, Telecommunication Systems \& Software Group, Waterford Institute of Technology, Ireland. email: {s.ghafoor,bnoureddine,adavy}@tssg.org. }
\thanks{Mubashir Husain Rehmani is with Cork Institute of Technology, Ireland. email: mshrehmani@gmail.com.}
}


\maketitle

\begin{abstract}

Terahertz communication is emerging as a future technology to support Terabits per second link with highlighting features as high throughput and negligible latency. However, the unique features of the Terahertz band such as high path loss, scattering and reflection pose new challenges and results in short communication distance. The antenna directionality, in turn, is required to enhance the communication distance and to overcome the high path loss. However, these features in combine negate the use of traditional \blue{m}edium access protocols \blue{(MAC)}. Therefore\blue{,} novel MAC protocol designs are required to fully exploit their potential benefits including efficient channel access, control message exchange, link establishment, mobility management, and line-of-sight blockage mitigation. An in-depth survey of Terahertz MAC protocols is presented in this paper. The paper highlights the key features of the Terahertz band which should be considered while designing an efficient Terahertz MAC protocol, and the decisions which if taken at Terahertz MAC layer can enhance the network performance. Different Terahertz applications at macro and nano scales are highlighted with design requirements for their MAC protocols. The MAC protocol design issues and considerations are highlighted. Further, the existing MAC protocols are also classified based on network topology, channel access mechanisms, and link establishment strategies as Transmitter and Receiver initiated communication. The open challenges and future research directions on Terahertz MAC protocols are also highlighted.

\end{abstract}


\IEEEpeerreviewmaketitle

\section{Introduction}\label{sec:intro}

The demand for wireless data traffic has increased significantly since the evolution of Internet and Mobile Technology and is projected to exceed Petabytes by 2021~\cite{ciscodata2018}. The existing wireless technology although reaching the capacity of wired technology, still it is not meeting the demands of future ultra-high bandwidth communication networks. The spectrum at and below 60 GHz still orders of magnitude below the targeted Terabits per second (Tbps) link. The Free-space optical (FSO) which operates at Infrared (IR) frequencies also has several issues that limit the practicality of these systems for personal wireless communications~\cite{hamza2016,Akyildiz2014}. In this perspective, the Terahertz (THz) band from 0.1 to 10 Terahertz has the potential to provide up to Tbps link speed to satisfy beyond fifth-generation (5G) communication requirements such as high throughput and low latency~\cite{Akyildiz2014,teranet2014, Elayan2018, Yilmaz2016}. The Terahertz bands offer much larger bandwidth (up to 1 THz) than the existing millimeter-wave (mmWave) systems (up to 10 GHz))~\cite{zhnagh2010,Yongsu2006,Kurner2014}.

While, the technology is rapidly advancing with new transceiver architectures, materials, antenna design, channel/propagation model, and physical layer techniques, there still exist several research challenges that need to be addressed before achieving the Tbps links. Among these different fields of interest, Medium Access Control (MAC) is least explored area of research in Terahertz communication networks. The existing MAC protocols of traditional networks cannot be directly applied, because they do not consider the unique features of Terahertz band like path and molecular loss, multipath, reflection, and scattering. Therefore, novel and efficient MAC protocols are required which should consider the features of Terahertz bands and antenna requirements. In this paper\blue{,} a comprehensive survey on Terahertz MAC protocols is presented with classification, design issues, and considerations, requirements for different application areas and challenges. The acronyms used commonly throughout this survey are shown in Table~\ref{tab:acronyms}.

\begin{table}[htbp]
  \centering
  \small
  \caption{\mage{Acronym definitions used throughout this survey.}}
    \begin{tabular}{ll} \hline \hline
    \multicolumn{1}{l}{\textbf{Acronyms}} & \multicolumn{1}{l}{\textbf{Definitions}} \\ \hline
    5G    & Fifth generation \\
    ACK   & Acknowledgement \\
    AP    & Access point \\
    BER   & Bit error rate \\
    CA	  & Collision avoidance \\
    CSMA & Carrier sensing multiple access\\
    CTS   & Clear to send \\
    DMDS  & Distributed \blue{m}aximum depth scheduling \\
    ESaware & Energy and spectrum aware \\
    FSO   & Free space optical \\
    FTDMA & Frequency and time division multiple access \\
    LOS   & Line of sight \\
    LTE-A & Long term evolution advanced \\
    MAC   & Medium Access Control \\
    MIMO  & Multiple input multiple output \\
    mmWave & Millimeter wave \\
    MRAMAC & Multiradio assisted MAC \\
    NLOS  & Non line of sight \\
    OFDM  & Orthogonal frequency division multiplexing \\
    PAM   & Pulse amplitude modulation \\
    PNC   & Piconet coordinator \\
    PPM   & Pulse position modulation \\
    QoS   & Quality of service \\
    RA    & Random access \\
    RD    & Rate division \\
    RTDs  & Resonant tunnelling diode \\
    RTR   & Request to receiver \\
    RTS   & Request to send \\
    SDN   & Software defined network \\
    SINR  & Signal to interference and noise ratio \\
    TABMAC & Terahertz assisted beamforming MAC \\
    Tbps  & Terabits per second \\
    TC    & Transmission confirmation \\
    TCN   & Terahertz communication network \\
    TDMA  & Time division multiple access \\
    THz   & Terahertz \\
    TLAN  & Terahertz local area network \\
    TR    & Transmission request \\
    TS-OOK & Time spread On-off Keying \\
    TTS   & Test to send \\
    TPAN & Terahertz personal area network \\
    UL    & Uplink \\
    UTC-PD & Uni-travelling carrier photo diodes \\
    UV    & Ultraviolet \\
    WLAN  & Wireless local area network \\
    WPAN  & Wireless personal area network \\ \hline \hline
    \end{tabular}%
  \label{tab:acronyms}%
\end{table}%

\subsection{Terahertz Communication: Related survey articles}

Table~\ref{tab:thz_general_survey} highlights and summarise the overall survey papers on Terahertz communication without MAC layer protocols. These survey papers cover different application areas, however covers mostly the device, antenna, channel, and physical layer aspects. These include the nano-communication networks~\cite{Akyildiz2011, Akyildiz2010, Dressler2012}, Internet of nano-things \blue{(IoNT)}~\cite{Jornet2012, Akyildiz2010b}, molecular communication network~\cite{Darchini2013, Malak2012}, nano-sensor network~\cite{Akyildiz2010a}, in-body nanonetworks~\cite{Abbasi2016, Dressler2015, Malak2012}, broadband communication~\cite{Busari2018}, vehicular networks~\cite{Mumtaz2017}, wireless indoor/outdoor communications which include the office/data-center or small cells deployment~\cite{Petrov2018, Huang2011} and Terahertz Communication Networks~\cite{Han2018, Elayan2018, Boulogeorgos2018, Dat2017, Yilmaz2016, Petrov2016, Akyildiz2016, Song2011, Huang2011, Hosako2007}. Table~\ref{tab:thz_relevant_survey} highlights the survey papers which discuss the MAC layer aspects to some extent and the differences with our work.

In~\cite{Busari2018,Han2018}, the joint impact of Ultra-Dense Network, Multiple Input Multiple Output (MIMO), mmWave and Terahertz communications is discussed for supporting the demands of mobile broadband services. Particularly, the indoor and outdoor environments are analyzed for noise levels and signal to interference and noise ratio (SINR). In~\cite{Mumtaz2017}, the challenges, and opportunities are mentioned but only with the perspective of Terahertz vehicular networks. It also describes briefly different aspects including transceiver design, MIMO antenna arrays, channel modeling and estimation, interference management, MAC layer design, and standardization. For Nano Communication Networks, the survey articles are discussed in~\cite{Abbasi2016,Darchini2013,Dressler2012,Akyildiz2011,Akyildiz2010}, with physical layer aspects, propagation models, security, in-body communication, biomedical applications, materials, and antenna design, and channel modeling. The Internet of nano-things for interconnecting devices at the nanoscale is discussed in~\cite{Jornet2012,Dressler2015,Akyildiz2010b}. Although, a brief discussion on architecture, channel modeling and challenges related to MAC and Network layer are mentioned, but only for a specific scenario of the IoNT. A network architecture for wireless nano-sensor networks is given in~\cite{Akyildiz2010a}, which discusses briefly the challenges related to channel modeling, information encoding and protocols for nano-sensor networks. The molecular communication network survey is given in~\cite{Malak2012} for the body area networks. 

In~\cite{Petrov2018}, the Terahertz recent progress is reviewed but only for the propagation models, antenna and testbed design, and an implementation \blue{roadmap} \blue{is} mentioned. For opportunities beyond 5G paradigm, an architecture is discussed with possible application areas in~\cite{Elayan2018}. A survey related to MIMO is given in~\cite{Akyildiz2016}. Some standardization related work is mentioned in~\cite{Petrov2016}. Some other survey papers on the usage of Graphene material, weather impact on Terahertz bands and Terahertz antenna are mentioned in~\cite{Hasan2016,Federici2016,Wu2012}. A guest editorial is also published recently on Terahertz communication~\cite{guest-editoral2018}. \blue{In~\cite{hadi2020sensing}, Terahertz band communication importance is highlighted in terms of the interaction between the sensing, imaging, and localization. In~\cite{hadi2020signal}, a tutorial is presented on Terahertz signal processing techniques highlighting ultra massive MIMO and reconfigurable intelligent surfaces to overcome distance problem.}


With these related survey articles, there are some articles that discuss the Terahertz MAC protocols but not with required detail, as shown in Table~\ref{tab:thz_relevant_survey}. In~\cite{HAN201967}, a survey on MAC schemes for mmWave and Terahertz wireless communication is presented. The MAC protocols overall related to Terahertz band communication are not fully covered and mostly MAC strategies related to mmWave are discussed. Only few challenges and design issues are mentioned. Whereas, in this survey paper, detailed work on existing Terahertz MAC protocols with classifications, band features, design issues and considerations, application requirements and challenges are discussed.

\subsection{Contributions of this survey}

The main contributions of this survey paper are:

\begin{itemize}
\item \blue{We Provide a} comprehensive survey of existing Terahertz MAC protocols.  
\item Classification of existing Terahertz MAC protocols based on network scale and topologies, channel access mechanisms and transmitter/receiver-initiated communication \blue{has been presented}.  
\item The unique features of the Terahertz band are highlighted to be considered for Terahertz MAC protocols. 
\item The design issues are highlighted which should be considered while designing efficient Terahertz MAC protocols with decisions that should be taken at the MAC layer for performance enhancements. 
\item The requirements and design challenges for Terahertz MAC protocols for different application areas are discussed. 
\item Different challenges and future research directions are also highlighted \blue{about THz MAC protocols}.
\end{itemize}

\subsection{Organization of survey}
The paper is organized as\blue{:} Section~\ref{sec:intro} presents the introduction and literature review. Section~\ref{sec:background}, presents the background on Terahertz band, technology, and MAC protocols. In Section~\ref{sec:applications}, different applications of Terahertz band communication are discussed with respect to macro and nanoscale communication with their requirements. Section~\ref{sec:requirements}, highlights the unique features of THz band, issues that needs to be considered at Physical and MAC layer with decisions while designing an efficient Terahertz MAC protocols. Section~\ref{sec:topologies}, mentions the topologies so far focused on Terahertz communication networks. Different channel access mechanisms are discussed in Section~\ref{sec:ch-access}. The transmitter and receiver-initiated communication are discussed in Section~\ref{sec:communication}. \blue{C}hallenges and future research directions are discussed in Section~\ref{sec:issues-challenges}. Finally, in Section~\ref{sec:conclusion}, the survey paper is concluded.

\begin{table*}[htbp]
  \centering
  \small
  \caption{General survey papers on Terahertz bands, devices, and communications.}
    \begin{tabular}{p{2.4em}p{3.4em}p{15em}p{30em}}
     \textbf{Year} & \textbf{Reference} & \textbf{Network Area/Type} & \textbf{Brief description of main topics covered} \\ \hline \hline
     
     2004 & \cite{fitch2004} & Terahertz Communication Network & An overview of communication and sensing applications is given with sources, detectors, and modulators for practical Terahertz Communication systems. \\ \hline      
     
     2007 & \cite{Hosako2007} & Terahertz Communication Network  &  The developments in the fields like Terahertz quantum cascade lasers, quantum well photodetectors, time-domain spectroscopy system and materials are discussed with measurements of atmospheric propagation.   \\  \hline
     
     2010 & \cite{Akyildiz2010} & Nano Communication Network  & Propagation models for molecular and nano-electromagnetic communications are discussed with challenges. \\ \hline
     
    \multirow{2}{*}{2011} & \cite{Huang2011}    & Terahertz Communication Network &  Different aspects of Terahertz communication are discussed with transistors, mixers, antennas, and detectors. \\
    	  				  &  \cite{Song2011}     & Terahertz Communication Network 	& Progress on Terahertz wave technologies is discussed. \\ \hline
     
    \multirow{5}{*}{2012} 	& \cite{Wu2012}  & Terahertz Communication Network & Terahertz antenna technologies are discussed with different substrate integrated antennas and beamforming networks.	  \\ 
     						& \cite{Dressler2012}  & Nano Communication Network  & An overview on biochemical cryptography is discussed with requirements related to security and challenges. \\
     						& \cite{kuner2012tftcs} & Terahertz Communication Network & An overview on demonstration of data transmission is given with standardization activities.\\
     						& \cite{truththz2012} & Terahertz Communication Network & The Terahertz technology is discussed with challenges for spectroscopy and communications. \\
     						& \cite{Malak2012}  	& Molecular communication networks & The elementary models for intra body molecular communication channel and their extensions are discussed with challenges. \\  \hline   
     
     \multirow{2}{*}{2013} & \cite{Darchini2013} & Nano Communication Network  & Issues of Nanonetworks are analyzed and discussed with particular focus on communication via microtubules and physical contact.  \\
     						 & \cite{sasib2013} & Molecular Communication Network &  A review on bacterial communication and neuronal networks are given with application areas in body area networks.  \\ \hline
     
    			2014		&	\cite{Kurner2014}  & Terahertz Communication Network & Summarizes the research projects, spectrum regulation, and standardization effort for the Terahertz band. \\ \hline
     
     \multirow{2}{*}{2015} & \cite{Dressler2015} & Internet of Nano-Things   & A survey is presented for connecting body area networks and external gateway for in-body nano communication. Network architecture, requirements, and simulation based performance evaluation are also discussed. \\  
          					& \cite{ahirata2015} & Terahertz Communication Network & A survey on Terahertz technology is presented including devices, antennas, and standardization efforts.  \\ \hline
          					
     \multirow{2}{*}{2016} 		&  \cite{Federici2016} 	& Terahertz Communication Network &	 A review is presented for impact of weather on Terahertz links, attenuation, and channel impairments caused by atmospheric gases like water vapor, dust, fog, and rain.  \\
     							&  \cite{Hasan2016}  	& Terahertz Communication Network &	A survey on \blue{G}raphene based devices for modulation, detection, and generation of Terahertz waves is discussed.   \\ \hline
        

     \multirow{3}{*}{2018} 	& \cite{Han2018}  & Terahertz Communication Network  & A review is presented for channel modelling for Terahertz band including  single antenna and ultra massive MIMO systems. \\
     						& \cite{nabilkhalid2018} & 5G Femtocell Internet of Things & A survey on low Terahertz band circuit blocks is presented with focus on energy consumption using best modulation schemes and optimizing hardware parameters. \\
     	 					& \cite{Elayan2018} & Terahertz Communication Network   & A review is presented for deployments of Terahertz wireless link and opportunities to meet future communication requirements. \\  \hline

	\multirow{1}{*}{2019} 	& \cite{lemic2019survey}  & Nano Communication Network  & \saim{A summary of current status of Nano communication network is presented with applications and different layers of protocol stack}. \\

\multirow{1}{*}{2019} 	& \cite{tsrapaport2019app}  & Terahertz Communication Network  & \saim{Discusses wireless communications and applications above 100 GHz bands with opportunities and challenges. } \\

\multirow{1}{*}{2019} 	& \cite{elayanhad2019}  & Terahertz Communication Network  & \saim{Recent activities on Terahertz development, standardization, applications and communications are reported.} \\

\multirow{1}{*}{2019} 	& \cite{Kursat2019}  & Terahertz Communication Network  & \saim{Discusses wireless communications and applications above 100 GHz bands with opportunities and challenges. } \\

\multirow{1}{*}{2019} 	& \cite{kko2019}  & Terahertz Communication Network  & \saim{A survey on Terahertz communications, applications and layers of protocol stack is presented. } \\

\multirow{1}{*}{2019} 	& \cite{kmshuq2019}  & Terahertz Communication Network  & \saim{A survey on Terahertz communications, applications and layers of protocol stack is presented. } \\
	
	\multirow{1}{*}{2019} 	& \cite{zchen2019}  & Terahertz Communication Network  & A review is presented on development towards Terahertz communications with key technologies. \\  \hline \hline         	 					
     	 									
    \end{tabular}%
  \label{tab:thz_general_survey}%
\end{table*}%

\begin{table*}[htbp]
  \centering
  \tiny
  \caption{\mage{Survey papers discussing the Terahertz MAC layer.}}
    \begin{tabular}{|p{9.78em}|c|c|p{3.39em}|p{1.445em}|p{1.445em}|p{2.61em}|p{2.61em}|p{2.61em}|p{1.445em}|p{1.445em}|p{1.445em}|p{1.445em}|p{1.335em}|p{1.445em}|p{1.445em}|p{1.445em}|p{1.445em}|p{1.445em}|p{1.445em}|}
    \toprule
    \multicolumn{6}{|c|}{}                        & \multicolumn{3}{p{7.83em}|}{\textbf{MAC protocols classification}} & \multicolumn{11}{p{18.395em}|}{\textbf{MAC functionalities}} \\
    \midrule
    \textbf{Network type} & \multicolumn{1}{p{2.11em}|}{\textbf{Year }} & \multicolumn{1}{p{5.72em}|}{\textbf{Reference}} & \begin{sideways}\textbf{MAC layer discussed}\end{sideways} & \begin{sideways}\textbf{Application areas}\end{sideways} & \begin{sideways}\textbf{Challenges}\end{sideways} & \begin{sideways}\textbf{Network topologies and scale}\end{sideways} & \begin{sideways}\textbf{Rx/Tx Initiatied communication}\end{sideways} & \begin{sideways}\textbf{Channel access/sharing}\end{sideways} & \begin{sideways}\textbf{Interference}\end{sideways} & \begin{sideways}\textbf{Error control}\end{sideways} & \begin{sideways}\textbf{Packet size}\end{sideways} & \begin{sideways}\textbf{Device discovery}\end{sideways} & \begin{sideways}\textbf{Handshaking}\end{sideways} & \begin{sideways}\textbf{Tx distance}\end{sideways} & \begin{sideways}\textbf{Data rate}\end{sideways} & \begin{sideways}\textbf{Antennas}\end{sideways} & \begin{sideways}\textbf{Beamforming}\end{sideways} & \begin{sideways}\textbf{Modulation}\end{sideways} & \begin{sideways}\textbf{Cross layer}\end{sideways} \\
    \midrule
    \multirow{9}[2]{*}{\textbf{Nanonetworks}} & 2008  & \multicolumn{1}{p{5.72em}|}{\cite{nanoparadigm2008}} & Partially & \checkmark     & \checkmark   & X     & X     & X   & X     & X     & X     & X   & X     & X     & X     & X     & X     & X     & X      \\
\multicolumn{1}{|c|}{} & 2010  & \multicolumn{1}{p{5.72em}|}{\textcolor[rgb]{ .376,  .376,  .376}{\cite{Akyildiz2010b}}} & Partially & X     & \checkmark   & X     & X     & \checkmark   & X     & X     & X     & \checkmark   & X     & X     & X     & X     & X     & X     & X     \\    
    \multicolumn{1}{|c|}{} & 2010  & \multicolumn{1}{p{5.72em}|}{\textcolor[rgb]{ .376,  .376,  .376}{\cite{Akyildiz2010a}}} & Partially & \checkmark   & \checkmark   & X     & X     & \checkmark   & X     & X     & X     & X     & X     & X     & X     & X     & X     & \checkmark   & X      \\
 \multicolumn{1}{|c|}{} & 2010  & \multicolumn{1}{p{5.72em}|}{\cite{akyljornnano2010}} & Partially & \checkmark     & \checkmark   & X     & X     & X     & X     & X     & X     & X     & X     & X     & X     & X     & X     & X     & X      \\    
    \multicolumn{1}{|c|}{} & 2011  & \multicolumn{1}{p{5.72em}|}{\textcolor[rgb]{ .376,  .376,  .376}{\cite{Akyildiz2011}}} & Partially & X     & \checkmark   & X     & X     & X     & X     & X     & X     & X     & X     & X     & X     & X     & X     & X     & X     \\
    \multicolumn{1}{|c|}{} & 2012  & \multicolumn{1}{p{5.72em}|}{\cite{imntthz2012}} & Partially & X     & \checkmark   & X     & X     & X     & X     & X     & X     & \checkmark   & X     & \checkmark   & \checkmark   & \checkmark   & X     & \checkmark   & \checkmark    \\
    \multicolumn{1}{|c|}{} & 2012  & \multicolumn{1}{p{5.72em}|}{\cite{Jornet2012}} & Partially & \checkmark   & \checkmark   & X     & X     & X     & X     & X     & X     & \checkmark   & X     & \checkmark   & \checkmark   & \checkmark   & X     & \checkmark   & \checkmark    \\
    \multicolumn{1}{|c|}{} & 2016  & \multicolumn{1}{p{5.72em}|}{\cite{Abbasi2016}} & Partially & \checkmark   & \checkmark   & X     & X     & X     & X     & \checkmark   & X     & X     & X     & X     & X     & X     & X     & X     & X      \\
    \multicolumn{1}{|c|}{} & 2017  & \multicolumn{1}{p{5.72em}|}{\cite{sivapriya2017}} & Partially & X     & \checkmark   & X     & X     & X     & X     & X     & X     & X     & X     & X     & X     & X     & X     & X     & X     \\   
    \midrule
    \textbf{Vehicular networks} & 2017  & \multicolumn{1}{p{5.72em}|}{\cite{Mumtaz2017}} & Partially & X     & X     & X     & X     & X     & X     & X     & X     & X     & X     & X     & X     & X     & X     & X     & X     \\
    \midrule
    \multirow{4}[2]{*}{\textbf{Terahertz networks}} & 2014  & \multicolumn{1}{p{5.72em}|}{\cite{Akyildiz2014}} & Partially & \checkmark   & \checkmark   & X     & X     & X     & \checkmark   & \checkmark   & X     & X     & X     & X     & X     & X     & X     & X     & X      \\
    \multicolumn{1}{|c|}{} & 2014  & \multicolumn{1}{p{5.72em}|}{\cite{teranet2014}} & Partially & X     & \checkmark   & X     & X     & X     & X     & \checkmark   & \checkmark   & X     & X     & X     & X     & X     & \checkmark   & X     & X     \\
    \multicolumn{1}{|c|}{} & 2016  & \multicolumn{1}{p{5.72em}|}{\cite{Akyildiz2016}} & Partially & X     & \checkmark   & X     & X     & X     & \checkmark   & X     & X     & X     & X     & X     & X     & \checkmark   & \checkmark   & X     & X     \\
    \multicolumn{1}{|c|}{} & 2016  & \multicolumn{1}{p{5.72em}|}{\cite{Petrov2016}} & Partially & \checkmark   & \checkmark   & X     & X     & X     & X     & X     & X     & X     & X     & X     & X     & \checkmark   & \checkmark   & X     & X      \\
   
    \multicolumn{1}{|c|}{} & 2019  & \multicolumn{1}{p{5.72em}|}{\cite{HAN201967}} & Partially & X   & \checkmark   & \checkmark     & X     & \checkmark     & X     & \checkmark     & X     & X     & X     & X     & X     & \checkmark   & \checkmark   & X     & X      \\
     \midrule
    
    \textbf{This work} & 2019  &       & Detailed & \checkmark   & \checkmark   & \checkmark   & \checkmark   & \checkmark   & \checkmark   & \checkmark   & \checkmark   & \checkmark   & \checkmark   & \checkmark   & \checkmark   & \checkmark   & \checkmark   & \checkmark   & \checkmark   \\
    \bottomrule
    \end{tabular}%
  \label{tab:thz_relevant_survey}%
\end{table*}%

\section{\mage{Background on Terahertz band, Technology and MAC protocols}}\label{sec:background}

\begin{figure}
\centering
\includegraphics[width=3.2in,height=1.2in]{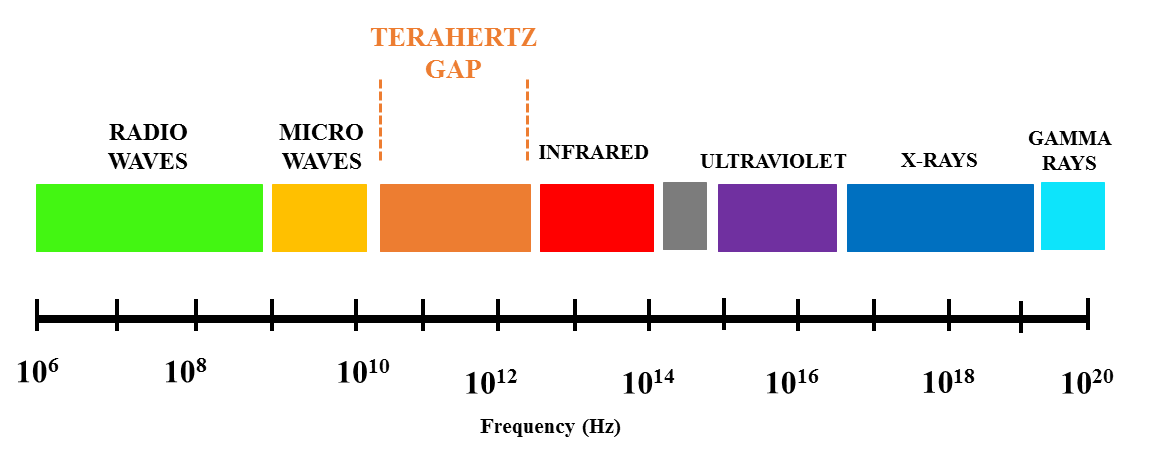}
\caption{Terahertz gap in the electromagnetic spectrum.}
\label{fig:electromagnetic-spectrum}
\end{figure}

\subsection{Terahertz bands}

The Terahertz can be termed as a unit of frequency (one trillion cycles per second or $10^{12}$ Hz) or electromagnetic waves within ITU designated band of frequencies. The Terahertz frequency range (0.1 - 10 THz) is the last span within the whole electromagnetic wave spectrum and is more commonly referred to as Terahertz Gap. They appear between the Microwave and Infrared bands, as shown in Figure~\ref{fig:electromagnetic-spectrum}. The wavelength of radiation in the Terahertz band range from 1 mm to 0.1 mm (or 100 $\mu$m). The bands from 100 GHz to 200 GHz are also referred as sub-Terahertz band~\cite{tsrapaport2019app}, as it begins at a wavelength of one millimeter and proceeds into shorter wavelengths. 

For nearly two decades, the Terahertz bands are been efficiently used for imaging applications because these waves are non-ionizing and able to penetrate through materials and absorbed by water and organic substances. Their properties allow them to be used in communication networks to provide higher data rates up to Tbps. The Terahertz Gap is still the least explored band for its potential use in communication networks and to achieve higher data rates. Table~\ref{tab:bands-features}, enlist the features of different frequency bands closest to Terahertz frequency bands. Its unique potentials motivate its usage for broadband wireless communications.

\subsection{Comparison between Terahertz band and other wireless technologies}  \label{sec-bkg:trad-comm}

A brief comparison between existing wireless communication technologies and Terahertz band communication including mmWave communication is presented below and shown in Table~\ref{tab:bands-features}.

\subsubsection{\saim{Comparison with mmWave band communications}}

The mmWave bands comprises frequencies from 30 GHz to 300 GHz. Due to higher frequencies\blue{,} these bands face severe attenuation due to oxygen absorption. Exceptionally, the propagation on 35 GHz, 94 GHz, 140 GHz, and 220 GHz experience relatively small attenuation which enables long-distance communications between peers~\cite{xwang2018}. Other bands like 60 GHz, 120 GHz, and 180 GHz attenuates up to 15dB/km and also experience poor diffraction due to blockages~\cite{xwang2018}. The range of mmWave and Terahertz can be reduced by path loss, molecular absorption and atmospheric attenuation. The channeling effect can be compensated by using high directional antenna for which mmWave is more mature than Terahertz band. Additional features such as modulation and coding schemes, massive MIMO and phased antenna can enhance the spectral efficiency and transmission range \blue{for THz}. An advantage of Terahertz frequency over mmWave is that communication windows for Terahertz wave are higher than for mmWave, and because of \blue{this} Terahertz frequencies seem to be more suitable for high data rate and low range communication.

\saim{The Terahertz and mmWave are neighboring bands but their properties are different. In comparison, the bandwidth at mmWave band is 10 GHz, which cannot support Tbps link speed, whereas the Terahertz band has distance varying transmission windows of up to Terahertz bandwidth. To reach the data rate up to 100 Gbps, the transmission schemes must reach the challenging spectral efficiency of 14 \blue{bits per second per hertz}~\cite{Kurner2014}. Moreover\blue{,} the link capacity required for few Gbps should be several times higher than the user required data rate for the timely delivery of data from multiple users. With increasing frequencies to Terahertz band, the Tbps links can attain moderate and realistic spectral efficiency of few bits per second per hertz. The Terahertz band can also allow higher link directionality compared to mmWave at same Transmitter aperture due to their less free space diffraction and shorter wavelength compared to mmWave. The transmitted power and interference between the antennas can also be reduced by using smaller antennas with good directivity in Terahertz communications~\cite{Ma2015TerahertzWC}. In Terahertz bands, the eavesdropping chances are also lower compared to mmWave band due to high directionality of Terahertz beams, in which the unauthorized users must also be on the same narrow beam to intercept messages.}

\saim{The difference between mmWave and Terahertz bands are summarised in Table~\ref{mmwave-thz-diff}. From a technical point of view, Terahertz band offers much higher bandwidth than mmWave band and therefore high data rate. The mmWave is deployed actually in WLAN as well as cellular network system by contrast to THz where researchers are striving to design new devices with high power generation and low receiving sensitivity. Due to the larger coverage and mobility support, mmWave communications can be used in backhaul communication and cellular communications. Whereas Terahertz can be used where high throughput and low latency are required with fixed infrastructure for now until mature devices will be available. MmWave antenna is much more mature than for THz, \blue{thus} it is possible to deploy antenna diversity and beam steering and tracking for mmWave. The channel model for mmWave is well developed as measurements are carried for many mmWave windows as well as for different scenarios. For Terahertz band few measurement campaigns are performed particularly for indoor scenarios around 300 GHz and 100 GHz. The free space attenuation increases as a function of frequency and molecular absorption loss occur due to oxygen molecules in mmWave, whereas in Terahertz band\blue{,} it occurs due to water vapors. The reflection loss is high for both mmWave and Terahertz band which results in severe loss of NLOS path \saim{compared} to LOS path. The scattering effect also becomes severe when the wavelength decreases below 1 mm which results in increase of multipath components, angular spreads, and delay. Due to much smaller wavelength many antennas can be packaged together to generate narrower beams. However, the stronger directivity increases the difficulties and overhead of beam alignment and tracking but reduces the interference. The difference between mmWave and Terahertz band communication can also be found in~\cite{B5G2018,HAN201967} for further reading. \blue{A large scale open source testbed for Terahertz and mmWave is in~\cite{polesetestbed2019}.}}
 

\subsubsection{\saim{Other technologies}}
The traditional 802.11 protocol is mainly designed for 2.4 GHz WiFi, which uses frequency-hopping spread spectrum and direct sequence spread spectrum. It provides a simple data rate of up to 2 Mbps. After that 802.11 (a and b) were published, operating at 5 and 2.4 GHz bands. The 802.11a is based on Orthogonal Frequency Division Multiplexing (OFDM) and can provide a data rate of up to 54 Mbps, whereas 802.11b supports only 11 Mbps. The 802.11ac aimed at providing the data rate up to more than 100 Mbps. Other than those, the 802.11ad is developed for the carrier frequency of 60 GHz and it belongs to mmWave frequency bands. The detailed description and comparison are provided in Table~\ref{tab:bands-features}. \blue{S}mart technologies like OFDM and communication schemes like large-scale MIMO can be used for frequencies below 5 GHz to achieve higher spectral efficiency. In Long-Term Evaluation Advanced (LTE-A), the peak data of up to 1 Gbps is possible only when a 4x4 MIMO scheme is used over a 100 MHz aggregated bandwidth~\cite{akylidiz-lte2014}.

The frequency bands above 10 Terahertz cannot support Tbps links. Although very large bandwidth is available in FSO communication system which operates at IR frequencies, it still holds some issues which limit its use for personal wireless communication like the atmospheric effects on the signal propagation (fog, rain, pollution and dust); high reflection loss; misalignment between transmitter and receiver; and low power link budget due to health safety which limits both transmission range and achievable data rates for FSO communication. It can support up to 10 Gbps of data rate with a proper line of sight (LOS) for Wireless Local Area Network (WLAN)~\cite{Glushko2013}. For non-LOS\blue{,} much lower data rate has been reported~\cite{xli2012}. For longer distance, an FSO system was demonstrated in~\cite{eciaramella2009} to support 1.28 Tbps, however, requires typical fiber optical solution to generate and detect high capacity optical signals which are injected in the optical front-end and also does not include the signal generation, detection, modulation, and demodulation blocks. These constraints limit the overall feasibility to achieve higher data rates for 5G and beyond networks. A comparison of wireless and optical technologies is presented in~\cite{hamza2016} for wireless indoor environments. It is mentioned in~\cite{hamza2016}, wireless communication has overall better chances for penetration through obstacles in comparison with FSOs. \blue{In~\cite{Taherkhan2020}, it is shown that with increasing gain the BER performance of Terahertz links is much better than the FSO link.} Further, the IR and ultraviolet are not considered as safe for human and the Visible Light communication requires the visibility of light at all times. \saim{A comparison of visible light and sub-Terahertz band communication is discussed in~\cite{calvanese2019} for 5G and 6G wireless systems.}

\begin{table*}[htbp]
  \centering
  \scriptsize
  \caption{Different types and features of wireless communication technologies\protect\cite{Elayan2018}.}
    \begin{tabular}{p{4.7em}p{4.7em}p{5em}p{12em}p{8em}p{5em}p{4.5em}cp{4.5em}p{2.5em}}
    \textbf{Technology} & \textbf{Frequency range} & \textbf{Wavelength} & \multicolumn{1}{p{11.335em}}{\textbf{Data rate}} & \textbf{Transmission range} & \textbf{Power consumption} & \textbf{Topology} & \multicolumn{1}{p{4.78em}}{\textbf{LOS/nLOS}} & \textbf{Noise source} & \textbf{Weather effect} \\ \hline \hline
    \textbf{mmWave} & 30GHz-300GHz & 3cm-1mm & \multicolumn{1}{p{11.335em}}{\saim{100 Gbps~\cite{zhnagh2010,Yongsu2006,Kurner2014}}} & \saim{High} & Medium & PTP, PTmP & \multicolumn{1}{p{4.78em}}{both} & Thermal & Sensitive \\ \hline
    \multirow{6}{*}{\textbf{Terahertz}} & \multicolumn{1}{p{4.8em}}{100-300 GHz (sub-THz)\newline{}300GHz-10THz} & \multirow{6}{*}{1mm-30$\mu$m} & \multicolumn{1}{p{12em}}{upto 240 GHZ: 10 Gbps~\cite{minoru2015}\newline{}upto 300 GHz: 64 Gbps~\cite{ingmar2015}\newline{}300-500 GHz: 160 Gbps (single channel)~\cite{xyu2016}\newline{}300-500GHz: > 160 Gbps (multiple channels)~\cite{xpang2016} } & \multirow{6}{*}{Short/Medium (1-10m)}  & \multirow{6}{*}{Medium} & \multirow{6}{*}{PTP, PTmP} &  \multirow{6}{*}{both} & \multirow{6}{4em}{Thermal and\newline{} molecular noise} & \multirow{6}{*}{Sensitive} \\ \hline
    \textbf{Infrared} & 10THz-430THz & 30$\mu$m-3 $\mu$m & \multicolumn{1}{p{11.335em}}{2.4 kbps to 1 Gbps} & Short, upto 1 m & Low   & PTP   & \multicolumn{1}{p{4.78em}}{LOS} & Sun/Ambient & Sensitive \\ \hline
    \textbf{Visible Light} & 430THz-790THz & 0.3$\mu$m & \multicolumn{1}{p{11.335em}}{100 Mbps to Gbps~\cite{phpathak2015,sraja2012}} & Short & Low   & PTP   & \multicolumn{1}{p{4.78em}}{\saim{both}} & Sun/Ambient & \multicolumn{1}{c}{} \\ \hline
    \textbf{Ultra Violet} & 790THz-30PHz & 100-400 nm &       & Short & Low   & PTmP  &       & Sun/Ambient & Sensitive \\ \hline \hline
    \end{tabular}%
  \label{tab:bands-features}%
\end{table*}%

\subsection{\mage{Background and motivation for Terahertz MAC protocols}}
 In this Section, \blue{we present} the definition of MAC layer protocols with motivation for the need of different MAC protocols for Terahertz communication networks is presented.

\subsubsection{\mage{Medium Access Control background}}
The MAC layer is mainly responsible for controlling the hardware to interact with wireless transmission medium by flow control and multiplexing. It serves the interaction between the Physical and Upper layers. It provides a method to identify the frame errors and sends the packet to \blue{the} Physical layer when channel access is permitted or scheduled. It also decides when and how much to wait before sending a packet to avoid collisions among the nodes. Different wireless technologies require different MAC protocols to serve the transmission purpose. For example, the MAC protocols for LTE and GSM standard has different user requirements \blue{compared to a MAC protocol required for Wireless sensor network.}

\mage{As the user demands and network requirement are enhancing, efficient MAC protocols are in demand to assist the network operations and provide adaptive solutions. The MAC although suppose to provide same functionalities mentioned above, but due to different band features and requirements (cf. Section~\ref{sec:requirements}), new mechanisms are required to facilitate user demands and networks requirements (cf. Section~\ref{sec:applications}).}


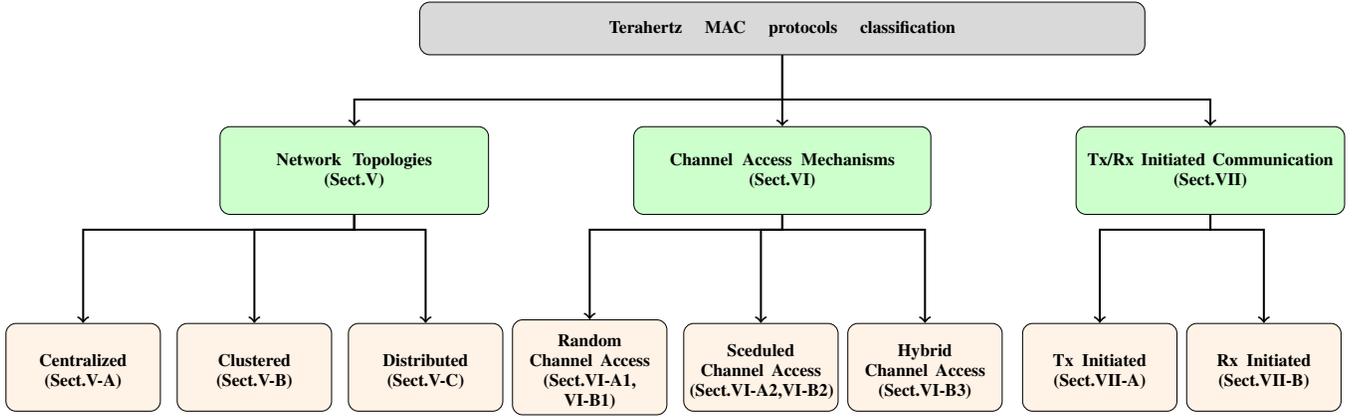
\begin {figure*}[!hbtp]
\centering
\scriptsize
\begin{adjustbox}{width=\linewidth}

\tikzstyle{block} = [rectangle, draw, text width=4cm, text centered, rounded corners, minimum height=5em,fill=blue!20]
\tikzstyle{line} = [draw,thick, -latex']
\tikzstyle{cloud} = [draw, ellipse, text width=4.5cm, text centered]
\tikzstyle{edge from parent}=[->,thick,draw]
\begin{tikzpicture}[auto,edge from parent fork down]
\tikzstyle{level 1}=[sibling distance=60mm,level distance=18ex]
\tikzstyle{level 2}=[sibling distance=25mm,level distance=15ex]
\tikzstyle{level 3}=[sibling distance=30mm,level distance=34ex]
\tikzstyle{level 4}=[sibling distance=35mm,level distance=34ex]
\node [block,text width=10cm,minimum height=3em,,fill=black!15] (cst) {\textbf{Terahertz MAC protocols classification  }}
{
child{node [block,text width=3.6cm,fill=green!20] (rnm) {\textbf{Network Topologies \\ (Sect.\ref{sec:topologies})   } }
		child{node [block,text width=2cm,fill=orange!10,xshift=-1.3cm, yshift=-1.1cm] (pmt) {      \\ 
  		{\bf Centralized  \\ (Sect.\ref{subsec-topo:centralized})  } \\
  		}} 
  	    child{node [block,text width=2cm,fill=orange!10,xshift=-1.4cm, yshift=-1.1cm] (pmt) {      \\ 
  		{\bf Clustered \\ (Sect.\ref{subsec-topo:clustered})  } \\ 
  		}}
  		child{node [block,text width=2cm,fill=orange!10,xshift=-1.5cm, yshift=-1.1cm] (pmt) {      \\ 
  		{\bf Distributed \\ (Sect.\ref{subsec-topo:distributed})  } \\ 
  		}}  
}
child{node [block,text width=4cm,fill=green!20] (opm) {\textbf{Channel Access Mechanisms \\ (Sect.\ref{sec:ch-access})  } }
		child{node [block,text width=2cm,fill=orange!10,xshift=-0.2cm, yshift = -1.1cm] (pmt) {  \\
		{\bf Random Channel Access \\ (Sect.\ref{subsubsec-ch:nano-random}, \ref{subsubsec-ch:macro-random})  } \\
		}}
		child{node [block,text width=2cm,fill=orange!10,xshift=-0.3cm, yshift = -1.1cm] (pmt) {  \\
		{\bf Sceduled Channel Access \\ (Sect.\ref{subsubsec-ch:nano-scheduled},\ref{subsubsec-ch:macro-scheduled})  } \\
		}}
		child{node [block,text width=2cm,fill=orange!10,xshift=-0.5cm, yshift = -1.1cm] (pmt) {  \\
		{\bf Hybrid Channel Access \\ (Sect.\ref{subsubsec-ch:macro-hybrid})  } \\
		}}		
}
child{node [block,text width=3.6cm,fill=green!20] (rnm) {\textbf{Tx/Rx Initiated Communication \\ (Sect.\ref{sec:communication}) } }
		child{node [block,text width=2cm,fill=orange!10,xshift=-0.3cm, yshift=-1.1cm] (pmt) {      \\ 
  		{\bf Tx Initiated  \\ (Sect.\ref{subsec-comm:tx})  } \\
  		}} 
  	    child{node [block,text width=2cm,fill=orange!10,xshift=-0.5cm, yshift=-1.1cm] (pmt) {      \\ 
  		{\bf Rx Initiated \\ (Sect.\ref{subsec-comm:rx})  } \\ 
  		}} 
}
};
\end{tikzpicture}
\end{adjustbox}
\caption{\protect \mage{Classification of Terahertz MAC protocols based on network topologies, channel access mechanisms and Tx/Rx initiated communication. They are further discussed based on Terhaertz applications and network scale as macro and nano.}}
\label{fig:classification}
\end{figure*}

\subsubsection{\mage{Review of various Terahertz MAC protocols}}

\blue{The existing THz MAC protocols can be broadly classified, as shown in Figure~\ref{fig:classification}.} 

The network topologies \blue{for} MAC protocol design are discussed in Section~\ref{sec:topologies}. In centralized networks, a controller or an access point is used to provide coordination among the nodes, whereas in distributed networks each node takes its own decision and coordinates in a distributed manner. In cluster-based networks, nodes form a group to transmit their information via a single group head. 

\mage{The channel access mechanisms are presented in Section~\ref{sec:ch-access} and they include mainly random, scheduled and hybrid channel access mechanisms. In random channel access, each node contends for a shared channel to improve channel utilization and can introduce collisions. The directional antenna usage in Terahertz frequency band in carrier sense based multiple access can provide spatial reuse of spectrum which can improve the throughput of the network. Whereas, the schedule based channel access methods like TDMA assigns channel access slots in advance without spatial reuse. This provides a contention-free environment at the cost of high synchronization overhead. In~\cite{Ghadikolaei2015}, the scheduled based mechanism is shown to provide worse throughput and latency performance compared to random schemes. The hybrid channel access mechanism combines the functionalities of random and scheduled channel access mechanisms. In this scheme, the channel access is performed using random scheme followed by scheduled data transmission. }

\mage{The initial access mechanisms are presented in Section~\ref{sec:communication}. The unique features of the Terahertz band (discussed in Section~\ref{sec-req:features}) adds different challenges for Terahertz wireless communications like directional antenna and high path loss. These unique features, network scale, and application requirements demand new initial access mechanisms. The receiver initiated communication is mainly followed in networks with severe energy limitation, in which a receiver triggers the communication when it has sufficient energy to receive a packet. Whereas, transmitter initiated communication is mostly followed in wireless communication, in which a sender when having some data to transmit triggers the communication. }

\mage{Moreover, the Terahertz MAC protocols can also be classified based on applications as nano and macro scale applications. These applications are discussed and their design challenges are identified in Section~\ref{sec:applications}. The nano applications involve applications with range up to a few centimeters, whereas applications with range higher than a meter can be considered as macro scale applications. }

\subsubsection{\mage{Difference between Terahertz MAC protocols and protocols for other communications}}

Terahertz is characterized by high bandwidth comparing to lower frequency band. \blue{I}t is also characterized by its high-frequency attenuation. Therefore, traditional concepts of MAC should be modified or extended in most cases to \blue{consider} application requirements such as ultra high throughput, coverage and low latency. The Terahertz communication allows concurrent narrow beam links as compared to the traditional networks. The unique features of the Terahertz band, and Terahertz MAC protocol design considerations are discussed in detail in Section~\ref{sec:requirements}. 

\mage{\textit{Link preparation for access, ACK and Data: } The traditional WiFi networks use in-band data and ACKs with short inter-frame spacing gap of 20 ms between data and ACK message. In Terahertz using the same link for ACK and data involves many challenges including, beam switching and steering, and additional synchronization overhead, which can increase the delay.}

\mage{\textit{Deafness issue: } The traditional networks operates in both omni and directional antenna modes, whereas in Terahertz networks directional antennas are required at both the sender and receiver, which can introduce deafness and collision problem. Because the nodes can sense only in one direction at a time which makes it difficult to capture accurate and timely information of their neighbors. }

\textit{Noise and interference :} Terahertz band is sensitive to thermal and molecular noise, interferences come generally from receiver side lobe and interferers beam alignment or from multipath reflections. It can be neglected if directional antennas are used, but still can reduce system performances, mmWave is much more mature regarding interference and noise modeling, techniques to reduce impact on the system are available using adequate equalizers and filters as well as high-efficiency coding schemes. From MAC point of view designing efficient scheduling algorithm can improve system performance.

\textit{Antenna directionality and beam management: } In traditional networks mostly Omni-directional antennas are used. Although directional antennas are used for point-to-point connectivity, they cover high transmission distance. However, using directional antennas for Terahertz band, introduce many challenges including beam management for initial access and data transmission, frequently than in traditional networks. Fast beam tracking mechanisms are required mainly at Terahertz band involving applications with mobility.

\textit{Frame length and duration: } Long frame can increase throughput however frame error rate can increase, and an efficient error control mechanism should be used to mitigate frame losses. For Terahertz system, the frame duration is very low compared to microwave and mmWave system. The system can accept more users without affecting overall delay\blue{. F}or example, the scheduling time for LTE at microwave frequency is equal to 1 ms whereas for Terahertz it can be much lower, and more nodes can join system within 1 ms, if we consider it as a threshold for network delay. In order to increase nodes in the system, fast beam switching and steering is then required. For frame size, as Terahertz system is still immature, frames should be of low size, as they are prone to channel errors. Therefore\blue{,} robust channel coding should be implemented to detect and correct bits and reduce retransmissions and delays. Comparing to traditional techniques, it is possible to re-use traditional mechanisms for frame error handling such as frame retransmission.

\textit{Channel access and huge bandwidth availability: } \blue{T}he bandwidth available for tradition networks, as discussed in previous subsection, is few MHz or up to a few GHz in case of mmWave band communication. The huge bandwidth availability means less contention and collision at these bands with low transmission times. The main functionalities required at Terahertz band are coordination and scheduling, whereas in traditional networks contention and interference management are main requirements for channel access. Therefore, TDMA with highly directional antennas can be a good choice for Terahertz networks. 

\mage{\textit{High energy fluctuations: } The heavy energy fluctuations at nanoscale networks require efficient energy harvesting mechanisms in support of MAC layer to perform handshaking and data transmissions. Whereas in traditional networks\blue{,} energy fluctuations are not as high as in nanonetworks. }

\mage{\textit{Scheduling: } For macro network, MAC is based on time division techniques in most case as communication can occur only if beams are aligned between nodes, the scheduling is a crucial part of the THz MAC procedure that makes THz macro networks different from traditional MAC techniques.}

\mage{\textit{Throughput and latency requirements: } Traditional networks offer a limited amount of throughput, as discussed in previous subsection. Higher throughput is possible by using MIMO and higher-order modulation techniques, whereas, in Terahertz networks, low complexity modulation techniques are required for nanoscale networks. The latency requirements are also very tight in Terahertz networks.}

\begin{table*}[htbp]
  \centering
  \footnotesize
  \caption{\protect \mage{Comparison between mmWave and Terahertz wave technologies.}}
\begin{tabular}{|p{3.5cm}|p{5.2cm}|p{5.2cm}|}
\hline
           & {\bf \saim{Milimeter wave technology}} & {\bf \saim{Terahertz wave technology}} \\
\hline
\saim{Transceivers Device} & \saim{High performance mmWave devices Available~\cite{marnat2017}} & \saim{Available immature: UTC-PD, RTD and SBD~\cite{Akyildiz2014}} \\
\hline
\saim{Modulation and coding} & \saim{High order modulation techniques available~\cite{roh2014}}. \saim{For example, QPSK, LDPC 1/2 (264 Mbps)
16QAM, LDPC 1/2 (528 Mbps)} & \saim{Low order modulation techniques (OOK, QPSK)~\cite{Akyildiz2014}, LDPC, Reed Soloman, Hamming} \\
\hline
   \saim{Antenna} & \saim{Omni and high gain directional, MIMO supported~\cite{zhnagh2010}, antenna gain =18dBi when 8X8 antenna array used~\cite{roh2014}} & \saim{Omni and Directional, phased array with low number of antenna elements (4x1)~\cite{jalili2018}} \\
\hline
 \saim{Bandwidth} & \saim{7GHz @60GHz~\cite{tsrapaport2019app}} & \saim{69 GHz at 300GHz} \\
\hline
\saim{channel models~\cite{hemadeh2018}} &        \saim{Yes} &  \saim{Partially} \\
\hline
 \saim{Data rate} & \saim{Maximum of 100Gbps~\cite{Yongsu2006,zhnagh2010}} & \saim{100 Gbps~\cite{chinni2018,elayanhad2019} to few Tbps (theoretical)} \\
\hline
 \saim{Standards} & \saim{5G NR, IEEE 802.11 ad, IEEE 802.11 ay} & \saim{IEEE 802.15.3d} \\
\hline
  \saim{Mobility} & \saim{Supported~\cite{rupasinghe2016}} &    \saim{Not yet} \\
\hline
\saim{Beam management~\cite{giordani2019}} &       \saim{ Yes} &        \saim{ No} \\
\hline
\saim{Adaptive beam searching and switching time} & \saim{45 ms~\cite{roh2014}} & \saim{in progress} \\
\hline
\saim{Outdoor deployment} &        \saim{Yes} &         \saim{No} \\
\hline
\saim{Free space loss} &        \saim{Low} &      \saim{ High } \\
\hline
 \saim{Coverage} &  \saim{High~\cite{hwang2019}} &       \saim{ Low} \\
\hline
\saim{Radio Measurements~\cite{hemadeh2018}} & \saim{Available for many windows: 28GHz, 72GHz, 52 GHz, 60 GHz} & \saim{300 GHz indoor, example measurement carried at data center environment} \\
\hline
\saim{Device size} & \saim{Few millimetres} & \saim{Few micrometers} \\
\hline
\saim{End to end simulators} & \saim{Available on ns3 for 5G cellular network~\cite{mezzavillam2018,openairinter}} &   \saim{NS3 Terasim~\cite{terasimhus2018}} \\
\hline
\end{tabular}  
\label{mmwave-thz-diff}
\end{table*}

\section{Terahertz band applications and their requirements}\label{sec:applications}

\begin{figure*}
\centering
\includegraphics[width=5.3in,height=3.2in]{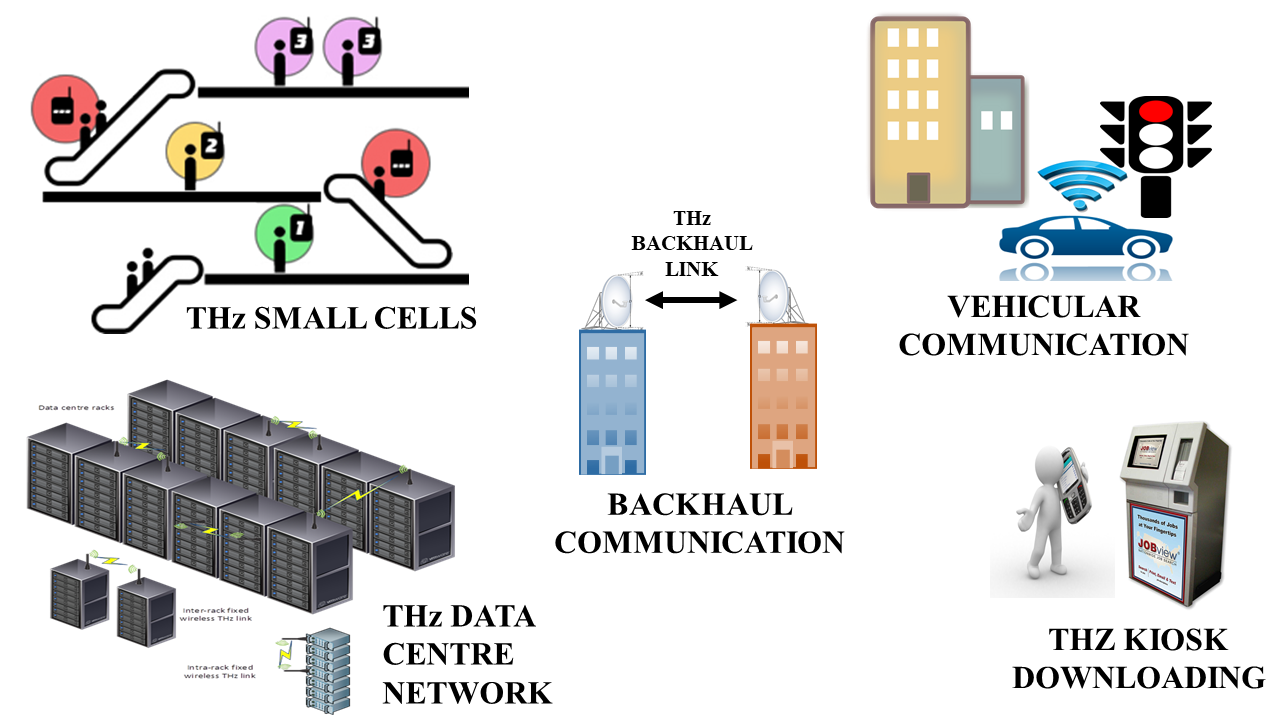}
\caption{Terahertz communication applications for macro scale networks.}
\label{fig:macroapp}
\end{figure*}

\begin{figure}
\centering
\includegraphics[width=3.2in,height=2.2in]{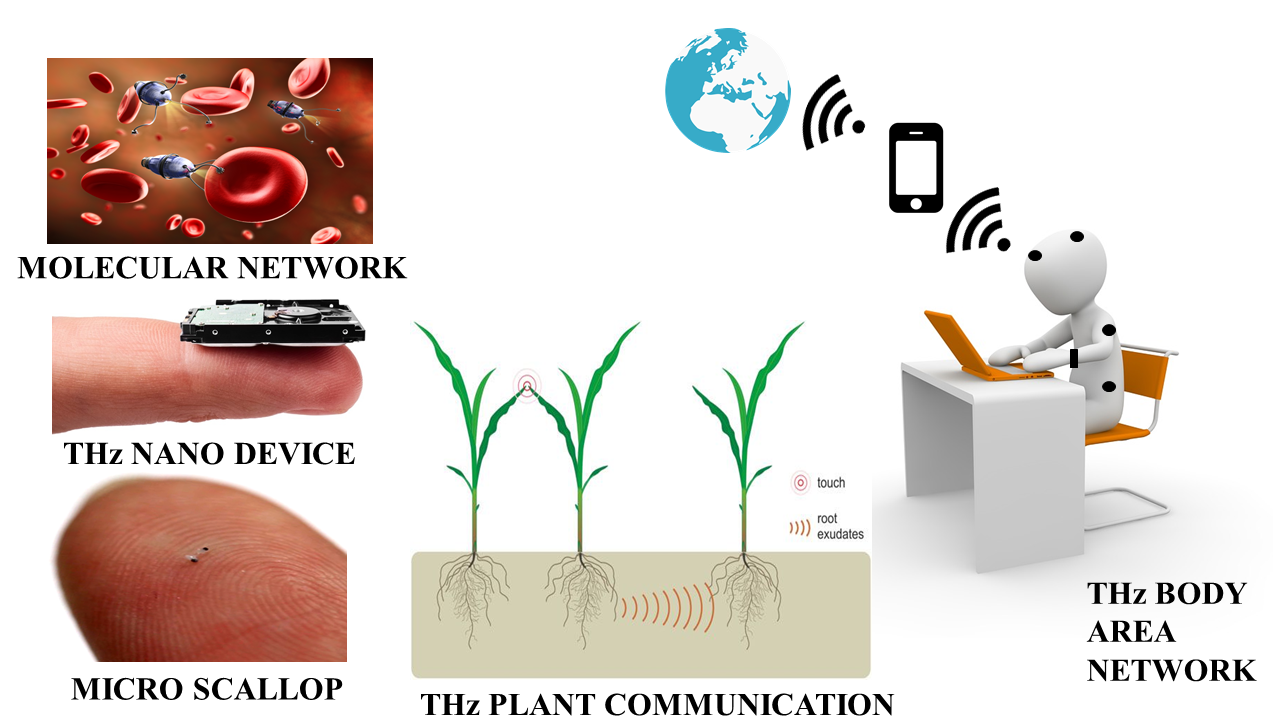}
\caption{Terahertz communication applications for nanoscale networks.}
\label{fig:nanoapp}
\end{figure}

\blue{The existing applications of Terahertz band are categorized into macro and nano applications.} Typically, these applications include outdoor as well as indoor applications that require speed from Gbps to Tbps. There are applications in which users require Gbps speed like small cells, WLAN, vehicle to vehicle communication and device to device communication. The applications which require Tbps speed and can not be satisfied using traditional bands include applications that use traffic aggregation and nano communication, for example backhaul communication, edge communication within a Data Centre and nanodevices communication which can utilize full bandwidth of Terahertz band due to small distance. The Terahertz link can be used to aggregate 5G traffic including control plane signaling, IoT traffic, \blue{Internet}, and mobile services at the backhaul and core network side to replace existing optical fiber links. It can be used also for traffic augmentation. The Tbps links can also be required for inter-chip communication, where chips can exchange ultra-high data rate in a flexible way using short-range, then THz communication will be feasible with the ultra-high data rate. These applications are also shown in Figures~\ref{fig:macroapp} and~\ref{fig:nanoapp}. Their design requirements are highlighted to emphasize their particular necessities to progress in the Terahertz MAC protocol design. Their performance target requirements are given in Table~\ref{tab:thz-app-chal}. Table~\ref{tab:paraaware}, presents the details of these protocols based on different communication aspects and parameter aware Terahertz MAC protocols\footnote{1) Channel aware: Nodes are aware of the spectrum information, 2) Physical layer: Nodes are aware of physical layer parameters like propagation loss and bit error rate, 3) Memory aware: Nodes are aware of the available memory at each node, 4) Position aware: Nodes are aware of the position of other nodes, 5) Nodes are aware of the bandwidth and adapt according to the available bandwidth.}. The MAC layer related design requirements, issues and considerations will be discussed in Section IV.

\begin{table*}[htbp]
  \centering
  \tiny
  \caption{Terahertz applications features, requirements and some general MAC related challenges.}
    \begin{tabular}{|p{2em}|p{2em}|p{6.1em}|p{3.0em}|p{4.0em}|p{3.3em}|p{3.485em}|p{5.0em}|p{6.0em}|p{5.07em}|p{4.0em}|c|p{22.715em}|} \hline
    \toprule
    \multicolumn{1}{|p{2em}|}{\textbf{Network scale}} & \multicolumn{1}{p{2em}|}{\textbf{Category}} & \textbf{Application Areas} & \textbf{Coverage} & \textbf{Mobility} & \textbf{Data rate} & \textbf{Latency} & \textbf{No of connections} & \textbf{Link availability and reliability} & \textbf{Connectivity} & \textbf{Energy efficiency} & \multicolumn{1}{p{3.215em}|}{\textbf{Target BER}} & \textbf{\blue{Terahertz MAC layer challenges \newline (Sect.~\ref{sec:applications}, Sect.~\ref{subsec:req-apps}, and Sect.~\ref{sec:issues-challenges} ) }} \\
    \midrule
    \multicolumn{1}{|c|}{\multirow{11}[22]{*}{\textbf{Macro}}} & \multicolumn{1}{c|}{\multirow{5}[10]{*}{\textbf{Indoor applications}}} & Data Centre networks~\saim{\cite{hamza2016,Petrov2018,TRD2015}} & < 20 m & No    & upto 100 Gbps & 0.1-0.2 ms & large & \multicolumn{1}{c|}{99.99\%} & P2P,P2MP & \saim{Low priority}   & $10^{-12}$   & \blue{In DCNs, Top of Rack (ToR) nodes require coverage in all direction to connect to other nearby ToR nodes. Due
to directional beam requirement at each node, a node cannot cover all direction which requires an efficient beam synchronization mechanism to establish reliable links.  } \\
\cmidrule{3-13}          &       & Terahertz Local Area Networks~\cite{tyilmaz2014,cwang2014,Petrov2018,Boulogeorgos2018} & < 50 m & Yes   & upto 100  Gbps & < 1 ms & medium & High  & P2P, P2MP, Adhoc & \saim{Medium priority}   & $10^{-6}$   & \blue{In a local area scenario, many users can access THz access point placed in indoor environment such as office. The medium access layer at base station should guarantee access to the multiple users based on their distance and the available bandwidth. Another challenge that should be addressed is the tracking of mobile users’ position and to adapt link parameters as users are moving within the network range. }  \\
\cmidrule{3-13}          &       & KIOSK downloading system~\saim{\cite{Petrov2018,cwang2014,ieee802153d2017,hsong2018}} & 0.1m  & No    & 1 Tbps & < 1 ms & small & \multicolumn{1}{c|}{99.99\%} & P2P   & \saim{Low priority}   & $10^{-6}$     & \blue{The MAC layer should support the transfer of huge amount of data in a short period. Fast link establishment should also be provided in an efficient way with authentication mechanism.} \\
\cmidrule{3-13}          &       & Terahertz Personal Area Networks~\cite{tyilmaz2014,Petrov2018,IHTLD-MAC2017,Boulogeorgos2018} & < 20 m & Yes   & upto 100 Gbps & < 1 ms & meidum & High  & P2P, P2MP, Adhoc & {Medium priority}   & $10^{-6}$    & \blue{High bandwidth availability at Terahertz bands for WPANs require efficient distance aware radio resource management. Another challenge is to discover the nodes and achieve synchronization to inform nodes about the topology distribution and management. } \\
\cmidrule{3-13}          &       & Information \saim{Broadcast}~\cite{vpyk2016,ARD2015} & 0.1-5m & Yes   & 1 Tbps & upto few sec & small to medium & \multicolumn{1}{c|}{99.99\%} & P2P,P2MP & \saim{Low priority}   &      & \blue{This include applications which supports prefetching of data (file transfer, video streaming etc.). The maximum transfer rate at the Terahertz band and the number of users for an efficient data offload can be handled at the MAC layer. Other challenges will be to accommodate the short range Terahertz links and mobility.  } \\
\cmidrule{2-13}          & \multicolumn{1}{c|}{\multirow{6}[12]{*}{\textbf{Outdoor applications}}} & Small and ultra dense cell technology~\cite{ptdat2018,Petrov2018,barrosmall2017,Busari2018,vpetrovsinr2015,SDNC2017}    & 10-15m & Yes   & > 100Gbps & upto few ms & medium to large & High  & P2MP  & \saim{High priority}   & $10^{-10}$    & \blue{Due to high density of cells, user mobility, and shorter range of Terahertz link, small cells coordination should be addressed while considering pencil beam width for link establishment and transmissions. Additional challenges can be proactive blockage detection and avoidance at MAC layer.} \\
\cmidrule{3-13}          &       & Vehicular networks and driver-less cars~\cite{Guank2017,SDNC2017,ATLR2018,pkumari2018,vpetrovv2v2019,Mumtaz2017} & > 100 m  & Yes   & > 100Gbps & upto few ms & medium & medium & P2P, P2MP, Adhoc & \saim{High priority}   &    & \blue{In V2X scenarios the Terahertz short range and directional antenna usage includes challenges where mobile vehicles needs to establish and manage the link and transfer huge amount of data (maps and road views etc.). These challenges require efficient MAC layer solutions.  Similarly, when a mobile vehicle needs to connect to a static infrastructure, the node needs beam tracking and management techniques to communicate and transfer huge amount of data with intelligent blockage management. } \\
\cmidrule{3-13}          &       & Military applications~\cite{woolard2007,sonmez2015,krzysztof2012} & > 100 m  & Yes   & 10-100Gbps & upto few ms & medium & High  & P2P, P2MP, Adhoc & \saim{For fixed units: Low priority. For sensors: High priority}  &     & \blue{In a battlefield and harsh environments, the military vehicles and planes may require secure bulk data transfer to assist other vehicles. This require secure and adaptive mechanisms to transfer data. Additionally, highly directional antennas with long range communication will be required. MAC layer should be designed to provide secure and reliable links with energy efficient mechanisms for these challenges. } \\
\cmidrule{3-13}          &       & Space Applications~\cite{suhwu2013,sdong2011} & kms   & Yes   & 10-100Gbps & upto few sec & small to medium & High  & P2P,P2MP & \saim{Low priority}   &     & \blue{Due to specific Terahertz channel characteristics and noise in space, the inter and intra satellite communication requires strict LOS communication. At MAC layer, adaptive resource allocation and link reliability should be addressed with blockage detection and mitigation.  } \\
\cmidrule{3-13}          &       & Backhaul connectivity~\saim{\cite{Petrov2018,narytnyk2014author,cwang2014,Boulogeorgos2018}} & 1 km  & No    & > 100Gbps & upto few ms & small & High  & P2P   & \saim{Low priority}    & $10^{-12}$    & \blue{THz wireless backhauling links can replace wired links for a specific environment. At MAC, link parameters can be monitored to achieve reliable high data rate links. Because, the Terahertz channels can be easily affected by environment and blockages which requires highly directional beams. In backhaul networks, nodes switching can be used to avoid blockages and change communication paths.  } \\
    \midrule
    \multicolumn{1}{|c|}{\multirow{5}[10]{*}{\textbf{Nano}}} & \multicolumn{1}{p{5.785em}|}{\textbf{In/On-body applications}} & Health monitoring~\cite{FELICETTI201627,zarepour2015,nanoparadigm2008} & 0.5-2.5mm & Yes   & 100Gbps & 1 ms  & large & High  & adhoc, P2MP & \saim{High priority}   &       & \blue{Disease monitoring and early detection is crucial which requires energy efficient mechanisms and nodes mobility to carry data from one place to another where MAC layer should provide energy efficient and interference mitigation mechanisms. }  \\
\cmidrule{2-13}          & \multicolumn{1}{c|}{\multirow{2}[4]{*}{\textbf{Outdoor applications}}} & Defence applications~\cite{wang2008} & \multicolumn{1}{c|}{} & No    & 100Gbps & 1 ms  & large & High  & adhoc, P2MP & \saim{High priority}   &       & \blue{In defence applications, nano devices can be used to monitor a specific area or on-body communication. The nodes will require efficient access and scheduling mechanism to transmit data efficiently while balancing energy consumption and harvesting for massive connectivity. Additionally, discovery and coordination among the nodes will also be required for adaptive network management. 
}  \\
\cmidrule{3-13}          &       & Agricultural applications~\cite{mathankar2013,azahid2018,dynamic_CH2016} & 4mm   & No    & 100Gbps & 1 ms  & large & High  & adhoc, P2MP & \saim{High priority}   &       & \blue{Application of THz radio communication for agriculture application is challenged by many factors such as a specific channel characteristic including attenuation of leaves, soil, dust and water vapour. These factors should be monitored with efficient data transmission scheduling and frame structure. UAVs can also be used to monitor a field by fetching data from a gateway node which will require beam alignment and efficient resource allocation. }   \\
\cmidrule{2-13}          & \multicolumn{1}{c|}{\multirow{2}[4]{*}{\textbf{Indoor applications}}} & Internet of Nano-Things~\cite{nkhalid3002017,Dressler2015} & 2m    & No    & 90Gbps & 1 ms  & massive & High  & adhoc, P2MP & \saim{High priority}   &       & \blue{Nano nodes will be deployed in high density to perform specific tasks such as sensing and relaying. MAC layer should perform scheduling and channel sensing for link adaptation and efficient resource allocation. Additionally, node buffer should be optimized for this massive connectivity scenario with efficient path selection. 
 } \\
\cmidrule{3-13}          &       & Intra chip network~\saim{\cite{Petrov2018,narytnyk2014author,cwang2014,vpetrovIC2017,oyal2018,skimaz2016}} & 0.03m & No    & 100Gbps & 1 ms  & small to medium & High  & Ad-hoc & \saim{Medium priority}   & $10^{-12}$   & \blue{The intra chip communication will require bulk data transmission which requires effective memory synchronization and access. Optimization of number of cores supported for high throughput and low delay is a challenge to address. Further, Hybrid MAC protocols and multiplexing mechanisms are also required.
 } \\ \hline
    \bottomrule
    \end{tabular}%
  \label{tab:thz-app-chal}%
\end{table*}%

%

\subsection{Applications for Macro Scale Terahertz Networks}\label{subsec-app:macro}

\saim{The macro-scale communication involve applications in which the transmission range is higher than 1 meter and up to a kilometer. The Terahertz bands although has huge bandwidth availability, the transmission distance can vary due to high path and absorption loss. The indoor application differs from outdoor applications mainly due to scattering and reflection effects. Therefore\blue{,} requires different channel models for different environment which should be considered while designing MAC protocols for these applications. These applications are shown in Figure~\ref{fig:macroapp}. Table~\ref{tab:thz-app-chal}, is showing the technical requirements for different Terahertz applications. It includes fixed point-to-point, point-to-multipoint, and mobility scenarios.}
 
The applications which require Tbps links include wireless backhaul~\cite{wiback2013} and Data Centres~\cite{seandell2019}. Wireless Backhaul typically involves point-to-point connections for information transmission to the base stations of the macrocell, especially where the fiber optic is not available. Small cell communication can be another application scenario in which the backhaul part will require Tbps links to transfer huge amount of data. High power antennas can be used for larger distance coverage but should consider environmental factors and frequent user association while designing a MAC protocol. The atmosphere loss is higher, however can be compensated by high antenna directivity. For capacity enhancement and larger bandwidth with Tbps transmission, the Terahertz band between 200 and 300 GHz has shown low atmospheric losses~\cite{koenig2013}. The wireless fiber extender is also an interesting scenario to extend the communication range and capacity of existing backhaul communication set up to provide reliable data communication with Tbps throughput for distance up to 1 Km in outdoor environments. A very large antenna array or Massive MIMO techniques can be used to transfer information between cells. The use of massive MIMO arrays can be used in an adaptive manner to modify the transmit and receive beams to accommodate the changes in the environment than to physically readjusting the arrays~\cite{narytnyk2014author}. They can be used to communicate with multiple backhaul stations by electronic beam steering.  

\saim{At the macro scale, another promising application is wireless Data Centres~\cite{hamza2016,ykatayama2011}. In an attempt to increase the user experience and high-speed network, the cloud applications have introduced the competition between the Data Centres. This resulted in extensive extension of servers and required bandwidth to support numerous applications to manage bursty high traffic. The Terahertz band can be augmented to support Tbps links especially at the edge where the traffic aggregates rather than using the cables with limited capability. The Terahertz links can be used in parallel with existing architecture to provide backup links, failover mechanisms and high-speed edge router links with SDN support~\cite{mollahasani2016}. This can improve the user experience and can also reduce the cost of deployment within Data Centres.}

\saim{Using the high capacity wireless Terahertz links can also help in re-designing the Data Centres geometry~\cite{Wukaishun2012}. However, require careful communication protocol design like Physical, MAC and network layer, to be efficiently utilized for the Data Centre network. The top-of-rack (ToR) Terahertz devices can connect using point-to-point and multipoint wireless links. However, requires directional antennas for inter rack communication for enhanced coverage for more than a meter. The intra rack communications can also use omni-directional antennas due to short distance between the routers. A fair and efficient channel access scheme is required for both inter/intra rack communication with scheduling (for directional antennas) and with collision avoidance (CA) techniques (for Omni-directional antennas) due to multi-user interference. To connect different ToR devices, the link establishment is very important however becomes challenging with directional antennas and energy minimization constraint.}

\saim{The other point-to-point connectivity applications include the KIOSK downloading system which can be used for instant transfer of bulk amount of data, as shown in Figure~\ref{fig:macroapp}. Its a peer to peer communication system between stationary transmitter and a mobile receiver with limited mobility. It can be imagined as a stationary terminal or point, which may be connected with a data center using fiber-optic and can provide the bulk amount of data to various users in few seconds, typically in GBs. This type of application can use Tbps links to satisfy user demands. The challenges include rapid user association and link establishment, beam management, error detection, and correction strategies, and secure authentication on public systems. An experimental demonstration of a prototype for KIOSK Downloading system using 300 GHz band is presented in~\cite{hsong2018} and~\cite{ntt}, which includes the channel and LOS analysis with comparison of error correction algorithms. The approximate beam size is mentioned in~\cite{hsong2018} as 22 cm with 30 dBi as antenna gain for 1-meter distance. }

\saim{Vehicle to infrastructure and its backhaul network can also utilize Tbps links to improve vehicular communication networks. For example, Google's auto-driver cars generate the sensor data at the rate of 750 MBps~\cite{google2013} and are expected to generate 1 TB of sensor data in a single trip~\cite{sas}. The delivery, processing, and optimization of such huge data to assist vehicles can require Tbps links to improve the efficiency and latency of communication. As an alternative solution for fiber-based backhauling, the vehicles can also serve as digital mules to reduce the deployment cost and to migrate the data~\cite{kanno2014,greenberg2008}. }

\saim{The point-to-point communication can also be utilized in military applications between air to air and ground machines~\cite{sonmez2015,krzysztof2012}. For many years, space agencies such as NASA and ESA, are developing sensors and instruments for space technology~\cite{suhwu2013},~\cite{Ding2016,sdong2011,wdou2014}. Using THz link for intra satellite application can be feasible outside the atmospheric region, where the only free space loss is considered. A possible scenario can be an inter-satellite link within the same constellation or between low orbit and geostationary satellite.}

\saim{The interesting scenarios for TLAN are the indoor home networks and LAN~\cite{yilmaz22015,tyilmaz2014}. Mainly, the backhaul part of this home network to core networks can use Tbps links to transfer aggregated data, which can facilitate multiple users at a time to download huge data. However, currently, the short distance within an indoor home or office environment requires point-to-point communication or efficient beam management strategies. The high speed and instant connectivity between multiple personal devices are possible using Terahertz communication links~\cite{tyilmaz2014}. }

\subsubsection{Summary of macro scale applications}


\saim{The Terahertz applications range from indoor to outdoor from short to large distances. The applications with very short communication distance like KIOSK downloading system and intra rack communication within a Data Centre does not involve mobility challenge. However, a delay can be involved depending upon the amount of data transfer and channel access schedule. Mainly, they require point-to-point connectivity between the systems. The last mile access and backhaul point-to-point and multipoint are also very interesting scenarios that require high throughput and low latency for longer distances up to a km. Currently, it has reached 10 Gbps and is realized to reach beyond 100 Gbps using Terahertz band with high bandwidth availability~\cite{thor,teranovareport2018}. }


\saim{The atmospheric losses affect both indoor and outdoor type of applications differently and therefore the appropriate channel and propagation models should be considered. The indoor environments like Data Centres require fixed links between the racks \blue{with} very limited mobility. The outdoor environments like backhaul links involve fixed point-to-point links, however differ from indoor environments in terms of atmosphere, distance, reliability and link-budget requirements. The scenarios like vehicular and small cells require high mobility and therefore involve frequent handovers and must support high user density and scalability for new MAC protocols. For both scenarios, efficient channel access mechanisms, reliable connections with error detection and correction are required with efficient beam management techniques \blue{for the MAC design}.}

\subsection{Applications for NanoScale Terahertz Networks}\label{subsec-app:nano}

\saim{Nanotechnology enables the nano-components which are able to perform simple specific tasks, like data storage, computation, sensing, and actuation. These nano-components can be integrated into a device with only a few cubic meters in size and can enable the development of more advanced nano devices~\cite{Akyildiz2014}. These nanodevices can exchange information to achieve complex tasks in a centralized or a distributed manner~\cite{drih-mac2015}, which enable unique and interesting applications in the field of biomedical, industrial, military, health monitoring, plant monitoring, nanosensor networks, chemical and biological attack prevention and system-on-chip wireless networks~\cite{Akyildiz2014,FELICETTI201627,gkwalia2016}. It mainly involves communication upto few centimeters. This also includes the networks using electromagnetic radiation like nanonetworks and molecular networks in which transmission occur using the flow of molecules~\cite{FELICETTI201627,gkwalia2016}. }

\saim{The nano devices, due to their smaller size and communication range can benefit with larger bandwidth and so can utilize Tbps links. Due to smaller range, the path loss remains at the lowest which enables high throughput in nano communication. These applications are shown in Figure~\ref{fig:nanoapp} and their challenges are also mentioned in Table~\ref{tab:thz-app-chal}. The nanosensors can be used to detect an early disease by using molecular communication by gathering heart rate. The gathered information that can be transmitted over the Internet using a device or mobile phone to a healthcare provider. Other applications are nano-scallop to swim in a biomedical fluid~\cite{Qiu2014}, a bio-bot which is powered by skeletal muscles ~\cite{Cvetkovic2014}, on board drug delivery system~\cite{syim2012}, a magnetic helical micro-swimmer to deliver single cell gene to the human embryonic kidney in a wireless way~\cite{qiuf2015}. \blue{THz} MAC layer protocols should consider energy consumption and harvesting trade-off, error recovery, scheduling of transmissions among different nodes and efficient channel access.}

\saim{Terahertz band can be used for wireless on-chip networks~\cite{sabadal2013,oyal2018}. It can be used at the very small scale to connect two chips due to its high bandwidth and low area overhead. The MAC should consider supporting maximum number of cores by addressing MAC performance by specifying input traffic and interface characteristics~\cite{vpetrovIC2017}. The tolerable delay should be analysed among different layer architecture and analysis of maximum cores supported for throughput delay~\cite{vpetrovIC2017}. The challenges include efficient channel access mechanism for intra chip communication with scheduling, efficient inter-core communication, small-scale antennas to provide high bandwidth and low delay. }

\subsubsection{Summary of Nanoscale applications}

\saim{The nano applications, especially on chip communications can utilize high bandwidth Tbps links. In nano communications\blue{,} the communication range is very small, due to which the path loss remains at the lowest. Mainly, these applications requires efficient energy consumption and harvesting mechanisms to address limited energy issues while considering nonbatteries/nanogenerators/nontransreceivers architecture and performance enhancements. The antenna technology and new channel/propagation and noise models are required with tools to estimate path loss for different nanoscale network environments. Efficient communication protocols such as modulation and coding techniques, power control, routing and MAC strategies for nanoscale communications. Their MAC protocols also requires to support scalability and connectivity among large number of devices. Due to limited capacity of devices, energy efficient mechanism are required with harvesting and transmission balance for efficient communication. Link establishment and scheduling the communication among devices is also a challenge. Some architectures for nanonetworks to handle complex task by combining it with current technologies like SDN, IoTs, virtual network and fog computing are presented in~\cite{Abbasi2016}.}

\subsection{\saim{Other applications for Terahertz communications}}

\saim{The Terahertz band is emerging as the most promising way to realize Gbps link, due to increase in the current traffic demand~\cite{narytnyk2014author}. A radio link over 120 GHz band for data transmission at the rate of 10 Gbps to the range of over 800 m is presented in~\cite{amro2012}. There are applications mostly at the end users, which requires high speed data connections upto a more than 100 Gbps~\cite{vpyk2016,Petrov2016}. That can be fulfilled by using low Terahertz frequency bands with higher order modulation schemes. These applications include a broadcast system at public places like metro stations, airports, and shopping malls to transfer \blue{bulk amount of} data \blue{with Gbps speed}. However, the transmission or data transmission currently can be possible only for the short distance. The short distance, mobility and user density per coverage area should be considered while designing a MAC protocol. The benefits of this application is also recognised by IEEE 802.15.3d (TG - 100 Gbps wireless) as one of the use cases for the Terahertz communication~\cite{ARD2015}. An information broadcast case with area density, number of users and mobility pattern is discussed in~\cite{vpyk2016}. Further, the front ends of applications like Terahertz LAN, vehicle to vehicle communication, small cells to mobile users are all applications where Terahertz bands can be used for high throughput. However\blue{,} will require extreme antenna directionality management and rapid user association and disassociation mechanisms. }

\saim{The fronthaul is between the base station and the end user radio equipments which requires Gbps links. These radio equipments can be mobile as in small cells and wireless LAN scenarios and can be fixed users. The critical parameters for these applications other than high data rate (upto 1 Tbps) is distance which should be above a meter to a kilometre. Future applications which will include massive deployment of small cells for cloud radio access network which may increase the data rates for end users at front end and back end which will require Tbps links. The small cells deployment can utilize the huge bandwidth available in Terahertz band and can free up the lower frequency bands which leads to several Tbps of data transfer~\cite{Petrov2016}. One of the possible and upcoming applications of Terahertz band is the small cell communication for mobile cellular networks, in which ultra-high data rate can be provided to mobile users within transmission range up to 20 m~\cite{SDNC2017,akylidiz-lte2014}. The Terahertz small cell can be a fixed point installed to serve multiple mobile users. The mobility of users with higher data volume offloading needs to be supported. The users moving from cell to cell requires seamless handover for uninterrupted communication. The Terahertz directional antenna usage can increase challenges in user association and tracking with scheduled channel access. Therefore, the Terahertz MAC protocol should consider these requirements and the target performance to ensure the user satisfaction.}

\saim{The virtual reality (VR) device is an interesting application which requires at least 10 Gbps data traffic transfer. However, currently it relies on wired cord and needs to be shifted to wireless transfer with more than 10 Gbps data rate. The VR applications mainly requires ultra high reliability and low latency with high data rates for their services and fast data processing~\cite{christina2019}. Using Terahertz band for VR services requires transmission and processing delay to be low\blue{~\cite{Chaccour2020CanTP}}.}

\begin{table*}[htbp]
  \centering
  \tiny
  \caption{Summary of existing Terahertz MAC protocols with MAC aspects and parameter awareness.}
    \begin{tabular}{|p{1.1em}|p{0.42em}|p{0.33em}|p{0.34em}|p{0.34em}|p{0.34em}|p{0.34em}|p{0.34em}|p{0.34em}|p{0.34em}|p{0.34em}|p{0.34em}|p{0.34em}|p{0.34em}|p{0.34em}|p{0.34em}|p{0.34em}|p{0.34em}|p{0.34em}|p{0.34em}|p{0.34em}|p{0.34em}|p{0.34em}|p{0.34em}|p{26em}|p{11em}|}

\cmidrule{3-24}   \multicolumn{2}{c|}{} & \multicolumn{5}{|p{8.6em}|}{\textbf{Parameter aware MAC protocols}} & \multicolumn{17}{p{23.5em}|}{\textbf{                  MAC layer aspects}}  & \multicolumn{1}{l}{} \\
    \midrule
    \begin{sideways}\textbf{Paper}\end{sideways} & \multicolumn{1}{p{1em}|}{\begin{sideways}\textbf{Year}\end{sideways}} & \begin{sideways}\textbf{Channel}\end{sideways} & \begin{sideways}\textbf{Physical layer}\end{sideways} & \begin{sideways}\textbf{Energy}\end{sideways} & \begin{sideways}\textbf{Memory}\end{sideways} & \begin{sideways}\textbf{Position/Distance}\end{sideways} & \begin{sideways}\textbf{Bandwidth adaptive}\end{sideways} & \begin{sideways}\textbf{Hand shake}\end{sideways} & \begin{sideways}\textbf{Synchronization}\end{sideways} & \begin{sideways}\textbf{Neighbor discovery}\end{sideways} & \begin{sideways}\textbf{Channel access method}\end{sideways} & \begin{sideways}\textbf{channel selection}\end{sideways} & \begin{sideways}\textbf{carier sensing}\end{sideways} & \begin{sideways}\textbf{Scheduling}\end{sideways} & \begin{sideways}\textbf{Cross layer MAC design}\end{sideways} & \begin{sideways}\textbf{Collision and congestion}\end{sideways} & \begin{sideways}\textbf{Interference}\end{sideways} & \begin{sideways}\textbf{Packet size and structure}\end{sideways} & \begin{sideways}\textbf{Data transmission}\end{sideways} & \begin{sideways}\textbf{Error Control}\end{sideways} & \begin{sideways}\textbf{Delay and throughput}\end{sideways} & \begin{sideways}\textbf{Multiplexing}\end{sideways} & \begin{sideways}\textbf{Beam forming and scanning}\end{sideways} & \textbf{Protocol Description}& \textbf{\blue{THz challenge addressed}} \\
    \midrule
    \cite{LLModelling2011} & 2011  & X     & X     & X     & X     & X     & X     & X     & X     & X     & X     & X     & X     & X     & X     & \checkmark     & X     & \checkmark     & X     & X     & X     & X     & X     & Effects of congestion and traffic generation intensity are analysed for nano-networks through competition among bacteria for conjugation at nano gateways. & \blue{Traffic congestion at nano gateways and link layer performance analysis.} \\ [-0.5em]
    \midrule
    \cite{phlame2012} & 2012  & \checkmark     & X     & \checkmark     & X     & X     & X     & \checkmark     & \checkmark     & X     & X     & \checkmark     & X     & X     & \checkmark     & \checkmark     & \checkmark     & \checkmark     & \checkmark     & X     & \checkmark     & X     & X     & The communication and coding schemes are jointly selected to maximise the decoding probability and minimise the interference while considering energy limitations. & \blue{MAC protocol with low weight coding scheme and pulse-based communication.} \\[-0.5em]
    \midrule
    \cite{EESR-MAC2012} & 2012  & X     & X     & X     & X     & X     & X     & X     & \checkmark     & X     & \checkmark     & X     & X     & \checkmark     & X     & \checkmark     & X     & \checkmark     & X     & X     & X     & X     & X     & An energy efficient, scalable and reliable MAC protocol is proposed for dense nanonetworks with control and data packet structures. & \blue{MAC protocol for Nanonetworks with scalability.} \\[-0.5em]
    \midrule
    \cite{esaware2013} & 2013  & \checkmark     & \checkmark     & \checkmark     & X     & X     & \checkmark     & X     & X     & X     & \checkmark     & X     & X     & \checkmark     & \checkmark     & \checkmark     & X     & X     & X     & X     & \checkmark     & X     & X     & An energy and spectrum aware MAC protocol is proposed to achieve fair throughput and optimal channel access by optimising the energy harvesting and consumption in nano-sensors. & \blue{A sub-optimal symbol compression scheduling algorithm with throughput and lifetime balance.} \\[-0.5em] 
    \midrule
    \cite{MAC-TUDWN2013} & 2013  & X     & X     & X     & X     & X     & X     & X     & \checkmark     & X     & \checkmark     & X     & X     & \checkmark     & X     & X     & X     & \checkmark     & X     & X     & \checkmark     & X     & X     & A MAC protocol based on IEEE 802.15.3c is proposed for Terahertz ultra high data rate wireless networks is proposed with super frame structure and timeslot allocation scheme. & \blue{MAC protocol for high data rate support.} \\[-0.5em]
    \midrule
    \cite{smart-mac2013} & 2013  & X     & X     & X     & X     & X     & X     & X     & X     & X     & X     & X     & X     & X     & \checkmark     & X     & X     & X     & \checkmark     & X     & X     & X     & X     & A MAC protocol is proposed for health monitoring for nanosensor network with  anlaysis of node density and Tx range with routing strategies. & \blue{Simulator for nano networks, Node density and Transmission analysis.} \\[-0.5em]
    \midrule
    \cite{MAC-TC2013} & 2013  & X     & X     & X     & X     & X     & X     & X     & X     & X     & \checkmark     & X     & X     & X     & X     & X     & X     & X     & \checkmark     & X     & X     & X     & X     & A MAC protocol for Terahertz communication is proposed with channel access and data rate analyses with superframe structure. & \blue{MAC protocol with super frame structure.} \\[-0.5em]
    \midrule
    \cite{TRPLE2014} & 2014  & X     & X     & \checkmark     & X     & X     & X     & X     & X     & \checkmark     & X     & X     & X     & \checkmark     & X     & X     & X     & X     & X     & X     & X     & \checkmark     & \checkmark     & A MAC design is proposed for macro scale communication at 100 Gbps for pulse-level beam switching and energy control with focus on neighbor discovery and scheduling. & \blue{MAC protocol with pulse level beam switching and energy control.} \\[-0.5em]
    \midrule
   \cite{MDP2014} & 2014  & X     & X     & \checkmark     & X     & X     & X     & X     & X     & \checkmark     & X     & X     & X     & \checkmark     & X     & X     & \checkmark     & X     & X     & X     & X     & X     & X     & A technique to utilize the harvested energy in wireless nano=networks is presented with focus on optimal energy consumption for transmission and reception with packet scheduling. & \blue{Energy harvesting and consumption with low energy storage.} \\[-0.5em]
    \midrule
    \cite{TSN2014} & 2014  & X     & X     & X     & X     & X     & X     & X     & X     & X     & X     & \checkmark     & X     & X     & X     & X     & X     & X     & X     & X     & X     & X     & X     & A frequency hopping scheme is presented to overcome the problems of molecular absorption. & \blue{Frequency selection based on channel noise and absorption.} \\[-0.5em]
    \midrule
    \cite{RIH-MAC2014} & 2014  & X     & X     & \checkmark     & X     & X     & X     & \checkmark     & X     & X     & X     & X     & X     & \checkmark     & X     & X     & X     & X     & X     & X     & X     & X     & X     & A receiver initiated MAC protocol is proposed based on distributed and probabilistic schemes for adaptive energy harvesting nanonodes with scheduling. & \blue{Rx initiated communication with energy harvesting.}  \\[-0.5em]
    \midrule
    \cite{drih-mac2015} & 2015  & X     & X     & \checkmark     & X     & X     & X     & \checkmark     & X     & X     & X     & X     & X     & \checkmark     & X     & X     & X     & X     & X     & X     & X     & X     & X     & A distributed receiver initiated MAC protocol is proposed with scheduling scheme to minimize collisions and maximize the utilization of energy harvesting. & \blue{Distributed Scheduling with energy harvesting.}  \\[-0.5em]
    \midrule
    \cite{LLsynch2015} & 2015  & X     & X     & X     & X     & \checkmark     & X     & \checkmark     & \checkmark     & X     & X     & X     & X     & X     & X     & X     & X     & X     & X     & X     & X     & X     & X     & An Rx initiated handshake based link layer synchronization mechanism is proposed to maximise the channel utilization with analysis of delay, throughput and packet transmission rate. & \blue{Handshaking and synchronization mechanisms.}  \\[-0.5em]
    \midrule
    \cite{TCN2015} & 2015  & \checkmark     & X     & X     & \checkmark     & X     & X     & X     & X     & X     & X     & X     & X     & X     & X     & \checkmark     & X     & X     & \checkmark     & X     & X     & X     & X     & A scheme with logical channel information is proposed in which information is encoded in timings of channel. It supports low rate communication in an energy efficient and reliable manner. & \blue{Information encoding over logical communication channels.} \\[-0.5em]
    \midrule
    \cite{joint_error2016} & 2015  & X     & X     & X     & X     & X     & X     & X     & X     & X     & X     & X     & X     & X     & X     & X     & X     & X     & X     & \checkmark     & \checkmark     & X     & X     &  A cross layer analysis of error control strategies is presented for nanonetworks with trade-off between bit error rate, packet error rate, energy consumption and latency. & \blue{Error control analysis.} \\[-0.5em]
    \midrule
    \cite{Design-WNSN2015} & 2015  & X     & X     & \checkmark     & X     & X     & X     & X     & X     & X     & \checkmark     & X     & X     & X     & X     & X     & X     & X     & \checkmark     & X     & X     & X     & X     & An intra-body disease detection is proposed for wireless nanosensor network using on-off keying and TDMA framework for analysing the data transmission efficiency. & \blue{Multiple channel access scheme.}  \\[-0.5em]
    \midrule
    \cite{DMDS2016} & 2016  & X     & X     & \checkmark     & X     & X     & X     & X     & X     & X     & X     & X     & X     & \checkmark     & X     & X     & X     & X     & \checkmark     & X     & \checkmark     & X     & X     & A fully distributed low-computation scheduling MAC protocol is proposed for maximising network throughput by jointly considering the energy consumption and harvesting. & \blue{Scheduling with energy harvesting and consumption.} \\[-0.5em]
    \midrule
    \cite{TAB-MAC2016} & 2016  & X     & X     & X     & X     & \checkmark     & X     & X     & X     & \checkmark     & X     & X     & X     & X     & X     & X     & X     & \checkmark     & \checkmark     & X     & \checkmark     & X     & \checkmark     & An assisted beam-forming and alignment MAC protocol is proposed with neighbor discovery, data transmission, delay and throughput analysis. & \blue{MAC protocol with assisted beam forming.}  \\ [-0.5em]
    \midrule
    \cite{GMAC2016} & 2016  & X     & X     & \checkmark     & X     & X     & X     & X     & \checkmark     & X     & X     & X     & X     & X     & X     & X     & X     & X     & X     & X     & X     & X     & X     & A synchronization mechanism is proposed for nano sensor network based on TS-OOK with analysis of consumed energy, collision probability, delay and throughput.  & \blue{Synchronization with energy consumption and collision analysis.}  \\[-0.5em]
    \midrule
    \cite{MGDI2016} & 2016  & X     & \checkmark     & X     & X     & X     & X     & X     & X     & X     & X     & X     & X     & X     & X     & X     & X     & X     & \checkmark     & X     & X     & X     & X     & A networking approach for static and dense topologies is presented with flooding, network density, data dissemination and broadcast analysis. & \blue{Scalability and coverage, flooding based communication.} \\[-0.5em]
    \midrule
   \cite{pkt_size2016} & 2016  & X     & X     & X     & X     & X     & X     & X     & X     & X     & X     & X     & X     & X     & X     & X     & X     & \checkmark     & X     & X     & \checkmark     & X     & X     & A link throughput maximization problem is discussed. An optimal packet size is derived with combined physical and link layer peculiarities. & \blue{Link throughput maximization and optimal packet size.}  \\[-0.5em]
    \midrule
    \cite{NS-MAC2016} & 2016  & X     & X     & X     & X     & X     & X     & X     & X     & X     & X     & X     & X     & X     & X     & \checkmark     & X     & X     & \checkmark     & X     & X     & X     & X     & Different MAC protocols are compared and analysed in terms of transmission distance, energy consumption and collision probability. & \blue{Transmission, energy and collision performance analysis.} \\[-0.5em]
    \midrule
    \cite{dynamic_CH2016} & 2016  & X     & X     & X     & X     & X     & X     & X     & X     & X     & X     & \checkmark     & X     & X     & X     & X     & X     & X     & X     & X     & X     & X     & X     & Four frequency selection schemes are analysed for throughput and energy utilization. & \blue{Frequency selection with sensitivity to moisture level.} \\[-0.6em]
    \midrule
    \cite{HLMAC2016} & 2016  & \checkmark     & X     & X     & X     & X     & X     & X     & X     & X     & X     & X     & X     & X     & X     & X     & X     & \checkmark     & X     & X     & \checkmark     & X     & X     & A high throughput low delay MAC is proposed to reduces the delay with super-frame structure. & \blue{Low delay MAC with super frame structure.} \\[-0.6em]
    \midrule
    \cite{DLLC2017} & 2017  & \checkmark     & X     & X     & X     & X     & X     & X     & X     & X     & X     & X     & X     & X     & X     & X     & X     & \checkmark     & X     & \checkmark     & X     & X     & X     & A hardware processor for 100 Gbps wireless data links is presented. A light weight FEC engine, BER, frames fragmentation retransmission protocol is also presented. & \blue{Hardware processor for 100 Gbps wireless data link layer.}\\[-0.5em]    
     \midrule
    \cite{MAADM2017} & 2017  & X     & X     & X     & \checkmark     & X     & X     & X     & \checkmark     & X     & X     & X     & X     & X     & X     & X     & \checkmark     & X     & \checkmark     & X     & \checkmark     & \checkmark     & X     & A memory assisted MAC protocol with angular division multiplexing is proposed with multiple antennas for service discovery, coordination, data communications and interference. & \blue{Medium access with angular division multiplexing and memory consideration.} \\[-0.5em]
       \bottomrule
    \end{tabular}%
  \label{tab:paraaware}%
\end{table*}%

\begin{table*}[htbp]
  \centering
  \tiny
  \caption*{(TABLE VII Continued.) Summary of existing Terahertz MAC protocols with MAC aspects and parameter awareness.}
    \begin{tabular}{|p{1.1em}|p{0.42em}|p{0.33em}|p{0.34em}|p{0.34em}|p{0.34em}|p{0.34em}|p{0.34em}|p{0.34em}|p{0.34em}|p{0.34em}|p{0.34em}|p{0.34em}|p{0.34em}|p{0.34em}|p{0.34em}|p{0.34em}|p{0.34em}|p{0.34em}|p{0.34em}|p{0.34em}|p{0.34em}|p{0.34em}|p{0.34em}|p{26em}|p{11em}|}

\cmidrule{3-24}   \multicolumn{2}{c|}{} & \multicolumn{5}{|p{8.6em}|}{\textbf{Parameter aware MAC protocols}} & \multicolumn{17}{p{23.5em}|}{\textbf{                  MAC layer aspects}}  & \multicolumn{1}{l}{} \\
    \midrule
    \begin{sideways}\textbf{Paper}\end{sideways} & \multicolumn{1}{p{1em}|}{\begin{sideways}\textbf{Year}\end{sideways}} & \begin{sideways}\textbf{Channel}\end{sideways} & \begin{sideways}\textbf{Physical layer}\end{sideways} & \begin{sideways}\textbf{Energy}\end{sideways} & \begin{sideways}\textbf{Memory}\end{sideways} & \begin{sideways}\textbf{Position/Distance}\end{sideways} & \begin{sideways}\textbf{Bandwidth adaptive}\end{sideways} & \begin{sideways}\textbf{Hand shake}\end{sideways} & \begin{sideways}\textbf{Synchronization}\end{sideways} & \begin{sideways}\textbf{Neighbor discovery}\end{sideways} & \begin{sideways}\textbf{Channel access method}\end{sideways} & \begin{sideways}\textbf{channel selection}\end{sideways} & \begin{sideways}\textbf{carier sensing}\end{sideways} & \begin{sideways}\textbf{Scheduling}\end{sideways} & \begin{sideways}\textbf{Cross layer MAC design}\end{sideways} & \begin{sideways}\textbf{Collision and congestion}\end{sideways} & \begin{sideways}\textbf{Interference}\end{sideways} & \begin{sideways}\textbf{Packet size and structure}\end{sideways} & \begin{sideways}\textbf{Data transmission}\end{sideways} & \begin{sideways}\textbf{Error Control}\end{sideways} & \begin{sideways}\textbf{Delay and throughput}\end{sideways} & \begin{sideways}\textbf{Multiplexing}\end{sideways} & \begin{sideways}\textbf{Beam forming and scanning}\end{sideways} & \textbf{Protocol Description}& \textbf{\blue{THz challenge addressed}} \\
    \midrule
  \cite{MRA-MAC2017} & 2017  & X     & \checkmark     & X     & X     & X     & X     & X     & \checkmark     & \checkmark     & X     & X     & X     & X     & X     & X     & X     & X     & \checkmark     & X     & X     & X     & \checkmark     & A distributed multi radio assisted MAC protocol is proposed with multiple antennas for signal control mechanism with beam-forming. & \blue{Assisted beamforming for data transmission.}  \\ [-0.5em]
    \midrule
\cite{IHTLD-MAC2017} & 2017  & \checkmark     & X     & X     & X     & X     & X     & \checkmark     & \checkmark     & X     & X     & X     & X     & X     & X     & X     & X     & X     & X     & X     & X     & X     & X     & A high throughput low delay MAC is proposed with on-demand retransmission mechanism based on verification, reserved timeslots based channel condition and adaptive retransmission mechanism. & \blue{Retransmissions and reserved timeslots mechanisms based on THz channel conditions.} \\[-0.5em]    
    \midrule
    \cite{RBMP2017} & 2017  & X     & X     & X     & X     & X     & X     & X     & X     & \checkmark     & X     & X     & X     & X     & X     & X     & X     & X     & \checkmark     & X     & X     & X     & X     & A relay-based MAC is presented which considers communication blockage and facing problem. It further presents a neighbor discovery mechanism and data transmission. & \blue{Relay based MAC for obstacle effect mitigation.}  \\[-0.5em]
    \midrule
    \cite{APIS2017} & 2017  & X     & \checkmark     & \checkmark     & X     & X     & \checkmark     & X     & X     & X     & X     & X     & X     & \checkmark     & X     & X     & X     & X     & X     & X     & X     & X     & X     & An adaptive pulse interval scheduling mechanism based on pulse arrival pattern is presented. & \blue{Pulse level access scheduling.}  \\[-0.5em]
    \midrule
    \cite{OPTRS2017} & 2017  & X     & X     & X     & X     & \checkmark     & X     & X     & X     & \checkmark     & X     & X     & X     & X     & \checkmark     & X     & X     & X     & X     & X     & \checkmark     & X     & \checkmark     & Optimal relaying strategies with cross layer analysis. & \blue{Cross layer design for high throughput.} \\[-0.5em]
    \midrule
    \cite{SDNC2017} & 2017  & \checkmark     & X     & X     & X     & X     & X     & X     & X     & X     & X     & \checkmark     & X     & \checkmark     & X     & X     & X     & X     & \checkmark     & X     & X     & X     & X     & Channel handoff mechanism for mmWave and Terahertz channels, high bandwidth data transfer, scheduling and channel capacity modelling. & \blue{Spectrum switching and scheduling for mmWave and THz bands.} \\[-0.5em]
    \midrule
    \cite{EEWNSN2017} & 2017  & X     & X     & X     & X     & X     & X     & X     & X     & X     & X     & X     & X     & X     & X     & X     & X     & X     & X     & X     & X     & X     & X     & An energy efficient MAC with clustering and TDMA scheduling for mobility and collisions. & \blue{TDMA scheduling with mobility.} \\[-0.5em]
    \midrule
    \cite{ATLR2018} & 2018  & X     & X     & X     & X     & \checkmark     & X     & X     & X     & X     & X     & X     & X     & X     & X     & X     & X     & X     & X     & X     & X     & X     & X     & An autonomous relay algorithm is presented for vehicular networks. & \blue{Short directional links for autonomous vehicular communication.} \\[-0.5em]
    \midrule
    \cite{ISCT2018} & 2018  & \checkmark     & X     & X     & X     & X     & X     & X     & X     & X     & X     & \checkmark     & X     & X     & X     & X     & X     & X     & X     & X     & X     & X     & X     & A secure and intelligent spectrum control strategy is presented with fixed channel. & \blue{Security in Terahertz band with coding and interference consideration.} \\[-0.5em]
    \midrule
    \cite{B5G2018} & 2018  & X     & X     & X     & X     & \checkmark     & X     & X     & X     & X     & X     & \checkmark     & X     & X     & X     & X     & X     & X     & X     & \checkmark     & \checkmark     & X     & X     & Channel switching based on distance, signalling overhead, throughput maximization and error recovery. & \blue{Spectrum switching with distance, signalling overhead, error and throughput consideration.} \\[-0.5em]
    \midrule
    \cite{2stateMAC2018} & 2018  & X     & X     & \checkmark     & X     & X     & X     & X     & X     & X     & X     & X     & X     & X     & \checkmark     & X     & X     & X     & X     & X     & \checkmark     & X     & X     & Throughput maximization with molecular absorption, interference, energy harvesting, and link capacity. & \blue{Throughput maximization and energy harvesting.} \\[-0.5em]
    \midrule
    \cite{slottedCSMAMAC2018} & 2018  & X     & X     & X     & X     & X     & X     & X     & \checkmark     & \checkmark     & X     & \checkmark     & X     & \checkmark     & X     & X     & X     & \checkmark     & \checkmark     & X     & X     & X     & X     & Performance of energy consumption with dynamic super-frame durations and packet lengths. & \blue{Packet length and energy consumption analysis.} \\[-0.5em]
    \midrule
    \cite{MAC-Yugi2018} & 2018  & \checkmark     & \checkmark     & X     & X     & X     & X     & X     & X     & X     & X     & \checkmark     & X     & X     & X     & X     & X     & X     & X     & X     & X     & X     & \checkmark     & A MAC Yugi-Ada antenna is presented for frequency and beam direction reconfigurability. & \blue{Reconfigurable antenna and programmable physical layer for MAC protocol.} \\[-0.5em]
    \bottomrule
    \end{tabular}%
  \label{tab:paraawaree}%
\end{table*}%

\section{Design issues and considerations for Terahertz MAC protocols}\label{sec:requirements}

\blue{This section discusses the feature, design issues, and challenges which needs to be considered while designing efficient THz MAC protocols.}

\subsection{Feature of Terahertz band communication related to Terahertz MAC protocol design}\label{sec-req:features}

By using frequencies above $0.1$ THz, new propagation phenomena can appear such as reflection, wave absorption by some molecules and channel noise generation~\cite{Salous2013}. The understanding of the Terahertz band seems to be crucial to design systems exploiting this frequency, hence, researchers are focusing on the behavior of Terahertz wave traveling under different environments and in the presence of items such as walls, concrete or grass. Following are the Terahertz band features which can affect the MAC-layer performance including throughput and delay.

\subsubsection{Path loss}\label{sec-req:pathloss}

To realize the Terahertz band and its characteristics, it is important to understand its propagation phenomenon and to analyze the impact of molecular absorption on the path loss and noise~\cite{chmodel2011,GORDON20173,haixia2011,khaledn2019}. The current efforts are mainly focused on channel characterization at 300 GHz band~\cite{hsong2010,spriebe2011,spriebe2013,spriebekannicht2013,tsujimura2018}. As Terahertz wave propagates, it suffers from different types of attenuation due to absorption, scattering, and refraction~\cite{aafshari2015}. It can follow different paths at the receiver as the sum of non-/line of sight. The path loss includes the spreading as well as the absorption loss. The spreading loss occurs due to the expansion of waves as it propagates through the medium, whereas the absorption loss occurs when a Terahertz wave suffers from the molecular absorption at the medium~\cite{joonas2016}. These losses make a Terahertz band a frequency selective. The spreading loss can be given as~\cite{itufspl2016},

\begin{equation}\label{eq:ploss1}
a_{1}(f,d)=\left( \frac{c}{4\pi f d}\right)^{2}
\end{equation}

\noindent \saim{whereas the absorption loss depends upon the parameters such as the temperature, pressure, and distance, and can be demonstrated as, }

\begin{equation}\label{eq:ploss2}
a_{2}(f,d)=e^{-K(f)d}
\end{equation}

\noindent \saim{where $K(f)$ is the total absorption coefficient and $d$ is the distance between transmitter and receiver~\cite{radiowave2013}. $K(f)$ can be calculated using the High-Resolution Transmission Molecular Database (HITRAN) database~\cite{GORDON20173}.}

For a particular transmission distance, the path loss increases with frequency due to spreading loss. For a few meters distance, the path loss can increase up to 100 dB. Further, the molecular absorption defines several transmission windows depending upon the transmission distance. For a few centimeters distance, the transmission window behaves like a 10 Terahertz wide transmission window due to negligible absorption loss. However, for distance more than 1 meter the absorption becomes significant which narrows down the transmission window. Such extreme path loss results in reduced bandwidth and only a few transmission windows. Different transmission windows are marked as feasible in~\cite{Akyildiz2014} showing up to less than 10 dB of Path loss due to negligible impact of molecular absorption. However, due to the spreading loss, the path loss remains higher, which motivates the usage of highly directional antennas and MIMO techniques~\cite{Akyildiz2014}. The Terahertz wave can be absorbed by raindrops, ices, foliage and grass and any medium containing water molecule~\cite{bhardwaj2013}.

\subsubsection{Noise}\label{sec-req:noise}

Within the Terahertz band, the molecules presented in the medium are excited by \saim{electromagnetic waves} at specific frequencies. These excited molecules vibrate internally where the atom vibrates in periodic motion and the molecule vibrates in a constant translational and rotational motion. Due to the internal vibration, the energy of the propagating waves is converted into kinetic energy partly. From the communication perspective, it can be referred to as a loss of signal. Such molecule vibration at given frequencies can be obtained by solving the Schrodinger equation for particular molecular structure~\cite{miller2008}. A model for computation of attenuation by gases in the atmosphere is also described by International Telecommunication Union, which considers the water vapor and oxygen molecules over Terahertz band from 1-1000 GHz~\cite{ituattenu2016}. A HITRAN database is also found useful for the computation of attenuation due to molecular absorption in Terahertz band~\cite{GORDON20173}. 

The molecular absorption is an important issue to consider along with free space path loss, as it also causes the loss to the signals due to partial transformation of electromagnetic energy into internal energy~\cite{chmodel2011,painescott2018}. Such transformation in the Terahertz band can introduce noise which can be due to atmospheric temperature or the transmission in the radio channel. The noise occurs due to atmosphere temperature (such as Sun) can be referred as Sky-noise~\cite{smithernest1982,itunoise2016,fbox1986}, \saim{and} the noise introduced due to transmission in the radio channel can be referred as the molecular absorption noise~\cite{joonas2016,chmodel2011,bronin2014}. A noise model for Terahertz wireless nanosensor networks with individual noise sources that impact intra-body systems is presented in~\cite{Elayan2018Ete}, with noise contributions of Johnson-Nyquist, black body, and Doppler-shift induced noise. 

\textit{Molecular absorption noise:} The molecular noise is the result of radiation of absorbed Terahertz energy by molecules which depends on the propagation environment. The fundamental equation of molecular noise under different assumptions, such as medium homogeneity or scattering properties, can be directly derived from radiative transfer theory~\cite{raditationbrown2003}. The absorption is generally caused when the transmitted EM wave shifts the medium to higher energy states, where the difference between the higher and lower energy state of a molecule determines the absorption energy which is drawn from the EM wave. It has a direct impact on the frequency as the absorbed energy is $E = hf$, where $h$ is the Planck's constant and $f$ is frequency~\cite{miller2008}. It can also be described stochastically using the absorption coefficient $K_{a}(f)$, which describes the average effective area of molecules per unit volume and depends upon frequency due to which the Terahertz band has a unique frequency-selective absorption profile. Similarly, the amount of radiation capable of penetrating through the absorption medium is known as transmittance, which can also be defined by the Beer's Lambert's Law~\cite{chmodel2011,painescott2018,raditationbrown2003} (\ref{eq:ploss2}). Further details on the calculation of molecular absorption coefficient and model can be found in~\cite{chmodel2011,painescott2018,boronin2015, radiative2015}.

\textit{Sky noise:} The Sky-noise is independent of the transmitted signals and can be known as background noise. It is caused by the temperature of the absorbing atmosphere and can be termed as an effective blackbody radiator or grey body radiator for a non-homogeneous atmosphere medium. Several papers have described the Sky-noise, like~\cite{ippolito1985,raditationbrown2003,smithernest1982,fbox1986,straiton1975}. It is identified in satellite communication and is mostly affected by the antenna temperature which is an additional temperature that accounts for the radiation from the absorbing temperature. The atmosphere can be considered as a dynamic medium with decreasing temperature and pressure as a function of elevation. In general, it depends upon the absorption coefficient and the distance due to the variable temperature and pressure in the atmosphere~\cite{joonas2016}. When the distance is small and the atmosphere is more likely homogeneous\blue{,} the absorption coefficient can be \blue{calculated as} as $K_{a}(s,f) = K_{a}(f)$ where $s$ represents the distance.

\textit{Black body noise:} A body with temperature T radiates energy, the energy can reach its maximum value for a given wavelength according to the Wien displacement law~\cite{feymenbook2015}. This phenomenon is known as black body radiation and it contributes for a specific range of temperature to the total noise of the Terahertz system~\cite{raditationbrown2003,Elayan2018Ete}.

%
%

%
%

\subsubsection{Terahertz scattering and reflection}\label{sec-req:scatt-ref}

Reflection and scattering are two physical properties that characterize electromagnetic wave, the region between transmitter and receiver can contain a large number of scatters with different sizes and are distributed randomly. There are two types of scattering: elastic scattering in which only the direction of the wave is changed, and inelastic scattering in which the scatter introduces a change to the energy. The scattering processes include Rayleigh scattering which occurs when the dimension of scatter diameter is larger than the Terahertz wavelength and Mie scattering otherwise. Mie and Rayleigh scattering can affect received Terahertz signal~\cite{kokko2015,mwu2015,ywang2016}. 
 In~\cite{skim2015}, a statistical model for Terahertz scatters channel is proposed, based on indoor measurements of a point-to-point link and transmitter and receiver were equipped with directional antennas, at $300 GHz$ window and a bandwidth equal to $10 GHz$. \saim{A scattering power radiated from different surfaces is analyzed in~\cite{Ju2019} for spectrum ranging from 1 GHz to 1 THz. Two scattering models are analyzed, the direct scattering model and radar cross-section model. The scattering can be an important propagation mechanism and for frequencies above 100 GHz, \saim{it} can be treated as a simple reflection~\cite{Ju2019}.} 
 
 Radio wave reflections occur commonly in indoor scenarios. The reflected ray depends on the electromagnetic properties of the reflector, the surface roughness and the location of the reflectors with respect to the transmitter and receiver. The received signal at the $R_X$ side is the sum of direct ray and all reflected rays. In~\cite{Jianjun2018}, a demonstrator is set up for four frequency windows: $100$, $200$, $300$ and $400$ GHz, to characterize reflections in each window. The reflection coefficient is given by:

 \begin{equation}
 r=\frac{Zcos(\theta_{i})-Z_{0}cos(\theta_{t})}{Zcos(\theta_{i})+Z_{0}cos(\theta_{t})}
 \end{equation} 
 
\noindent \saim{where, $Z_{0}=337\Omega$ is the wave impedance in free space and $Z$ the impedance of the reflector. $Z$ depends on frequency, material relative index and absorption coefficient. $\theta_i$ is the angle of incidence and $\theta_t$ is the angle of transmission. The reflected wave can be reduced using phased array antenna~\cite{skim2015,Jianjun2018}.}

\subsubsection{Multi-path}\label{sec-req:multipath}

In the presence of reflectors and scatters, a non-line of sight (NLOS) can be generated by the channel for Terahertz waves. In Terahertz communication, where the line of sight (LOS) and NLOS exist together, the NLOS can interfere with the main signal in LOS at the receiving side~\cite{Xiao18}. The advantage of NLOS component is when the LOS is obstructed, the receiver can still decode the transmitted signal. The magnitude of the received signal at the receiver depends on parameters such as reflector permittivity which characterizes the material, reflector roughness coefficient, incidence angle, and wave polarization and finally its position toward transmitter and receiver~\cite{radiowave2013,esaware2013}. The magnitude of the NLOS signal is also affected by the antenna properties, the distance between the source and receiver, and the plane containing the reflector.

Both LOS and NLOS propagation scenarios exist in the indoor environment~\cite{Moldovan2014} where the presence of NLOS is mainly due to scatters and reflectors. The channel attenuations and delays can be estimated using NLOS and LOS components of channel impulse response $h(f,t)$ by:


 \begin{equation}
 h(f,t)=\sqrt{l(d_{0},f)}\delta(t-t_{0})+\sum_{j=1}^{N_{NLOS}}\sqrt{l^{'}(d_{j},f)}\delta(t-t_{j})
 \end{equation}
 
\noindent \saim{where, $N_{NLOS}$ is the number of NLOS paths, $d$ is the distance, $f$ is frequency, $\delta$ is the Dirac function, and $l$ is the total attenuation and can be written as, }
 
 
 \begin{equation}
 l(d_0,f) = a_1(d_0,f) * a_2(d_0,f)
\end{equation}  

\noindent \saim{and,}

\begin{equation}
 l^{'}(d_j,f) = r^{2} a_1(d_j,f) * a_2(d_j,f)
\end{equation}  
 
\noindent \saim{Delay parameters in Equation 6 affect some of MAC decisions such as modulation and coding selection module, antenna beam steering module, then the estimation of delay parameters can help to select or switching to the path that gives the lowest attenuation for the link. Presence of NLOS and LOS components can be used as an alternative for link communication outage in which if LOS is blocked the NLOS can be used as an alternate path.}

\subsubsection{Terahertz transmission windows}\label{sec-req:trwindow}

The path losses which occur in Terahertz wave communication give these bands a frequency selective behavior in which some chunks of bands can be used to provide higher bandwidth due to less amount of losses. Terahertz windows for communication depends on many parameters, such as communication range and technology requirements. The distance-dependant bandwidth is given by~\cite{esaware2013}:

\begin{equation}
B_{3dB}(d)=\{f|\frac{a_{1}(f,d)a_{2}(f,d)}{N(f,d)}\geq \frac{a_{1}(f_0,d)a_{2}(f_0,d)}{N(f_0,d)}-3dB \}
\end{equation}

\noindent \saim{where, $f_{0}$ is the central frequency, $a_{1}$ is the spreading loss, $a_{2}$ is the absorption loss and $N(f,d)$ is the total molecular noise.}

Typically, four Terahertz windows can be exploited within the band $[0.1-1 THz]$ for a communication range of $1-10 m$. The optimal compact window with low attenuations and high bandwidth is the one centered around $0.3 THz$. The $300 GHz$ window is characterized by an available bandwidth of $69.12 GHz$, subdivided into separate channels or sub-bands. The supported channels for Terahertz communication for the frequency range from $F_{min}=252.72$ GHz to $F_{max}=321.84$ GHz were proposed by IEEE 802.15.3d wireless personal area networks (WPAN) working group and summarized in Table~\ref{tab:sub-bands}~\cite{ieee802153d2017,Salous2013}. \blue{In~\cite{chandtmc2014}, transmission windows are selected based on the distance between the nodes. This is due to higher attenuation in channel impulse response as a result of molecular absorption at longer distance.}

\begin{table}[htbp!]
	\centering
	\caption{Bandwidth and maximum achievable data rate for sub bands of 0.3 THz window for single career using high order modulation~\protect\cite{ieee802153d2017}.}
	\label{tab:sub-bands}
	\begin{tabular}{|c||c|c|}
		\hline
		 Bandwidth(GHz)& Index range & Data Rate (Gbps)\\
		\hline
		\hline
		 $2.16$&$1-32$ & 9.86   \\
		\hline
		  $4.32$&$33-48$  & 19.71 \\
		\hline
		 $8.64$&$49-56$  & 39.42  \\
		\hline
		 $12.96$&$57-61$ & 59.14    \\
		\hline
		 $17.28$&$62-65$ & 78.85 \\
		\hline
		 $25.92$&$66-67$  & 118.27 \\
		\hline
		 $51.84$&$68$  & 236.54\\
		\hline
		 $69.12$&$69$ & 315.39\\
		\hline
	\end{tabular}
\end{table}

\subsection{Design issues and considerations for Terahertz MAC protocols} \label{sec-req:design-issues}

The Terahertz band can provide high bandwidth for future high-speed networks. However, possesses unique features as discussed in the above section which can affect communication performance. These features do not affect the MAC performance directly but impact hugely the Physical layer design, antenna and link capacity, which affects the MAC-layer performance, throughput, and latency. The choice of physical layer functionalities can also affect the MAC layer design such as antenna technology, modulation and coding scheme, and waveform. MAC functionalities depends on channel characteristics, device technology, and physical layer features. There are several design issues related to Physical and MAC layer features which should be considered for designing an efficient MAC protocol for different applications. These issues and considerations are highlighted in Table~\ref{tab:summ-issue-features} with Terahertz features and decisions to be taken at MAC layer.

\begin{table*}[htbp]
 \centering
  \tiny
  \caption{Terahertz features, design issues and considered in existing Terahertz MAC protocols.}
    \begin{tabular}{|c|c|p{9.665em}|p{1.335em}p{1.335em}p{1.335em}p{1.335em}p{1.335em}|cccccccc|cccccccc|}
\cmidrule{9-24}    \multicolumn{1}{c}{} & \multicolumn{1}{c}{} & \multicolumn{1}{c}{} & \multicolumn{1}{c}{} & \multicolumn{1}{c}{} & \multicolumn{1}{c}{} & \multicolumn{1}{c}{} & \multicolumn{1}{c|}{} & \multicolumn{16}{p{22.24em}|}{\textbf{Terahertz MAC design issues and considerations}} \\
\cmidrule{4-24}    \multicolumn{1}{c}{} & \multicolumn{1}{c}{} & \multicolumn{1}{c|}{} & \multicolumn{5}{p{6.175em}|}{\textbf{THz band features}} & \multicolumn{8}{p{10.28em}|}{\textbf{Physical layer \& device related}} & \multicolumn{8}{p{14.16em}|}{\textbf{MAC layer related}} \\
    \midrule
    \multicolumn{1}{|p{3.28em}|}{\textbf{Network scale}} & \multicolumn{1}{p{2.11em}|}{\textbf{Ref.}} & \textbf{Application area} & \multicolumn{1}{p{1.135em}|}{\begin{sideways}\textbf{Path loss}\end{sideways}} & \multicolumn{1}{p{1.135em}|}{\begin{sideways}\textbf{Noise}\end{sideways}} & \multicolumn{1}{p{1.05em}|}{\begin{sideways}\textbf{Scattering \& reflections }\end{sideways}} & \multicolumn{1}{p{1.05em}|}{\begin{sideways}\textbf{Multi path}\end{sideways}} & \begin{sideways}\textbf{Transmission windows}\end{sideways} & \multicolumn{1}{p{1.135em}|}{\begin{sideways}\textbf{Channel model}\end{sideways}} & \multicolumn{1}{p{1.035em}|}{\begin{sideways}\textbf{Antenna}\end{sideways}} & \multicolumn{1}{p{1.035em}|}{\begin{sideways}\textbf{Interference and noise}\end{sideways}} & \multicolumn{1}{p{1.035em}|}{\begin{sideways}\textbf{THz device}\end{sideways}} & \multicolumn{1}{p{1.035em}|}{\begin{sideways}\textbf{Waveform}\end{sideways}} & \multicolumn{1}{p{1.035em}|}{\begin{sideways}\textbf{Modulation and coding}\end{sideways}} & \multicolumn{1}{p{1.035em}|}{\begin{sideways}\textbf{Link budget}\end{sideways}} & \multicolumn{1}{p{1.035em}|}{\begin{sideways}\textbf{Channel capacity}\end{sideways}} & \multicolumn{1}{p{1.035em}|}{\begin{sideways}\textbf{Channel access}\end{sideways}} & \multicolumn{1}{p{1.81em}|}{\begin{sideways}\textbf{Neighbor discovery and  link establishment}\end{sideways}} & \multicolumn{1}{p{2.01em}|}{\begin{sideways}\textbf{Mobility management and handovers}\end{sideways}} & \multicolumn{1}{p{2.01em}|}{\begin{sideways}\textbf{Collisions and multi user interference}\end{sideways}} & \multicolumn{1}{p{1.09em}|}{\begin{sideways}\textbf{Throughput and latency}\end{sideways}} & \multicolumn{1}{p{1.05em}|}{\begin{sideways}\textbf{Reliability}\end{sideways}} & \multicolumn{1}{p{1.05em}|}{\begin{sideways}\textbf{Coverage and connectivity}\end{sideways}} & \multicolumn{1}{p{1.05em}|}{\begin{sideways}\textbf{Energy considerations}\end{sideways}} \\
    \midrule
    \multirow{25}[2]{*}{nano} & \cite{pkt_size2016} & Wireless Nanosensor\newline{}Networks & \checkmark     & X     & X     & X     & X     & X     & X     & X     & X     & X     & \checkmark     & X     & X     & X     & X     & X     & X     & \checkmark     & X     & X     & \checkmark  \\
          & \cite{APIS2017} & Wireless Nanosensor\newline{}Networks & X     & X     & X     & X     & X     & X     & X     & X     & X     & X     & \checkmark     & X     & X     & X     & X     & X     & \checkmark     & \checkmark     & \checkmark     & X     & \checkmark  \\
          & \cite{MGDI2016} & Software defined metamaterials & \checkmark     & X     & \checkmark     & X     & X     & \checkmark     & X     & \checkmark     & X     & X     & X     & \checkmark     & X     & X     & X     & X     & \checkmark     & \checkmark     & \checkmark     & \checkmark     & \checkmark  \\
          & \cite{Liaskos2014} & SDM   & X     & X     & X     & X     & X     & X     & X     & \checkmark     & X     & X     & X     & X     & X     & X     & X     & X     & \checkmark     & X     & X     & X     & \checkmark  \\
          & \cite{joint_error2016} & Nanonetworks & \checkmark     & \checkmark     & X     & X     & X     & \checkmark     & X     & \checkmark     & X     & X     & \checkmark     & \checkmark     & X     & X     & X     & X     & X     & \checkmark     & \checkmark     & X     & \checkmark  \\
          & \cite{EESR-MAC2012} & Nanonetworks & X     & X     & X     & X     & X     & X     & X     & X     & X     & X     & X     & X     & X     & \checkmark     & \checkmark     & X     & \checkmark     & X     & \checkmark     & X     & \checkmark  \\
          & \cite{MDP2014} & Nanonetworks & X     & X     & X     & X     & X     & X     & X     & X     & X     & X     & X     & X     & X     & X     & X     & X     & X     & X     & X     & X     & \checkmark  \\
          & \cite{RIH-MAC2014} & Nanonetworks & X     & X     & X     & X     & X     & X     & X     & X     & X     & X     & \checkmark     & X     & X     & \checkmark     & \checkmark     & X     & \checkmark     & \checkmark     & X     & X     & \checkmark  \\
          & \cite{Mabed2018} & Nanonetworks & X     & X     & X     & X     & X     & X     & X     & \checkmark     & X     & X     & \checkmark     & X     & X     & X     & X     & X     & \checkmark     & \checkmark     & \checkmark     & X     & \checkmark  \\
          & \cite{smart-mac2013} & Health monitoring & X     & X     & X     & X     & X     & X     & \checkmark     & X     & \checkmark     & X     & \checkmark     & X     & X     & \checkmark     & \checkmark     & \checkmark     & \checkmark     & X     & X     & X     & \checkmark  \\
          & \cite{Design-WNSN2015} & In-body Nanonetworks & \checkmark     & \checkmark     & X     & X     & X     & X     & X     & X     & X     & X     & X     & X     & \checkmark     & X     & X     & X     & X     & X     & X     & X     & \checkmark  \\
          & \cite{slottedCSMAMAC2018} & Health monitoring & X     & X     & X     & X     & X     & X     & X     & X     & X     & X     & \checkmark     & X     & X     & \checkmark     & \checkmark     & \checkmark     & \checkmark     & X     & X     & X     & \checkmark  \\
          & \cite{TSN2014} & Industrial monitoring & X     & \checkmark     & X     & X     & X     & \checkmark     & X     & \checkmark     & X     & X     & X     & X     & \checkmark     & \checkmark     & X     & X     & X     & X     & X     & X     & \checkmark  \\
          & \cite{dynamic_CH2016} & Agriculture monitoring & \checkmark     & \checkmark     & X     & X     & X     & \checkmark     & X     & \checkmark     & X     & X     & X     & X     & X     & \checkmark     & X     & X     & \checkmark     & \checkmark     & X     & X     & \checkmark  \\
          & \cite{MAC-Yugi2018} & Wireless Nanosensor\newline{}Networks & X     & X     & X     & X     & \checkmark     & X     & \checkmark     & X     & X     & X     & X     & X     & X     & \checkmark     & X     & X     & X     & X     & X     & X     & X \\
          & \cite{RBMP2017} & Nanonetworks & X     & X     & X     & X     & X     & X     & X     & X     & X     & X     & X     & X     & X     & X     & X     & X     & X     & \checkmark     & X     & X     & X \\
          & \cite{TCN2015} & Nanonetworks & X     & X     & X     & X     & X     & X     & X     & X     & X     & X     & X     & X     & X     & \checkmark     & X     & X     & \checkmark     & \checkmark     & \checkmark     & X     & \checkmark  \\
          & \cite{LLModelling2011} & Health monitoring & \multicolumn{1}{c}{X} & \multicolumn{1}{c}{X} & \multicolumn{1}{c}{X} & \multicolumn{1}{c}{X} & \multicolumn{1}{c|}{X} & X     & X     & X     & X     & X     & X     & X     & X     & X     & X     & X     & X     & \checkmark     & X     & X     & X \\
          & \cite{LLsynch2015} & Nano/Macro-networks & \checkmark     & \checkmark     & X     & X     & X     & \checkmark     & \checkmark     & X     & X     & X     & X     & X     & X     & \checkmark     & X     & X     & \checkmark     & \checkmark     & X     & X     & \checkmark  \\
          & \cite{NS-MAC2016} & Wireless Nanosensor\newline{}Networks & X     & \checkmark     & X     & X     & X     & X     & X     & X     & \checkmark     & X     & X     & X     & X     & \checkmark     & \checkmark     & X     & \checkmark     & \checkmark     & X     & X     & \checkmark  \\
          & \cite{DMDS2016} & Internet of Nano-Things & \checkmark     & \checkmark     & X     & X     & X     & X     & X     & X     & \checkmark     & X     & X     & X     & \checkmark     & \checkmark     & \checkmark     & X     & X     & \checkmark     & \checkmark     & X     & \checkmark  \\
          & \cite{drih-mac2015} & Health monitoring & \checkmark     & X     & X     & X     & X     & X     & X     & X     & \checkmark     & X     & X     & X     & X     & \checkmark     & \checkmark     & X     & \checkmark     & X     & \checkmark     & X     & \checkmark  \\
          & \cite{2stateMAC2018} & Nanonetworks & \checkmark     & \checkmark     & X     & \checkmark     & X     & X     & X     & X     & X     & X     & X     & X     & \checkmark     & \checkmark     & X     & X     & \checkmark     & \checkmark     & X     & \checkmark     & \checkmark  \\
          & \cite{phlame2012} & Nanonetworks & \checkmark     & \checkmark     & X     & X     & X     & X     & X     & \checkmark     & X     & \checkmark     & \checkmark     & X     & X     & \checkmark     & \checkmark     & X     & \checkmark     & \checkmark     & X     & X     & \checkmark  \\
          & \cite{esaware2013} & Wireless Nanosensor\newline{}Networks & \checkmark     & \checkmark     & \checkmark     & X     & X     & \checkmark     & X     & X     & X     & X     & X     & X     & \checkmark     & \checkmark     & \checkmark     & X     & \checkmark     & X     & X     & X     & \checkmark  \\
    \midrule
    \multirow{10}[2]{*}{macro} & \cite{MRA-MAC2017} & THz communication network & \checkmark     & \checkmark     & X     & X     & \checkmark     & \checkmark     & \checkmark     & \checkmark     & X     & \checkmark     & X     & X     & \checkmark     & X     & \checkmark     & X     & X     & \checkmark     & X     & \checkmark     & X \\
          & \cite{IHTLD-MAC2017} & THz Wireless Personal Area Networks & X     & X     & X     & X     & X     & X     & X     & X     & X     & X     & X     & X     & X     & \checkmark     & \checkmark     & X     & X     & \checkmark     & \checkmark     & X     & X \\
          & \cite{SDNC2017} & THz Vehicular networks and small cells, SDN & \checkmark     & \checkmark     & X     & X     & X     & \checkmark     & X     & \checkmark     & X     & X     & X     & X     & \checkmark     & \checkmark     & \checkmark     & \checkmark     & X     & \checkmark     & \checkmark     & \checkmark     & X \\
          & \cite{DLLC2017} & THz communication network & X     & X     & X     & X     & X     & \checkmark     & X     & X     & \checkmark     & X     & \checkmark     & X     & X     & X     & X     & X     & X     & \checkmark     & \checkmark     & X     & \checkmark  \\
          & \cite{OPTRS2017} & THz communication network & X     & \checkmark     & X     & X     & X     & \checkmark     & X     & \checkmark     & X     & X     & X     & X     & X     & X     & X     & X     & X     & \checkmark     & X     & \checkmark     & \checkmark  \\
          & \cite{MAC-TUDWN2013} & THz Wireless Personal Area Networks & X     & X     & X     & X     & X     & X     & X     & X     & X     & X     & X     & X     & X     & \checkmark     & \checkmark     & X     & X     & \checkmark     & \checkmark     & X     & X \\
          & \cite{TAB-MAC2016} & THz communication network & X     & \checkmark     & X     & X     & X     & \checkmark     & \checkmark     & \checkmark     & X     & X     & X     & X     & X     & X     & \checkmark     & X     & \checkmark     & \checkmark     & \checkmark     & \checkmark     & X \\
          & \cite{ATLR2018} & THz Vehicular network & X     & X     & X     & X     & X     & \checkmark     & X     & X     & X     & \checkmark     & X     & X     & \checkmark     & X     & X     & \checkmark     & X     & \checkmark     & \checkmark     & \checkmark     & X \\
          & \cite{MAADM2017} & THz communication network, indoor networks & \checkmark     & \checkmark     & X     & X     & X     & \checkmark     & \checkmark     & X     & X     & \checkmark     & X     & X     & X     & X     & X     & X     & X     & \checkmark     & X     & X     & X \\
          & \cite{TRPLE2014} & THz communication network, indoor networks & \checkmark     & \checkmark     & \checkmark     & X     & X     & \checkmark     & \checkmark     & X     & X     & X     & \checkmark     & X     & X     & \checkmark     & \checkmark     & X     & X     & \checkmark     & \checkmark     & X     & \checkmark  \\
    \bottomrule
    \end{tabular}%
  \label{tab:summ-issue-features}%
\end{table*}%

\subsubsection{\saim{Physical layer and device related issues and considerations}}\label{sec-req:phylayer} \hfill

\textit{Antenna technology: } The transmitted Terahertz signal undergoes several impairments due to the propagation through a medium, ranging from free space loss caused by high frequencies to molecular absorption noise and scattering. To cope with this issue, high gain antennas are required to strengthen the signal in one particular direction and to compensate losses~\cite{ningzhu2013}. In communication networks, many nodes try to access the shared channel, it will be challenging to simultaneously serve all of them if we assume antenna is directional. 
 
 The antenna technology endowed with fast beam switching capability can determine the way nodes access the shared channel using narrow beams and by assigning each node access to the channel for a given time slot assigned by the MAC layer~\cite{woongsoo2015}. MAC layer should include an antenna steering module to rapidly steer beams toward the receiver. For example, the switching can be performed at the pulse or packet level. However, the MAC and antennas of different nodes should be well synchronized in order to reduce errors and delays. With optimized antenna gain, it is possible to reach high data rate and good signal quality. Massive MIMO antenna is also envisioned for Terahertz applications \saim{which} can enhance MAC performance by increasing the data throughput and by serving more nodes in the network using spatial multiplexing techniques~\cite{chan2018mimo}, where the number of small antennas can reach 1024. In macro scale communication\blue{,} it is still an open challenge for deep investigations.   
 
\saim{In nano communication networks, nodes inter-distance is short, therefore antenna is assumed to be isotropic and non-complex.} Efforts to design terahertz antennas goes back to some few years, radiation is possible at this frequency band for antennas consisting of materials such as InGaAs and graphene taking advantages from their chemical and electronic properties~\cite{ssdhillon2017,jornetcabellos2015,esquius2014,graphenejoornakil2010}. 
Antenna dimensions are in the order of micrometers for THz frequencies. The second issue to be considered is materials to be used for antenna design and feeding. Some interesting results achieved for the nano-antenna industry~\cite{jornetcabellos2015,elayan2016,jornetakylidiz2013}, power consumption and radiated power, operating band and directivity are the main properties of antenna~\cite{mnaftally2017}. The mutual coupling is also an important challenge to address due to ultra-dense integration of multi-band nanoantenna arrays. A frequency selective approach is proposed in~\cite{zhangjornakil2019}, to reduce the coupling effects for multi-band ultra massive MIMO systems.
 
\saim{To mitigate high attenuation and multipath problem, fast beam switching and steering techniques can be managed by MAC, phased array antennas are the best choice to meet this requirement. The phased arrays can improve the link budget and also increase fairness among users via beam switching and steering\blue{~\cite{Kherani2018,yli2018,khalids2019,moshir2016}}. However\blue{,} the actual status of technology is still developing towards reducing the switching time and increasing the number of antenna elements in order to increase gain. Mathematically, the antenna gain for a uniform planar array at a specific angular direction $(\theta,\phi)$ describes both functionalities of beamforming and beam steering and can be calculated by~\cite{phasearr2005}:}

\begin{figure}[htbp]
\centering
\includegraphics[width=3.2in,height=2in]{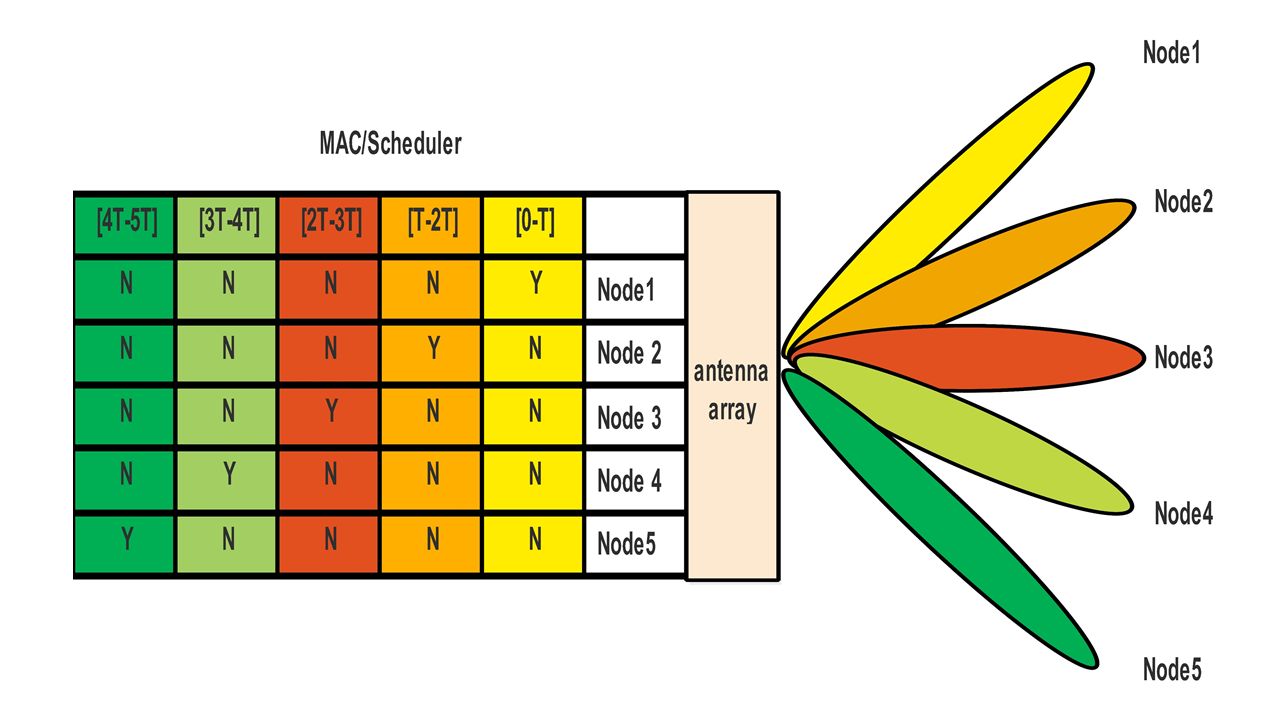}
\caption{MAC scheduler with an antenna array. The figure shows the MAC scheduler module and how it is linked to the antenna system, if the current node $i$ needs to send data to node $j$, then MAC sends a command to the antenna system, coded  to steer its beams toward node $j$ using a digital to analog module interfacing between MAC and the physical layer. The beam steering operation can be repeated for each frame period to schedule a new transmission. }
\label{fig:antenna_sched}
\end{figure}
 
\begin{equation}
\begin{split}
 G(\theta,\phi) & = G_{max} \frac{sin(Ma(sin(\theta)cos(\phi)-\nu_{0}))}{M sin(a(sin(\theta)cos(\phi)-\nu_{0}))} \\ 
 &\qquad\quad \frac{sin(Nb(sin(\theta)sin(\phi)-\nu_{1}))}{N sin(b(sin(\theta)sin(\phi)-\nu_{1}))} 
 \end{split}
\end{equation}

\noindent \saim{where $a$ and $b$ are two parameters related to vertical and horizontal separation between antenna elements and also a function of Terahertz frequency. $\nu_{0}$ and $\nu_{1}$ are two horizontal and vertical steering parameters. MAC layer should select properly $\nu_{0}$ and $\nu_{1}$ to establish a communication link with a node, $G_{max}$ denotes the maximum antenna gain. }
 
The Terahertz MAC scheduler maps traffic data to each antenna beam as depicted in Figure~\ref{fig:antenna_sched}, the diagram shows an example of mapping between user data traffic and antenna beams. MAC selects, based on traffic requirements, for each transmission time interval $[(n-1)T, nT]$, a destination node and its associated beam to transmit data traffic. Many scheduling algorithms can be used such as Round Robin, maximum throughput or minimum delay algorithms. The switching operation of the beam can be performed at a pulse, symbol or frame level.

\textit{Interference model and SINR: } Interference exists in the Terahertz communication system and it can be generated from nodes using the same frequency band at the same time or from the signal itself. Interferences can be caused also by reflected and scattered signals~\cite{vpetrovmmwave2017,vpetrovsinr2015} for either fixed or mobile users. Research works on interference modeling are not sufficient, as a signal to interference ratio depends mainly on the channel model. The interference level affects signal quality and leads to higher bit error rate (BER). The design of MAC should be aware of interference level by enhancing nodes synchronization or adopting channelization methods to access the channel. Access methods define the way each node transmits its data, then elaborating an interference model will help deeply selecting the right access technique.

\textit{Link budget and capacity: } A communication link is characterized by link budget and system capacity. Link budget includes transmitting power, all gains, and losses. Link quality is good when the link budget value is higher than the receiver signal to noise ratio threshold. \saim{This} threshold characterizes the receiver device as well as the bandwidth. Link budget gives the information of all power gains and losses that should be present in one Terahertz link, Terahertz link budget depends on many factors such as antenna gains, atmospheric attenuations, available bandwidth, distance, temperature,  total noise, and the THz source output power. The link budget should \saim{be} higher than a fixed threshold \saim{which} mainly depends on device technology, to guarantee reliable Terahertz communication. Enhancement of link budget leads to increase in reachability, and reduce data loss. It is used as a reference metric to determine the link range. Shannon capacity is derived from the mutual information maximization between sender and receiver for a particular channel model, it indicates how much data can be transmitted for a given bandwidth and SINR for different scenarios.

Most of the studies on Terahertz capacity analysis derived for the nanoscale Terahertz network~\cite{jmjor2011,chmodel2011,chcapacity2010}, and macro-scale networks~\cite{fytampanis2017} are based on theoretical assumptions and deterministic propagation channel. In realistic scenarios, additional properties of the Terahertz wave should be considered such as scattering, dispersion, atmospheric factors, and Terahertz statistical model.

 Capacity increases with bandwidth and SINR, as a result, MAC layer design should take into consideration the achievable channel capacity for different channel models, as throughput is bounded by the maximum data rate.  
MAC layer should be aware of link quality: link budget and capacity; frames should be protected against errors\blue{. M}oreover, frame length and transmission duration should be tuned. For MAC design, link requirements should be considered, for example, in the Datacenter use case, the data rate can exceed $100Gbps$, then efficient tracking of link fluctuation is required.
Knowledge of channel capacity and link budget enhances MAC awareness of the channel and the physical layer via frame optimization and transmission scheduling. In~\cite{fytampanis2017}, the impact of the outdoor channel on fixed link capacity is studied at a 1 km distance. The channel capacity and BER performance for data transmission are analyzed in~\cite{Akkas2017}.

\saim{\textit{Modulation and coding: }} \saim{The Terahertz channel is characterized by high free space loss, molecular absorption, and noise, as well as limitations of transceivers capabilities. To mitigate the signal quality issue, modulation and coding are the main features envisioned. The modulation guarantees an adaptive data rate for fluctuating channel, high order modulation for low bit error probability can increase data rate~\cite{kliui2018}\blue{. In}~\cite{xpang2017}\blue{,} it is possible to reach $100Gbps$ using 16-QAM optical modulation and BER equal to $10^{-3}$ and using UTC-PD source for heterodyning; coding helps to reduce errors. MAC layer can be designed to support variable throughputs and fits frame length to channel conditions via information from the physical layer. More works should be addressed to modulation and coding techniques to enhance main MAC layer performance and to design MAC protocol's physical layer aware. For nanoscale communication, basic modulation techniques were used, such as On-Off Keying (OOK) and pulse position modulation (PPM) together with femtosecond pulses~\cite{krishne2014,jonetakyil2014,vavo2018}. For improved spectral efficiency, a spatial modulation technique is proposed and recommended in~\cite{sarieddeen2019} by using the dense packet arrays of sub-arrays nanoantennas while achieving acceptable beamforming performance.}

\saim{Selection of modulation scheme depends on Terahertz device's capabilities such as output power, bandwidth, and signal sensitivity. For short-range communication such as nano communication, it is possible to use larger bandwidth and a low complexity scheme such as OOK~\cite{2stateMAC2018} and QPSK~\cite{cwang2018yu} can be used. For macro communication, where the range is higher, the Terahertz gap is split into functional windows and transmission should be performed using carriers. High order modulation can be deployed such as 16-QAM~\cite{seandell2019} along with the directional antenna. The channel coding helps to detect and reduce errors at the receiving side, hence reducing the data loss. However it introduces computational complexity, for nano communication. Therefore, simple coding schemes can be implemented such as hamming. For macro communication Reed Solomon and Low-Density Parity Checks schemes can be added~\cite{TRD2015,ARD2015}.}

\textit{First generation Terahertz devices: } Due to high attenuation, devices with good performance such as higher output power and low noise level are required to optimize the link budget and increase the link data rate. Terahertz transceivers based on electronics were developed, for example, Silicon-Germanium (SiGe) based heterojunction bipolar transistor and Gallium-Nitride (GaN) based monolithic mmWave integrated circuit (MMIC) and also transceiver based on photonics such as quantum cascade laser QCL for high frequencies applications. 
 Resonant Tunnelling Diode (RTD) based on InGaAs/AlAs are also promising for Terahertz applications~\cite{sgmutlak2018}, RTDs convert mm-wave to Terahertz wave.
 Two kinds of devices that perform conversion to terahertz signal: electronic devices such as E-RTD and photonic devices such as Uni-Travelling Carrier photo-diodes (UTC-PD). Works on both technologies, photonic and electronic devices, are in progress to choose the best one of each scenario based on the required data rate, distance, and sensitivity. 
 For some applications,  replacing the high capacity wire, such as optical link, by Terahertz bridges is promising as it adds more flexibility to the network and reduces deployment cost. 
 Terahertz devices are responsible for signal emission and reception. They can affect the link quality such as link budget and the received power spectral density, and can also affect BER and outage probability. Therefore, to design a MAC protocol with a high data rate, devices with low system noise levels and variable output power are required while maintaining the SINR in the network. SINR of the network is the result of device’s transmitted power, channel and system noise. MAC layer, aware of device technology capability, can monitor power emission, antenna pattern shape, and beam orientation. The antenna technology and device performance can also enhance the node discovery functionality and reduce delay caused by this operation.

\subsubsection{\saim{MAC layer related issues and considerations}}\label{subsubsec-reqdes:MAC-issues}\label{sec-req:maclayer}\hfill

The issues and considerations related to the MAC layer are discussed below. These issues are also highlighted based on Terahertz applications in Table~\ref{tab:summ-issue-features}. 

\textit{Channel access, scheduling, and sharing: } \saim{For short-range coverage such as Nano communication, antenna is assumed to be isotropic, techniques for random channel access mechanism such as CSMA can be then deployed using the same frequency band. However, techniques to reduce interferences should be implemented as received signals can collide with other signals. For macro-scale communication, it is required to transmit over distances higher than one meter, antenna should be directional to overcome channel attenuation effects, different procedures should be executed before link establishment such as beam alignment and advanced nodes synchronization. Because, if they are not aligned and facing each other, they cannot receive the transmission, which could introduce the deafness problem. In centralized networks, a central controller is required to manage the beam alignment with scheduled transmissions and for Adhoc networks TDMA based approaches can be used to avoid collisions and manage beam alignment so that each node should know when to transmit, to which node to transmit and to which direction. A shared channel can be used among different nodes, but interference can be increased. An alternate band can be used, which can provide synchronization and coordination among different nodes to access the channel. }

\textit{Neighbor discovery and Link establishment: }\label{subsubsec-req:design-mac-nbrdisc} For any communication to occur, a link must be established first to discover the nodes, which needs to be considered while designing an efficient MAC protocol for nano and macro-scale networks. For nanoscale networks, due to constraints energy storage and generation, mechanisms with low message overhead are required. For macro-scale networks, the antenna directionality and mobility adds extra challenges to establish a link between the nodes, which requires a node to track and locate other nodes for stable link maintenance. For seamless communication, a stable link is required at all times. Efficient beam steering mechanisms are required to reduce the handshake time and to reduce the overall neighbor discovery time in Adhoc networks. Further, due to short-range and mobility, frequent link association and handovers may be required, which should also be supported in MAC protocol design. 

\saim{Nodes discovery occurs before the communication phase in which each node needs to inform its neighbors about its availability and identity. The discovery phase is constrained by the lack of information related to node's position. For nanonetwork, node can send identity messages to its neighbors and any node receiving a discovery message adds the sender to its list. Generally for dynamic networks, when new nodes can enter or leave the neighboring subset, the discovery procedure can be more frequent. For static networks, discovery procedure can be neglected as each node is aware of its neighbors from the deployment phase and discovery message can be exchanged after some critical events. For example one of nodes is out of service or a new node is introduced. Applying node discovery in a macro-scale network is challenging if mobility is added to the system, synchronized beam turning can be applied during this procedure, again a node receives the discovery message once beams are aligned, this procedure is time and energy-consuming, optimizations should be applied to reduce time and energy required to discover neighbors.   }


\textit{Mobility management and handovers: } Mobility and coverage are two mutually correlated concepts, for mobile Terahertz system~\cite{Guank2017,railguan2017} in which radio coverage should be guaranteed to decrease link outage probability. MAC layer should support mobility management functionality to guarantee service continuity. The handover is a technical concept for mobile networks to describe the changing of the serving base station without interruption of the traffic flow. 

\saim{One issue rises from nodes mobility is THz localization determination. In~\cite{elabsi2018}\blue{,} authors propose THz RFID techniques for device location using specific channel modeling and node's localization correlated to the handover procedure. For example, handover can be triggered in some positions where received power from a serving node is weak. For THz network, localization is more feasible for nanonetworks, however it becomes more challenging for macro networks where directional antennas are deployed. Solving the localization problem can help to accelerate the handover execution. }

\textit{Collision avoidance and interference management: } Due to high bandwidth availability and antenna directionality, it is unlikely that a collision might occur in Terahertz communication. However, it can occur when two node pairs beam directions crosses each other and perform frequent and long transmission. The multi-user interference can also occur in a scenario with large number of nodes with mobility. Therefore, a collision detection and avoidance mechanism should be considered while designing an efficient Terahertz MAC protocol. New interference models are required to capture the effect of Terahertz band features and multi user interference~\cite{Jornet2016LINKAN}. The directional communication can decrease the multi-user interference but it requires tight synchronization between Tx and Rx. Further, reduced channel code weights can also result in lower channel error probability and can also help in avoiding the multi-user interference and molecular absorption~\cite{JORNET201435}.  

\textit{Reliability: } Most of wireless systems require a reliable communication, where the degree of reliability defers from one application to another. The problem becomes more complicated when the channel conditions changes with time and causing time varying absorption~\cite{JORNET201435,zarepour2015}. For Terahertz systems, mainly low frame loss and high throughput are required. In Terahertz systems, the eror control module is mainly responsible for frame protection and retransmission to reduce frame losses where frame error depends on channel model as well as on frame length. Error control module is required especially for harsh channel conditions such as outdoor channel with dynamic conditions~\cite{aaboug2020}. For nano sensor network, a cross optimization method is proposed in~\cite{pkt_size2016}, to adapt with the frame transmission and size with the channel conditions. Further, for efficient usage\blue{,} reliable wireless links and beam tracking should be considered in a MAC design. 

\textit{Throughput and latency: } Terahertz band is endowed with large bandwidth, due to which it is possible to reach a throughput exceeding 100 Gbps. For some applications like Terahertz data center scenario, bandwidth is shared between many nodes and therefore a MAC should support and guarantee high data rate and low delay. Fast scheduling algorithms, appropriate MAC techniques and buffering should be implemented to meet application specific QoS requirements.

\saim{\textit{Energy efficiency and harvesting:} Energy efficiency means using less energy to achieve the required performance. For some scenarios\blue{,} it is hard to provide energy in a continuous way such as devices using low capacity battery, mobile nodes. Therefore, energy efficiency becomes a priority for certain Terahertz applications such as nano-communication~\cite{MDP2014,joint-energy2012}, body area network, fronthaul communication and THz wireless sensor network used in biomedical and military fields~\cite{jxu2018}. To cope with the lack of energy provision, techniques such as energy harvesting, lower modulation order techniques can be used to extend the battery lifecycle. It is also possible to design new data link layer techniques to increase energy efficiency~\cite{sverma2018}. For applications such as backhauling, Datacenter and information broadcast where the source of energy is always available, energy efficiency is less prioritized than the previous applications. In Table~\ref{tab:thz-app-chal}, energy efficiency requirement for different applications is highlighted based on the priority of energy-efficient techniques requirement.}

In some applications, the power of nodes is a limiting factor to transmit continuously, a power management module should be implemented on MAC to reduce power consumption without degradation of the system quality of service. For example\blue{,} switching from active to idle state if the node has no data to transmit and applying the power control strategy depending on channel conditions and target QoS. The second alternative to save power and guarantee the battery life is to harvest and manage energy\blue{. F}or example for nano-sensors the energy harvesting is applied for more active nodes. Due to low storage constraint of nanodevices, the tradeoff must be considered for energy harvesting and utilization in an efficient way~\cite{Akyildiz2011}. 
 
\textit{Coverage and Connectivity:} Terahertz communication is characterized by low range connectivity and high available bandwidth. The coverage or range can be optimized using a directional antenna, enhanced Terahertz devices, high output power, and optimized sensitivity. MAC layer can also contribute to coverage and connectivity enhancement by utilizing data link relaying, path diversity and spectrum switching. For example, in vehicular Terahertz network, a mobile node can coordinated to more than one node. In nanosensor network, for short-range communication, each node can transmit to any node out of its range using relaying capabilities. Path diversity is also an alternative solution to increase connectivity when a LOS is temporarily unavailable. With no direct LOS link and to support the seamless communication with less delay, reflectors can also be used to reach out far nodes with no LOS link~\cite{barrosmall2017}. A coverage and achievable rate performance analysis for multi-user Terahertz with single frequency is presented in~\cite{amoldovan2017akil}.

\subsection{MAC layer decisions}\label{sec-req:macdecisions}
MAC layer is responsible for traffic adaptation with the physical layer, adding sophisticated modules to optimize link performance which can be promising. Following are some of the decisions which can be taken at the MAC layer to enhance further the system performance. 
 \begin{itemize}
 \item {\bf Bandwidth and frequency selection:} MAC layer should actively sense the physical channel as well as be aware of service requirements of each traffic flow, then, the selection of the appropriate bandwidth and carrier frequency can adapt the traffic to the channel as well as reduce interferences. Most of the actual Terahertz system uses a single frequency, a multiband antenna is also worth considering.
 \item {\bf Modulation and coding selection:} The Terahertz channel is generally time-dependent, the transmitted signal undergoes impairments leading to high bit error rate, to mitigate this issue, an adaptive modulation, and coding scheme can be adopted and controlled by the MAC layer. High order modulation selection can increase throughput and low order modulation is required to reduce the bit error rate.
 \item {\bf Power management:} This module selects the appropriate power to increase coverage as well as reduce interferences when nodes coordinate between each other. Monitoring nodes using power control can reduce interferences as well as maintain an acceptable energy consumption value. The power management module, can adapt to its environment, for instance, the mean consumed power value in a humid environment will be different from a dry one.
 \item {\bf Beam steering:} when using a directional antenna for Terahertz communication, beams should be steered appropriately to the receiving node, selection of the beams orientation coordinates can be performed at MAC layer based on the inputs from the physical layer beam parameters such as phases between elements. 
 \end{itemize}

  Implementing the aforementioned modules in the MAC layer, will increase awareness of the physical layer and channel fluctuation as well as to adapt the Terahertz link to the upper layers. \saim{Different} physical layer functionalities can be monitored at MAC layer level, for instance, it is possible to change the modulation scheme from high order 16-QAM to low order QPSK to reduce bit error rate and from QPSK to 16-QAM to increase data throughput when channel condition is good, switching operation can be triggered using link quality statistics. The module responsible for beam steering can be also included in the MAC layer, for example using 3 bits to monitor 8 beams and establish a link with 8 neighbors. Monitoring frequencies and bandwidth can be also included in the MAC layer, for multi-band wideband antenna, to reduce interferences and increase data throughput. Finally, the power management module allows monitoring transmitter output power to enhance the link budget if the link breakdown or channel attenuation increases. The power management module can take a decision based on collected measurement from the physical layer and also from other nodes to control signal to interference ratio. Modulation scheme, beams orientation, frequency, and power can be updated at frame level based on collected statistics from a physical layer as well as reports from the networking layer.

\subsection{Discussion on Terahertz application scenarios}{\label{subsec:req-apps}

  In Table~\ref{tab:summ-issue-features}, different Terahertz MAC protocols are mentioned with their application areas. Terahertz band features and their MAC design issues and considerations are also highlighted. Mainly, these applications are categorized in nano and macro scale scenarios which are also discussed in Section III. Due to unique band features, each MAC protocol of different applications requires novel MAC mechanisms to accommodate the high bandwidth availability, path loss, and noise. Table~\ref{tab:summ-issue-features}, mentions only those Terahertz applications for which MAC protocol work is available. For nanoscale networks, mostly omnidirectional antenna are assumed due to the short-range and low path loss. For higher transmission range the path loss can severely damage the communication and affect the distance. The Terahertz MAC protocols for nanoscale networks still do not consider the unique features like path loss, molecular absorption noise, multipath effect. For Physical layer functionalities, the MAC protocols are there as mentioned in the Table~\ref{tab:summ-issue-features}, but they are not considering the antenna design, channel, propagation, and interference model. 

For macro-scale applications, each indoor and outdoor application has different requirements and therefore require different MAC mechanisms. Due to short-range constraint, the Terahertz band suits the indoor applications like TLAN and TPAN, these involve mobility with communication over short distances. The scenarios like Data centers involve static links between different racks and so require point-to-point/-multipoint communications. These scenarios require different channel models and scattering and multipath phenomena can affect communication in a different way. Therefore, while designing an efficient Terahertz MAC protocols, the features and design issues mentioned in Table~\ref{tab:summ-issue-features} should be considered. To enhance the communication range, directional antenna should be used which requires novel mechanisms for beam management and tracking with MIMO support, reflectors to mitigate blockage and reach to more than one hop distance. The static points application like KIOSK downloading system needs to support quick link establishment and reliability. These applications also required new mechanisms to access the Terahertz channel and link establishment, especially when frequent link establishment is required and where node density is high.  

The outdoor scenarios like vehicular communication, backhaul, and small cell are interesting scenarios, which involves mobile and static scenarios. However, the channel can be affected due to different environmental factors like rain, wind, humidity, and dryness. Therefore, new channel and propagation models are required which should also incorporate blocking factors like trees, humans and other physical types of equipment. Massive MIMO can be used to relay information between cells or nearby networks. Adaptive beam management can be utilized by using cooperative massive MIMO and electronically steerable beams~\cite{Mumtaz2017}. Further, due to different environmental factors interference mitigation techniques are required for outdoor applications.

\subsection{Summary}

Due to unique features of the Terahertz band like noise and path loss, the Terahertz band communication can easily be interrupted compared to the interference phenomenon in other lower frequency bands like ISM or GSM. The molecular absorption noise or the atmospheric noise can easily affect the Terahertz communication link and the problem increases with increase in the distance between the transmitter and receiver. Further, the additional environmental noise factors like Sky-noise can result in underestimation of noise or interference figure at the receiver and transmitter which can also affect the MAC protocol performance seriously. The modeling of these factors is very important in the sense that these factors can behave differently in different environments like indoor or outdoor environments, and should be modeled carefully depending on the scenario. Therefore, in designing the MAC protocols for short, medium or long-range Terahertz communication, these environmental factors, and their modeling must be taken into account. The indoor and outdoor scenarios required different channels, propagation and interference models and they need to consider different physical and MAC layer design issues discussed in this section. \saim{To strengthen the reflected and scattered signals, a metal reflector with goof reflection properties can be embedded and to reduce the power absorption the temperature and humidity can be maintained at a certain level for a particular indoor environment. However, for outdoor environment, novel mechanisms are required to overcome the effect of absorption loss.}

\section{Terahertz MAC protocols for different network topologies}\label{sec:topologies}

In this section, the existing Terahertz MAC protocols are categorized mainly in network topology as centralized, clustered and distributed, as shown in Figure~\ref{fig:nw_topologies}. Each topology design is then further classified based on the network scale. Different topological designs are considered in the existing literature based on the application area and its requirements and are discussed below. In general, the existing Terahertz MAC protocols are characterized and summarized in Table~\ref{tab:mac-proto-characterization}.


\begin {figure*}
\centering
\begin{adjustbox}{width=\linewidth}

\tikzstyle{block} = [rectangle, draw, text width=4cm, text centered, rounded corners, minimum height=5em,fill=blue!20]
\tikzstyle{line} = [draw,thick, -latex']
\tikzstyle{cloud} = [draw, ellipse, text width=4.5cm, text centered]
\tikzstyle{edge from parent}=[->,thick,draw]
\begin{tikzpicture}[auto,edge from parent fork down]
\tikzstyle{level 1}=[sibling distance=60mm,level distance=18ex]
\tikzstyle{level 2}=[sibling distance=25mm,level distance=15ex]
\tikzstyle{level 3}=[sibling distance=30mm,level distance=34ex]
\tikzstyle{level 4}=[sibling distance=35mm,level distance=34ex]
\node [block,text width=10cm,minimum height=4em,,fill=black!30] (cst) {\textbf{THz MAC Protocols for differnt network topologies \\ (Sect.\ref{sec:topologies})  } }
{
child{node [block,text width=4cm,fill=green!20] (opm) {\textbf{CENTRALIZED \\ (Sect.\ref{subsec-topo:centralized})   } }
  		child{node [block,text width=3.2cm,fill=orange!20,xshift=-0.65cm, yshift=-1.2cm] (pmt) {      \\ 
  		{\bf Nanoscale networks \\ (Sect.\ref{subsubsec-topo:centralized-nano})  } \\
  		LL-Modelling \cite{LLModelling2011} \\
		CSMA-MAC \cite{slottedCSMAMAC2018}  \\
		DESIGN-WNSN \cite{Design-WNSN2015} \\
		TCN \cite{TCN2015} \\
		CEH-TDMA \cite{xujuan2019} \\
		SSA-MAC \cite{wangw2019} \\
  	    }} 
  	    child{node [block,text width=3.2cm,fill=orange!20,xshift=0.7cm, yshift=-1.2cm] (pmt) {      \\ 
  		{\bf Macro scale networks \\ (Sect.\ref{subsubsec-topo:centralized-macro})  } \\ 
  		MA-ADM \cite{MAADM2017} \\
  		IHTLD-MAC \cite{IHTLD-MAC2017}\\ 
  		MAC-TUDWN \cite{MAC-TUDWN2013}\\ 
  		TRPLE \cite{TRPLE2014}\\
  		HLMAC \cite{HLMAC2016}\\
	    MAC-TC \cite{MAC-TC2013}
  		}} 
        }
child{node [block,text width=3.6cm,fill=green!20] (rnm) {\textbf{CLUSTERED \\ (Sect.\ref{subsec-topo:clustered})   } }
		child{node [block,text width=3cm,fill=orange!20,xshift=-0.4cm, yshift = -1.0cm] (pmt) {  \\ 
		{\bf Nanoscale networks \\ (Sect.\ref{subsubsec-topo:clustered-nano})  } \\
		ES-Aware \cite{esaware2013}    \\
		EEWNSN \cite{EEWNSN2017} \\
		EESR-MAC \cite{EESR-MAC2012} \\
		DYNAMIC-FH \cite{dynamic_CH2016}
		}}
}
child{node [block,text width=3.8cm,fill=green!20] (opm) {\textbf{DISTRIBUTED \\ (Sect.\ref{subsec-topo:distributed})  } }
		child{node [block,text width=3.4cm,fill=orange!20,xshift=-1.4cm, yshift = -2.7cm] (pmt) {  \\
 		{\bf Nanoscale networks \\ (Sect.\ref{subsubsec-topo:distributed-nano})  } \\
		PHLAME \cite{phlame2012}\\
		DRIH-MAC \cite{drih-mac2015}\\
		RIH-MAC \cite{RIH-MAC2014}\\
		DMDS \cite{DMDS2016}\\
		2-state MAC \cite{2stateMAC2018}\\
		G-MAC \cite{GMAC2016}\\
		RBMP \cite{RBMP2017}\\
		APIS  \cite{APIS2017}\\
		MGDI \cite{MGDI2016}\\
		NS-MAC \cite{NS-MAC2016}\\
		MDP \cite{MDP2014}\\
		TSN \cite{TSN2014}\\
		SMART-MAC \cite{smart-mac2013} \\
		SSA-MAC \cite{wangw2019} \\
		}}
		child{node [block,text width=3.6cm,fill=orange!20,xshift=0.2cm, yshift = -2.0cm] (pmt) {  \\
		{\bf Macro scale networks \\ (Sect.\ref{subsubsec-topo:distributed-macro})  } \\
		\textbf{Terahertz networks:} \\
		\begin{itemize}[label={}]
		\item LL-Synch \cite{LLsynch2015} \\
		\item ISCT \cite{ISCT2018} \\
		\item TAB-MAC \cite{TAB-MAC2016} \\
		\item MRA-MAC \cite{MRA-MAC2017} \\
		\item OPT-RS \cite{OPTRS2017} \\
		\end{itemize}
		\hfill\\
		\textbf{Vehicular network:} \\
		\begin{itemize}[label={}]
		\item ATLR \cite{ATLR2018} \\	
		\item SDN-CONTR \cite{SDNC2017}\\
		\item B5G \cite{B5G2018} 
		\end{itemize}
		}}
}
};
\end{tikzpicture}
\end{adjustbox}
\caption {MAC layer classification based on network topologies.  }
\label{fig:nw_topologies}
\end{figure*}
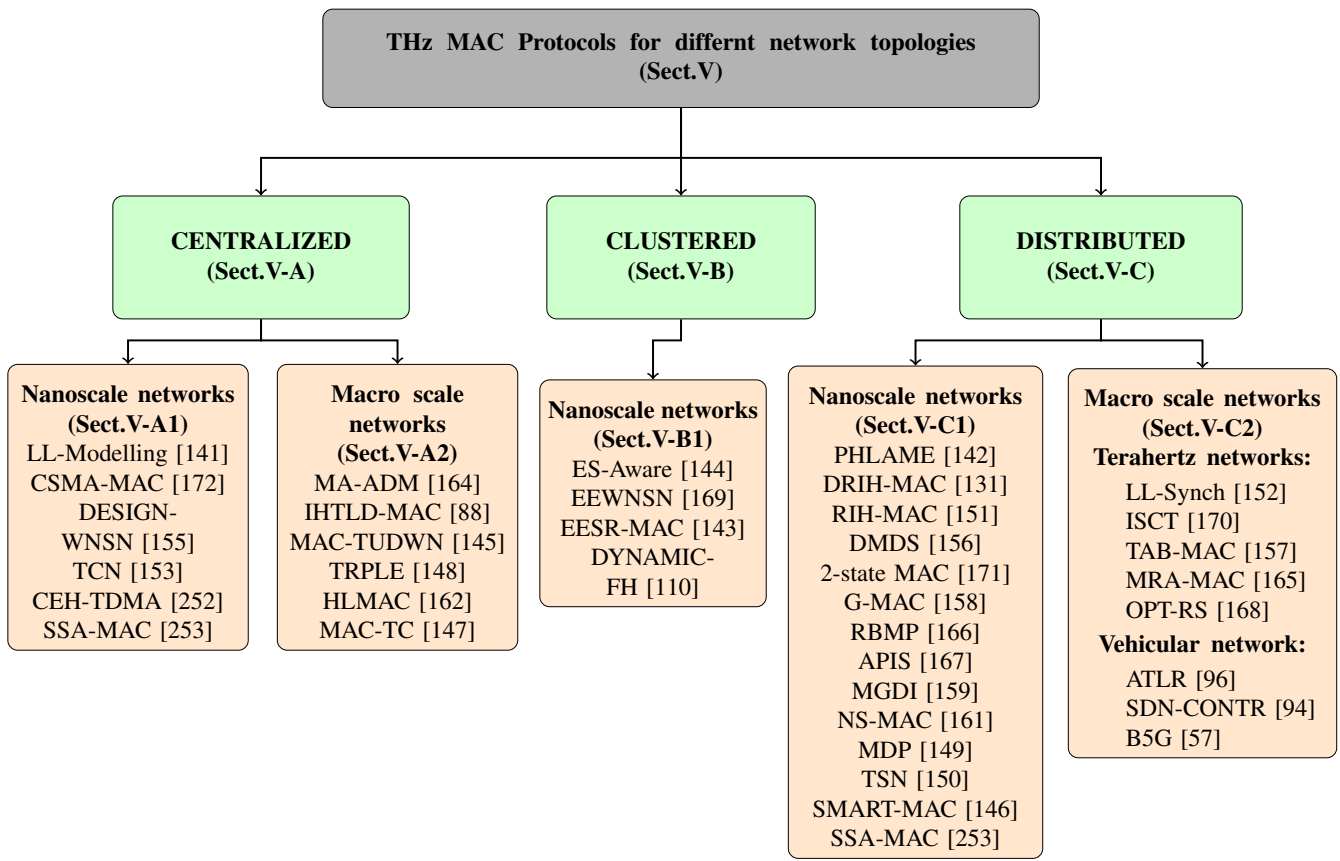

\begin{table*}[htbp]
  \centering
  \tiny
  \caption{Characterization of existing Terahertz MAC protocols.}
    \begin{tabular}{p{2em}lp{4em}p{6em}p{3.1em}p{5em}p{4em}p{15em}p{3.1em}p{8em}p{5em}p{5em}p{4em}p{4.1em}}
    \textbf{Year} & \multicolumn{1}{p{3.11em}}{\textbf{Paper}} & \textbf{Band} & \textbf{Network type} & \textbf{Network scale} & \textbf{Topology} & \textbf{Simulator} & \textbf{Simulation parameters} & \textbf{Analytical Model} & \textbf{Tx/Rx initiated communication} & \textbf{Modulation Scheme} & \multicolumn{1}{p{5.11em}}{\textbf{Channel access method}} & \textbf{Antenna} & \textbf{Complexity}\\ \hline \hline
   
   2011  & \cite{LLModelling2011} &  THz bands & Nanonetworks & Nano  & centralized & C++   & delay, throughput, performance analysis based on distance, propagation time, packet lifetime & \checkmark   & Transmitter initiated & \multicolumn{1}{l}{-} &  -     & nano & - \\  \hline
   
   \multirow{2}{*}{2012}  & \cite{phlame2012} &  0.1 - 10 THz & Nanonetworks & Nano & distributed & custom & Energy consumption, Delay, Throughput & \checkmark   & Transmitter initiated & RD-TS-OOK &  TDMA  & nano & - \\ 
   	& \cite{EESR-MAC2012} &  THz bands & Nanonetworks & Nano  & clustered & No    & X    & X    & Transmitter initiated & \multicolumn{1}{l}{-} & \multicolumn{1}{p{5.11em}}{TDMA} & nano & - \\ \hline
   
   \multirow{4}{*}{2013}  & \cite{esaware2013} &  0.1 - 10 THz & Wireless Nanosensor Networks & Nano  & Centralized & custom & Throughput, optimal channel access, network lifetime, critical transmission ratio & \checkmark   & Transmitter initiated & pulse based,TS-OOK & \multicolumn{1}{p{5.11em}}{TDMA} & nano & - \\
     & \cite{MAC-TUDWN2013} &  0.1 - 10 THz & Terahertz Wireless Personal Area Networks & Macro  & Distributed & OPNet & data trasmission rate, avg access delay, time for transmitting frames, access success rate & X    & Transmitter initiated & \multicolumn{1}{l}{} & Hybrid  & \multicolumn{1}{l}{-} & - \\
     & \cite{smart-mac2013} &  THz bands & Nanonetworks & Nano  & Distributed & NanoSim - NS3 & packet loss ratio, physical transmission & \multicolumn{1}{c}{} & Transmitter initiated & OOK   & Random   & nano & - \\
      & \cite{MAC-TC2013} & 0.1 - 10 THz & Terahertz Communication Network & Macro  & Distributed & \multicolumn{1}{l}{} & \multicolumn{1}{l}{-} & \multicolumn{1}{c}{-} & Transmitter initiated & \multicolumn{1}{l}{-} &  Hybrid    & \multicolumn{1}{l}{-} & - \\ \hline
   
   \multirow{4}{*}{2014}  & \cite{TRPLE2014} &  0.1 - 10 THz & Terahertz networks & Macro & Distributed & custom & Data rate, throughput & \checkmark   & Transmitter initiated & TS-OOK & TDMA & directional & - \\
     & \cite{MDP2014} &  THz bands & Nanonetworks & Nano  & Distributed & Matlab & energy efficiency, harvest rate, packet balance & \checkmark   & Receiver initiated & OOK   & TDMA   & nano & - \\
     & \cite{TSN2014} &  THz bands & Wireless Nanosensor Networks & Nano  & Distributed & custom & SNR, BER, capacity & X    & Transmitter initiated & pulse based & FTDMA & \multicolumn{1}{l}{-} & - \\
     & \cite{RIH-MAC2014} &  0.1 - 10 THz & Nanonetworks & Nano  & Distributed & NanoSim - NS3 & Prob collision, RTR, fairness,  & X   & receiver initiated & OOK   & TDMA & nano & - \\ \hline
  
   \multirow{4}{*}{2015}  & \cite{drih-mac2015} &  0.1 - 10 THz & Nanonetworks & Nano  & Distributed & NanoSim - NS3 & Delay, energy consumption, utilization capacity & X   & Receiver initiated  & Pulse based modulation & \multicolumn{1}{p{5.11em}}{TDMA} & nano & - \\
     & \cite{LLsynch2015} &  1.04 THz & Terahertz Communication Networks & Macro, Nano & Distributed & NS3   & Delay, throughput, Packet delivery ratio & \checkmark   & Receiver initiated & PSK, TS-OOK & \multicolumn{1}{p{5.11em}}{CSMA} & omni and directional & - \\
     & \cite{TCN2015} &  0.1 - 10 THz & Nanonetworks & Nano  & Distributed & custom & Failure probability, normalized energy per bot, number of retransmissions & \checkmark   & Transmitter initiated & \multicolumn{1}{l}{-} &  TDMA  & nano & - \\ 
     & \cite{Design-WNSN2015} &  THz bands & Wireless Nanosensor Networks & Nano  & Centralized & custom & Energy consumption, Delay, Throughput & X   & Transmitter initiated & OOK   & \multicolumn{1}{p{5.11em}}{TDMA} & nano & - \\ \hline
    
  \multirow{9}{*}{2016}  &  \cite{joint_error2016} &  0.1 - 10 THz & Nanonetworks & Nano  & Distributed & COMSOL multi-physics & BER, PER, enerhgy consumption, latency, throughput & \checkmark   & Transmitter initiated & OOK   & \multicolumn{1}{p{5.11em}}{-} & nano & - \\
     & \cite{DMDS2016} &  0.1 - 10 THz & Internet of Nano Things & Nano  & Distributed & custom & delivery ration, debt, throughput & \checkmark   & Transmitter initiated & \multicolumn{1}{l}{-} & CSMA & nano-antenna & - \\
     & \cite{TAB-MAC2016} &  2.4 GHz, 0.1-10 Thz & Terahertz Communication Networks & Macro & Distributed & custom & packet delay, throughput, failure probability,  & \checkmark   & Transmitter initiated & \multicolumn{1}{l}{-} & Random, multiple radios & omni and directional & - \\
     & \cite{GMAC2016} &  0.1 - 10 THz & Wireless Nanosensor Networks & Nano  & Distributed & Matlab & delay, throughput, collision probability & \checkmark   & Transmitter initiated & TS-OOK &  TDMA & nano & - \\
     & \cite{MGDI2016} &  100 GHz & Nanonetworks & Nano  & distributed & Any-Logic platform & Coverage, Packet Transmission rate, classification time, collision & X   & Transmitter initiated & OOK   & \multicolumn{1}{p{5.11em}}{-} & nano  & $O(x)$ \\
     & \cite{pkt_size2016} &  0.1 - 10 THz & Wireless Nanosensor Networks & Nano  & Distributed & COMSOL multi-physics & link efficiency, optimal packet length with distance & X   & Transmitter initiated & OOK   &   -    & nano-antenna & - \\
     & \cite{NS-MAC2016} &  THz bands & Wireless Nanosensor Networks & Nano  & Distributed & Matlab & collision probability, energy consumption, transmission distance & \multicolumn{1}{c}{} & Transmitter initiated & OOK   &   -    & nano & - \\
     & \cite{dynamic_CH2016} &  1-2 THz & Nanonetworks & Nano & Distributed & custom & Throughput, Transmission probability & \checkmark   & Transmitter initiated & OOK   & \multicolumn{1}{p{5.11em}}{FTDMA} & nano-antenna & - \\
     & \cite{HLMAC2016} &  0.1 - 10 THz & Terahertz Communication networks & Macro  & distributed & \multicolumn{1}{l}{-} & \multicolumn{1}{l}{-} & \multicolumn{1}{c}{-} & Transmitter initiated & \multicolumn{1}{l}{-} &  Hybrid  & \multicolumn{1}{l}{-} & - \\ \hline
   
   \multirow{9}{*}{2017}  & \cite{IHTLD-MAC2017} &  340 GHz & Terahertz Wireless Personal Area Networks & Macro   & Distributed & OPNet & delay,throughput, success rate, buffer overflow rate & \checkmark   & Transmitter initiated & \multicolumn{1}{l}{-} & \multicolumn{1}{p{5.11em}}{Hybrid} & omni   & - \\
     & \cite{MAADM2017} &  0.06 - 10 THz & Terahertz Communication Networks & Macro  & Centralized & custom & Throughput, Data rate, Delay, and outage probability & \checkmark   & Transmitter initiated & Pulse waveform modulation & Scheduled, multiple radios & omni and directional & - \\
    &  \cite{DLLC2017} &  240 GHz & Terahertz networks & Macro & \multicolumn{1}{l}{} & Matlab & probability of successful frame reception, goodput Gbps,  percentage of lost headers, ACK frame size, energy per bit & X   & Transmitter initiated & PSSS, PAM-16 &   -    & \multicolumn{1}{l}{-} & - \\
     & \cite{MRA-MAC2017} &  2.4 GHz, 0.1-10 Thz & Terahertz Communication Networks & Macro & Distributed & Monte Carlo & Delay, Throughput, outage probability & \checkmark   & Transmitter initiated & Pulse wave modulation & \multicolumn{1}{p{5.11em}}{Random, multiple radios} & omni and directional & - \\
     & \cite{RBMP2017} &  2.4 GHz, 0.1-10 Thz & Nanonetworks & Nano  & Distributed & NS3   & Throughput & X   & Transmitter initiated & \multicolumn{1}{l}{-} & \multicolumn{1}{p{5.11em}}{Random, mulitple radios} & omni and directional & - \\
     & \cite{APIS2017} &  100 GHz & Wireless Nanosensor Networks & Nano  & Distributed & NS3   & Bandwidth efficiency, pulse drop ratio, packet deliver ratio, fairness, energy consumption & X   & Transmitter initiated & OOK   & TDMA & nano & $O(N^{'})$ \\
     & \cite{OPTRS2017} &  1.0345 THz & Terahertz Communication Networks & Macro & Distributed &  custom & Throughout, optimal distance & \checkmark   & Receiver initiated  & -   & \multicolumn{1}{p{5.11em}}{CSMA} & directional antenna & - \\
     &  \cite{SDNC2017} & 73 GHz and 0.86 THz & Vehicular Network & Macro & Distributed & custom & data transfer & \checkmark   & Transmitter initiated & \multicolumn{1}{l}{-} & Scheduled, multiple radios & directional & $O(V^{k}) $ \\
     & \cite{EEWNSN2017} &  THz bands & Wireless Nanosensor Networks & Nano  & clustering & NS3   & pkt loss ratio, consumed energy, scalability & \multicolumn{1}{c}{} & Transmitter initiated & \multicolumn{1}{l}{-} & \multicolumn{1}{p{5.11em}}{TDMA} & \multicolumn{1}{l}{-} & -  \\ \hline
   
   \multirow{6}{*}{2018}  & \cite{ATLR2018} &  0.1 - 10 THz & Vehicular Network & Macro & Distributed & custom & Channel capacity, PSD, number of links & X   & Transmitter initiated & \multicolumn{1}{l}{-} &   -    & omni & - \\
     & \cite{ISCT2018} &  0.1 - 10 THz & THz Mobile Heterogeneous Network & Macro & Distributed & custom & Uniformity, Randomness, Hamming Correlation, throughput, BER & X   & Transmitter initiated & \multicolumn{1}{l}{} &  FTDMA  & \multicolumn{1}{l}{-} & - \\
     & \cite{B5G2018} &  mmWave, 0.1 - 10 THz  & Vehicular Network & Macro & Distributed & custom & data transmission rate & X   & Transmitter initiated & \multicolumn{1}{l}{-} &  Scheduled, multiple radios & directional & - \\
      & \cite{2stateMAC2018} & 0.1 - 10 THz & Nanonetworks & Nano  & Distributed & custom & Throughput capacity, interference power & \checkmark   & Transmitter initiated & Pulse based modulation (OOK) &  TDMA & nano-antenna & - \\
      &\cite{slottedCSMAMAC2018} &  0.1 - 10 THz & Wireless Nanosensor Networks & Nano  & Centralized & custom & slot assignment rate, energy consumption & X   & Transmitter initiated & TS-OOK & \multicolumn{1}{p{5.11em}}{CSMA} & nano-antenna & - \\
     & \cite{MAC-Yugi2018} &  2.3 THz & Terahertz Communication Networks & Macro, Nano & Centralized & custom & Antenna directivity, antenna controller overhead & X   & Transmitter initiated & DAMC  &   -    & omni and directional & -  \\ \hline 
     
      \multirow{2}{*}{\saim{2019}}  & \saim{\cite{seandell2019}} &  0.25 THz & Data Centre & Macro & Distributed & custom & FER, BER, Pkt loss, retransmission & X   & Transmitter initiated & 16 QAM &   -    & directional & - \\ 
       & \saim{\cite{xujuan2019}} &   & Wireless Nanosensor Networks & Nano & Centralized & custom & Average delay, remaining energy and transmitted packets & X   & Receiver initiated & OOK  &   -    & nano & $O(N), O(N log N)$  \\ \hline  \hline

    \end{tabular}%
  \label{tab:mac-proto-characterization}%
\end{table*}%

\subsection{Terahertz MAC protocols for Centralized networks}\label{subsec-topo:centralized}

The centralized architecture is mainly followed in nanoscale networks due to its limited energy, coverage area, and application in body area networks~\cite{LLModelling2011,slottedCSMAMAC2018,MAC-Yugi2018,Design-WNSN2015}. 

\subsubsection{Nanoscale networks}\label{subsubsec-topo:centralized-nano}

\saim{The nanonetworks include several nanodevices that work together to perform simple tasks. Due to limited energy capacity, centralized topology is used in different applications including In-body networks, air quality monitoring, and industrial applications. In these applications\blue{,} a nano-controller that is capable of performing heavy computation, scheduling and transmission tasks is used. Initially, nanodevices send their information to the controller, and controller can then process and schedule transmission and sends information to external networks via a gateway device. The works in~\cite{slottedCSMAMAC2018,LLModelling2011,Design-WNSN2015,TCN2015,xujuan2019} uses a centralized network topology for different nano communication network-related applications. A general figure for such architecture is shown in Figure~\ref{fig:nanobio}. }

\saim{A centralized approach is presented in~\cite{slottedCSMAMAC2018} in which nano nodes can be deployed in a specific area to detect problems like defects or pollution, where each nano node can perform computational tasks with limited memory and transmits small data over a short-range to the nano-controller. The gateway can collect the information and send to the Internet. A single hop delay-throughput performance is measured in~\cite{LLModelling2011} for the bio-nano communication network in which bacteria packets travel towards the nano-gateways following the attractant particles emitted by the conjugate nano-gateway. The centralized network topology is also used in~\cite{Design-WNSN2015,TCN2015}. A centralized TDMA and energy harvesting based protocol is presented in~\cite{xujuan2019} in which the nano-controller is responsible for channel access and time slot allocation for nano nodes. To cope with energy consumption\blue{,} nodes are also responsible for energy harvesting to increase node's life cycle. A centralized approach is then used to free some nodes from performing heavy computational tasks and channel access. Due to smaller distance, the path loss discussed in Section~\ref{sec-req:pathloss} remains low. However, the interference from near devices can affect the transmission opportunities and access to the channel (cf. Section~\ref{sec-req:phylayer}). In centralized topology, nodes are within a single hop distance from the controller node.  }

\mage{Depending on the application requirement, different topologies can be followed in nanonetworks. The works in~\cite{slottedCSMAMAC2018,LLModelling2011,Design-WNSN2015,TCN2015} follows a centralised approach. However, for higher nodes density, multi-hop communication is required. Although, in~\cite{wangw2019} high node density is considered. But, random arrangement of nodes with mobility is not considered. }

\begin{figure}[htbp!]
\centering
\includegraphics[width=3.2in,height=2.2in]{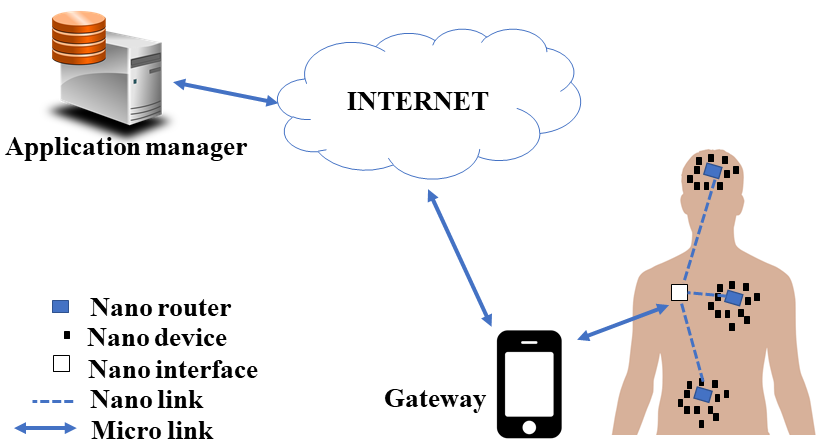}
\caption{Nano body communication network.}
\label{fig:nanobio}
\end{figure}

\subsubsection{Macro scale networks}\label{subsubsec-topo:centralized-macro}

Besides, the nanonetworks, centralized architecture is also used for Terahertz communication at larger networks such as macro-scale networks. In~\cite{MAADM2017}, a centralized network with Terahertz links is presented which consists of an AP and multiple nodes, where the AP coordinates and schedules the transmissions among the nodes and is able to communicate directly to each node. The interesting issue in such networks will be synchronization among the nodes, where each node using a directional antenna needs to point towards an AP, which also increases the interference or collisions probability from the neighbors. In~\cite{MAADM2017}, directional antennas are used for nodes discovery, initial access, and data transmissions phase without creating disparity problem~\cite{barati2015}.

The problem arises when the directional beams from the AP and the receiving node have to be well aligned, \saim{beams alignment is time and energy-consuming, a beam management module at MAC layer should be implemented for efficient beam steering.} For centralized macro-scale networks, the AP is assumed to be equipped with directional antennas whereas the other nodes are using the omnidirectional antennas and are assumed to switch to the directional antennas mode after the establishment of initial discovery which can incur more delays. A beam switching access technique is discussed in~\cite{MAADM2017}, beam alignment is performed periodically for the initial access and transmission period. The centralized topologies are also presented in~\cite{IHTLD-MAC2017,MAC-TUDWN2013,HLMAC2016,MAC-TC2013} in which a piconet coordinator is assumed to provide time synchronization information to nearby devices and handles the scheduling and access control. A MAC design for macro scale communication at 100 Gbps is discussed in~\cite{TRPLE2014} which considers an indoor picocell, where an AP communicates with a group of users using LOS and directed non-LOS.

\mage{An interesting challenge for centralized topology for distances higher than a meter will be to efficiently manage the beam steering and switching among nodes, which is discussed in Section~\ref{sec-req:phylayer}. For applications like TPAN and TLAN, mobility adds design challenges for MAC and the channel access and scheduling can be managed by a central controller (cf. Section~\ref{sec-req:maclayer}). In indoor environments with small distances and mobile nodes, multi-user challenges should be addressed which includes efficient mobility and interference management and resource scheduling. }


\subsection{Terahertz MAC protocols for Clustered networks}\label{subsec-topo:clustered}

\saim{In a clustered architecture of nodes, a cluster head is elected by nodes in the cluster, the cluster head is responsible for data processing and transmission to the gateway node and to other cluster heads. The clustered architecture is so far seen only in nanonetworks environment which also incurs low path loss due to short distance.  Energy efficiency can be improved using clustered architecture when using the central node. }

\subsubsection{Nanoscale networks}\label{subsubsec-topo:clustered-nano}

\saim{In hierarchical architecture, the network is mainly partitioned in a set of clusters where each cluster is locally coordinated by a nano controller. The nano controller is a device that has more processing capabilities of complex tasks and has high energy availability. Since nanosensor nodes are not capable of processing and handling complex tasks, these tasks are pushed towards the nano-controllers which then coordinate their tasks in an efficient manner. The MAC layer for nano-controller includes more functionalities such as link establishment and resource allocation. A clustered architecture is followed in~\cite{esaware2013}. The nanosensor nodes are battery limited devices, which only can store enough energy to perform few simple tasks. The clustered approach can be used to manage high node density where inter-cluster communication can be used to enhance network accessibility. In~\cite{esaware2013}, the transmission and harvesting slots are assigned among the different nanosensors within each cluster in a way that the harvested and consumed energy are balanced among nodes. }

A cluster-based nano-network is also discussed for dense networks in~\cite{EEWNSN2017,EESR-MAC2012}, in which inter and intracluster communications are carried out to reach gateway node. For plant communication, a clustered based architecture is followed in~\cite{dynamic_CH2016} which addresses the frequency selection problem in the Terahertz band which are considered as frequency selective bands. The nanodevice clusters are used to monitor the chemical reaction in plants that schedules the transmission among themselves and transmit the data to microdevices, which then transmit to the Internet via a gateway device.

\mage{The clustered architecture requires efficient MAC protocols to handle efficient data relaying among the inter and intra clusters. Nanonetworks can utilize very high bandwidth due to low path losses at small distances with basic modulation and coding schemes. In such environments, a challenge, related to efficient scheduling and channel access, rises from the fact of large number of nodes (cf. Section~\ref{sec-req:maclayer}). For higher data rates, the nano controller can be pushed towards physical layer synchronization (cf. Section~\ref{sec-req:maclayer}). In addition, efficient mechanisms are also required to perform grouping of nano nodes, the dynamic migration of nodes from one cluster to another, placement and selection of cluster heads constrained by energy consumption and efficient communication. }

\subsection{Terahertz MAC protocols for Distributed networks}\label{subsec-topo:distributed}

Depends upon the application requirements, the devices both in the nano and macro scale can perform communication tasks in a distributed manner. The details of the works following the distributed management are as follows\blue{:}

\subsubsection{Nanoscale networks}\label{subsubsec-topo:distributed-nano}

\saim{In distributed network architecture, nodes perform tasks individually and takes independent decisions for communication. Decisions can be easy when nodes are aware of the environment and physical layer parameters mentioned in Section~\ref{sec-req:phylayer}. For example, when nodes are aware of the channel access and scheduling states of neighbor nodes, overall delay can be minimized and throughput can be increased. In nanonetworks\blue{,} modulation schemes and schedules can be negotiated among the nodes to perform communication~\cite{phlame2012}. The scalable distributed networks are discussed in~\cite{drih-mac2015,RIH-MAC2014}.}


\saim{Due to the absence of controller in Adhoc nanonetworks, nodes need to schedule their transmissions and coordinate to access the channel without generating the interferences. A distributed scheduling mechanism is proposed~\cite{APIS2017} for nanosensor networks in which every node takes decision locally based on its incoming traffic and channel sensing results. The proposed protocol is shown to be reliable in data delivery with optimal throughput and addresses the fundamental challenge of limited memory nanodevices. An adaptive pulse interval scheduling scheme is \blue{also} proposed \blue{in this work} which schedules the arrival pattern of pulses transmitted by a nano sink. }

\blue{In Adhoc network architectures, since nodes positions can be random at times, therefore, nodes relaying can be essential to
guarantee connectivity between distant nodes.} A set of new functionalities should be taken into consideration such as updating the neighboring list for each node. A multihop network can emerge which needs further investigation~\cite{2stateMAC2018,smart-mac2013}. The work in~\cite{RBMP2017} is addressing the antenna facing problem, in which nanosensors that are in direct range can be communicated using an omnidirectional antenna, whereas to communicate with other message stations in presence of obstacles, relay nodes are used with directional antennas. The work is shown to improve the throughput, however, due to the limited capacity of nano nodes, it can increase the message and energy overhead. A flooding scheme for high node density ad-hoc nanonetworks is proposed in~\cite{MGDI2016}, in which a message from the external entity is broadcast in a nano-network with coverage in terms of percentage of receiver nodes. An internal node can also propagate the data towards the external entity of a gateway which is movable. A frequency hopping scheme is modeled in~\cite{TSN2014} to overcome attenuation and noise issues using multiple channels between two nanosensor nodes. For distributed networks, the energy problem is considered in~\cite{NS-MAC2016,MDP2014,GMAC2016}.

\mage{Different topologies require different solutions based on the application requirement, band limitations and design criteria. For example, the small transmission distance and size of nodes require the nodes to be close to each other. Depending on the application the number of the nodes can be huge. Although omnidirectional antennas can be used, high number of nodes can introduce collisions~\cite{2stateMAC2018}, which is not discussed in these works. Further, the band limitations and high node density with energy-efficient solutions are also not seen in combined work.}

\subsubsection{Macro scale networks}\label{subsubsec-topo:distributed-macro}

Besides nanoscale networks, the Terahertz band promises the ultra-high-speed wireless links for macro-scale networks. However, the free space path loss affects the throughput and results in a reduced coverage area. To extend the coverage and minimize the path losses the directional antennas have been encouraged to use. \saim{But, will require frequent beam switching and steering. An efficient beam scheduling mechanism can be helpful in providing continuous transmission with controlled delay. In mobile environments, the Adhoc nature can bring frequent topology changes that need to be sorted out using a MAC layer protocol. For example, nodes will require mechanism to facilitate nodes entering and leaving the network, relaying, handovers and efficient routes between source and destination with lowest delay (cf. Section~\ref{sec-req:design-issues}).}

The Terahertz communication network with high-speed Terahertz wireless link is presented in~\cite{LLsynch2015} for macro scale communication. It addresses the problem of handshaking with antenna speed considerations as an important factor to consider while designing a MAC protocol. Nodes are placed within an area of 10 m circular area in a distributed manner. The mobile heterogeneous architecture is presented in~\cite{ISCT2018,B5G2018} for Ad-hoc connectivity and WLAN to provide high-speed Terahertz links and broadband access using access points. In~\cite{ISCT2018}, an intelligent and secure spectrum control strategy is proposed for an indoor network with different access subnets and an anti-jamming strategy with adaptive frequency slot number selection.

\saim{The Omni-directional antennas can be used to perform the initial link establishment and for data transfer directional antennas can be used to reach further nodes. But, it introduces some challenges including alignment and synchronization to perform a task and extra antenna overhead.} A distributed Terahertz communication network using both directional and omnidirectional antennas is proposed in~\cite{TAB-MAC2016}, in which the anchor nodes are used with regular nodes. The anchor nodes are assumed to know their location in advance and regular nodes are equipped with beamforming antenna arrays. Similar work with Omni and directional antenna is described in~\cite{MRA-MAC2017} where control signals are used for beam alignment using 2.4 GHz link and for data transfer Terahertz links are used. Although using 2.4 GHz band for control signaling reduces the handshaking delay between the nodes, it limits the coverage area and can leave isolated areas in a network, which further requires multihop strategies to increase the reachability of the network. A relaying strategy is proposed in~\cite{OPTRS2017} in a network with randomly distributed nodes. However, only few dedicated relays are used to transfer data and nodes are assumed as switching between the transmission and receiving modes, which can increase the delays.

A software-defined network (SDN) based vehicular network is considered with distance-dependent spectrum switching, where mmWave and Terahertz band are used alternatively based on the data transfer usage~\cite{B5G2018}. It is argued that the universal coverage is not possible by using just Terahertz band and therefore a network architecture is proposed, which uses the microwave, mmWave, and Terahertz bands together to achieve the design goal of coverage and channel access. Although, it can extend the coverage, the switching delay increases as nodes number increases and traffic between nodes increases. Similar work is presented in~\cite{SDNC2017}, which discusses the handoff and MAC protocol to dynamically switch between the mmWave and Terahertz band for high bandwidth data transfer operations. Although, performance is shown to be improved the message overhead and switching delays are high. Further, the synchronization is not focussed on multiple vehicles. Another relay algorithm for autonomous vehicular communication using Terahertz band links to overcome short-range and unstable links is presented in~\cite{ATLR2018}.

\mage{The distributed arrangement of nodes can cause disconnected networks when directional antennas are used. It can also introduce the deafness problem. Synchronization among the nodes to align antenna and exchange neighbor information will be another challenge. To solve the synchronization issue works in~\cite{TAB-MAC2016,MRA-MAC2017} uses multiple bands, which can increase the hardware cost and switching delays. A work presented in~\cite{qxia2019,LLsynch2015} provides link layer synchronization while considering the Terahertz band features. However, the full-beam sweeping time can increase the synchronization delay when nodes will be unaware of other nodes and their beam direction. }

\subsection{Summary and discussion}

Each application has different topology requirements. In nano-communication networks, the nodes are placed at a very small distance from each other. Due to the near node placement, the path loss is less effective in a nano communication network. The omnidirectional antennas can be used in nano-network due to the near placement of nodes. The Omni-directional usage of such scenarios requires a MAC protocol to include collision avoidance methods with efficient sensing mechanism to detect interferences. The path loss increases with distance, therefore to mitigate free space attenuation effect, directional antennas are required. The antenna directionality requirement clearly impacts on link establishment and channel access mechanisms. The transmission schedule can be easily managed in a centralized scenario, in which a central controller is responsible for overall transmission schedules which also requires energy-efficient mechanisms. However, in distributed networks, scheduling the transmissions and resources is a challenge, especially when directional antennas are in use over the short distance. 

At macro scale, for Terahertz communication networks, the topology must account for many practical concerns like scalability, reconfigurability, LOS connectivity due to antenna directionality requirement, fault tolerance, and cost-performance index. In indoor scenarios\blue{,} the distance between the APs and users with mobility support should be covered in the MAC protocol design while providing fault-free and seamless communication. The distributed topology for Terahertz MAC protocol must accommodate the dynamic nature of the network while covering the whole network. The Terahertz Datacentre network, for example, requires top of rack nodes to transmit data among different racks. The short-range limits the connectivity of the nodes, therefore, novel mechanisms are required to approach far distance nodes within a Data Centre.

\section{Channel Access Mechanism for Terahertz communications}\label{sec:ch-access}


\begin {figure*}
\centering
\scriptsize
\begin{adjustbox}{width=\linewidth}

\tikzstyle{block} = [rectangle, draw, text width=4cm, text centered, rounded corners, minimum height=5em,fill=blue!20]
\tikzstyle{line} = [draw,thick, -latex']
\tikzstyle{cloud} = [draw, ellipse, text width=4.5cm, text centered]
\tikzstyle{edge from parent}=[->,thick,draw]
\begin{tikzpicture}[auto,edge from parent fork down]
\tikzstyle{level 1}=[sibling distance=60mm,level distance=18ex]
\tikzstyle{level 2}=[sibling distance=25mm,level distance=15ex]
\tikzstyle{level 3}=[sibling distance=30mm,level distance=34ex]
\tikzstyle{level 4}=[sibling distance=35mm,level distance=34ex]
\node [block,text width=10cm,minimum height=3em,,fill=black!30] (cst) {\textbf{Channel Access Mechanisms in THz MAC protcols \\ (Sect.\ref{sec:ch-access}) }}
{
child{node [block,text width=4cm,fill=green!20] (opm) {\textbf{Nanocale Networks \\ (Sect.\ref{subsec-ch:nano}) } }
  		child{node [block,text width=2.4cm,fill=orange!20,xshift=-1.3cm, yshift=-1.1cm] (pmt) {      \\ 
  		{\bf Random Channel Access Mechanisms \\ (Sect.\ref{subsubsec-ch:nano-random})  } \\
  		{\bf CSMA: } \\
  		CSMA-MAC \cite{slottedCSMAMAC2018}  \\
  		SMART-MAC \cite{smart-mac2013} \\
  		DMDS \cite{DMDS2016} \\
  		{\bf Multiple radios: } \\
  		RBMP \cite{RBMP2017} \\
  		}} 
  	    child{node [block,text width=2.5cm,fill=orange!20,xshift=-1cm, yshift=-2.6cm] (pmt) {      \\ 
  		{\bf Scheduled Channel Access Mechanisms \\ (Sect.\ref{subsubsec-ch:nano-scheduled})  } \\ 
  		{\bf TDMA: } \\
  		DESIGN-WNSN \cite{Design-WNSN2015} \\
  		TCN \cite{TCN2015} \\
  		ES-Aware \cite{esaware2013}    \\
  		EEWNSN \cite{EEWNSN2017} \\
  		EESR-MAC \cite{EESR-MAC2012} \\
  		PHLAME \cite{phlame2012}  \\
  		DRIH-MAC \cite{drih-mac2015} \\
		RIH-MAC \cite{RIH-MAC2014} \\
		2-state MAC \cite{2stateMAC2018} \\
		G-MAC \cite{GMAC2016} \\
		APIS  \cite{APIS2017} \\
		MDP \cite{MDP2014} \\
		SSA-MAC \cite{wangw2019} \\
		{\bf FTDMA: } \\
		DYNAMIC-FH \cite{dynamic_CH2016} \\
		TSN \cite{TSN2014} \\
  		}} 
  		%
        }
%
child{node [block,text width=4cm,fill=green!20] (opm) {\textbf{Macro Scale Networks \\ (Sect.\ref{subsec-ch:macro})  } }
		child{node [block,text width=2.4cm,fill=orange!20,xshift=-0.4cm, yshift = -1.03cm] (pmt) {  \\
		{\bf Random Channel Access Mechanisms \\ (Sect.\ref{subsubsec-ch:macro-random})  } \\
		{\bf CSMA/CA based:} \\
		LL-Synch \cite{LLsynch2015} \\
		OPT-RS \cite{OPTRS2017} \\
		{\bf Multiple radios:} \\
		TAB-MAC \cite{TAB-MAC2016}, \\ 
		MRA-MAC \cite{MRA-MAC2017}, \\
		}}
		child{node [block,text width=2.4cm,fill=orange!20,xshift=-0.2cm, yshift = -1.15cm] (pmt) {  \\
		{\bf Sceduled Channel Access Mechanisms \\ (Sect.\ref{subsubsec-ch:macro-scheduled})  } \\
		{\bf FTDMA:} \\
		ISCT \cite{ISCT2018}\\
		{\bf TDMA:}
		TRPLE \cite{TRPLE2014}\\
		{\bf Multiple radios:} \\
		SDN-CONTR \cite{SDNC2017}\\
		B5G \cite{B5G2018}, \\
		MA-ADM \cite{MAADM2017} \\
		}}
		child{node [block,text width=2.4cm,fill=orange!20,xshift=0.0cm, yshift = -1cm] (pmt) {  \\
		{\bf Hybrid Channel Access Mechanisms \\ (Sect.\ref{subsubsec-ch:macro-hybrid})  } \\
		{\bf CSMA/CA and TDMA:} \\
		HLMAC \cite{HLMAC2016}\\
  		IHTLD-MAC \cite{IHTLD-MAC2017}\\ 
  		MAC-TUDWN \cite{MAC-TUDWN2013}\\ 
  		MAC-TC \cite{MAC-TC2013}\\		
		}}		
}
};
\end{tikzpicture}
\end{adjustbox}
\caption {Terahertz Channel Access Mechanisms classification.}
\label{fig:ch_access_class}
\end{figure*}
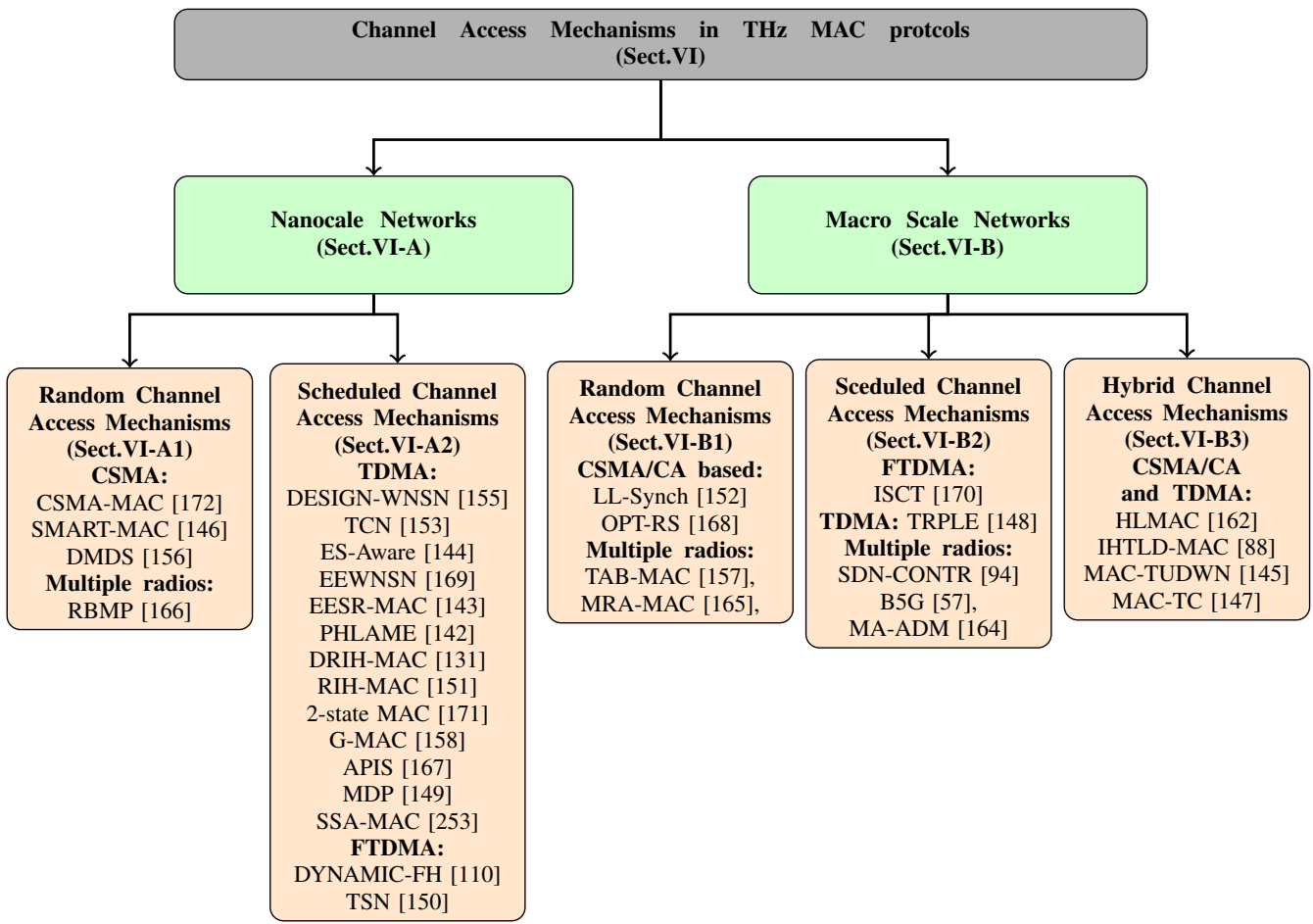

\saim{In this section, the existing channel access mechanisms for Terahertz band communications are presented. They are mainly classified based on macro and nanoscale Terahertz network. They are further classified as Random, Scheduled and Hybrid channel access mechanisms as shown in Figure~\ref{fig:ch_access_class} and discussed below. }

\subsection{Nanoscale Networks}\label{subsec-ch:nano}

\saim{The pulse-based communication in nanonetwork transfers information using short pulses which reduces the chance of having collisions (cf. Section~\ref{sec-req:phylayer}). To avoid possible collision, the duration between two pulses can be increased to allow different users stream at the same time. A nanodevice can send a packet when it has something to send without waiting in a random way, where receiver devices should be able to detect such pulse. A node can also be aware of or predict the next transmission from the received packet. }

In nanonetworks\blue{,} several nanosensor nodes, taking random positions, can be used to maintain the network connectivity for different applications, such as in-body sensing, toxic gas detection and control, and military fields. The sensing and communication capabilities limit their target area to few millimeters. \blue{Although, smaller size antennas can be integrated and massive data can be exchanged at high data rates, however, this requires a simple, robust and energy-efficient channel access mechanism for communication among the nano nodes.}


\subsubsection{Random channel access}\label{subsubsec-ch:nano-random}

\saim{In Random-access mechanisms, different nodes contend for the channel access or transmit packet in a random manner. The random access is not suitable for applications in which higher number of nano nodes are used. This is due to the limited sensing, computation, battery and memory capacity of nanodevices, which can allow only a few transmissions until another harvesting phase is required. }

\saim{The carrier-based channel access mechanisms are mostly unsuitable for nano communication due to extra sensing overhead and energy consumption. In~\cite{slottedCSMAMAC2018}, a slotted CSMA/CA-based channel access mechanism (CSMA-MAC) is proposed in which nodes contend for the channel. An energy harvesting model is also presented. The slots usage is found higher when slotted CSMA method is used. In addition, the super-frame duration and packet size are also mentioned to be considered for slotted CSMA protocols. In these types of networks, a beacon can be used to either synchronize the next transmissions or to ask for transmission of data packets directly. In direct beacon transmission for data transmission, collision can occur as two nodes can transmit at the same time and it also needs to address the energy constraint as frequent packet transmission can drain the nanodevice energy fairly quickly. Another work that uses carrier sensing is presented in~\cite{DMDS2016}, where sensing duration is used to optimize the transmission schedules. }

\saim{An simple Aloha based channel access mechanism is presented in~\cite{smart-mac2013}, as Smart MAC in which nodes perform handshake before sending a packet to know about one-hop neighbors.} \saim{When there will be no neighbor to transmit\blue{,} the node will apply a random backoff delay prior to start again another handshake mechanism. The receiving node verifies when there are any physical collisions. The collisions can occur when there is a higher number of nodes available, which is not addressed and the retransmission mechanism is not discussed. }



\saim{A random channel access with multiple radios is used for nanonetworks in~\cite{RBMP2017}. For control signal transmission\blue{,} 2.4 GHz band is used and \blue{THz band is used} for data transmission. The channel is accessed in a random manner in both phases. It also addresses the antenna facing problem in first phase by synchronizing the antenna directions for the data transmission phases. \blue{This is primarily because} the narrow beams can not cover the whole search space. It mainly overcomes the synchronization problem using directions antennas as in Terahertz band at the cost of multiple radios. The alignment between two phases is required to decide the time in each phase of transmission.}

\mage{When the number of nodes is higher, random access or packet transmission can pose several challenges including collisions, transmission delays, and higher energy consumption. However, the energy-efficient harvesting mechanisms are not considered in these works. The works in~\cite{slottedCSMAMAC2018,smart-mac2013} do not consider the limited memory and energy limitations of nano nodes. In~\cite{RBMP2017}, synchronization among the nodes is addressed but at the cost of multiple radios which increases the challenges. The random access mechanisms can increase the throughput but high message overhead and sensing require sufficient energy availability. Whereas, the scheduled mechanisms, solve the energy issue but at the cost of throughput. A work in~\cite{DMDS2016} provides the optimal schedules with random sensing at the beginning of the slot for distributed environment. It considers the limited memory and energy consumption of nano nodes, but not maximizing the overall throughput. The work is for distributed environment but only few nano nodes were used for implementation. }

\subsubsection{Scheduled channel access}\label{subsubsec-ch:nano-scheduled}

\saim{Nanodevices require a simple communication and medium access mechanism to effectively collect data from other nanosensor devices. Due to presence of large number of nanodevices, providing optimal schedules for nano nodes\blue{,} life long communication is a challenging task. In centralized network a nano controller is responsible mainly for single-hop nodes schedule. However, in distributed networks, managing schedules for channel access for more than hop distance nodes is a challenging task. The optimal solution should also consider the limited nano nodes capacity, huge bandwidth availability and energy fluctuations while designing an efficient and optimal channel access scheduling mechanism~\cite{TCN2015,sasiibala2013} (cf. Section~\ref{sec-req:design-issues}). }

\textit{TDMA based: } \saim{In TDMA based approaches, each node is assigned a time period to transmit its data. A scheduling mechanism based on a TDMA based approach is presented in~\cite{Design-WNSN2015,TCN2015,esaware2013} in which a nano-controller makes the decision for a nanosensor node to transmit the sensing data. In~\cite{TCN2015}\blue{,} a timing-based logical channel concept is used. In these logical channels information can be encoded in the silence period between two events. Although, these logical channels are shown to achieve synchronization among the nodes with energy efficiency, low rate, and collision avoidance. The unique features of Terahertz band are not considered (cf. Section~\ref{sec-req:features}).  }


\saim{A dynamic scheduling scheme based on TDMA is presented in~\cite{esaware2013} as ES-aware, in which a nanosensor dynamically assigns variable length transmission time slots which depend upon amount of data to be transmitted; the distance between the nanosensor and controller; and the energy of the nanosensor. To balance the trade-off between the throughput and a lifetime, an optimal scheduling strategy is proposed which aims to provide an optimal transmission order for the nanosensors to maximize the throughput. This algorithm utilizes the inter-symbol spacing for the pulse-based physical layer to allow a large number of nanosensors to transmits the packets in parallel without introducing any collisions. This works are mainly for single-hop networks. }

\saim{The TDMA based scheduling for multihop WNSN MAC (EEWNSN) protocol is presented in~\cite{EEWNSN2017}, which takes benefits of the clustering techniques to alleviate the mobility effects and transmission collision. After selecting a nano router, the nano router allocates the specific time slots to the nano node according to a systematic allocation pattern. The timeslots are considered fixed and due to the transmission to the closest nano router, the energy consumption can be decreased which can prolong the network lifetime. Another work following a cluster-based architecture for nanonetworks (EESR-MAC) is \blue{presented} in~\cite{EESR-MAC2012}, in which initially a master node is selected which then allocates the transmission schedules between the inter/intra-cluster using TDMA approach. The master node's roles are periodically changed among different nodes to avoid long-distance transmissions and to save energy.  }

\saim{In~\cite{esaware2013}, spectrum and energy parameters are considered for schedule assignment. In~\cite{phlame2012}, nodes select different Physical layer parameters, energy, and channel conditions, agreed by using a handshake process that can limit the performance due to the limited capacity of nano nodes. Although, parameters can be negotiated the dynamic and optimal parameter selection is still required for these networks to increase the lifetime and performance. A Rate division Time-Spread On-Off Keying (RD TS-OOK) is also proposed which is based on asynchronous exchange of femtosecond long pulses spread over time. To minimize the probability of multiple sequential symbol collisions in a packet, the time between the symbol $T_s$ and symbol rate $\beta = T_s / T_p$ are chosen differently for different nanodevices for different packets. When all nanodevices are transmitting at the same symbol rate, a catastrophic collision can occur which can cause collision for all symbols in a packet. The orthogonal time hopping sequences can be used to avoid this condition~\cite{lpan2016}. The symbol collisions are unlikely to occur due to a very short length of the transmitted symbols $T_p$ and because the time between symbols $T-S$ is much longer than the symbol duration $T_p$. By allowing different nanodevices to transmit at different symbol rates, collision in a given symbol does not lead to multiple consecutive collisions in the same packet. As an example, an RD TS-OOK illustration is shown in Figure~\ref{fig:rdtsook} in which two nanodevices transmit to a common receiver with different initial transmission times as ${\tau}^1$ and ${\tau}^2$. A short pulse represents a logical $1$ and a silence represents a logical $0$. The device 1 plot shown a sequence ``10100" and device 2 plot shows a sequence of ``11100". }


\begin{figure*}[htb!]
\centering
\includegraphics[width=4in,height=2.2in]{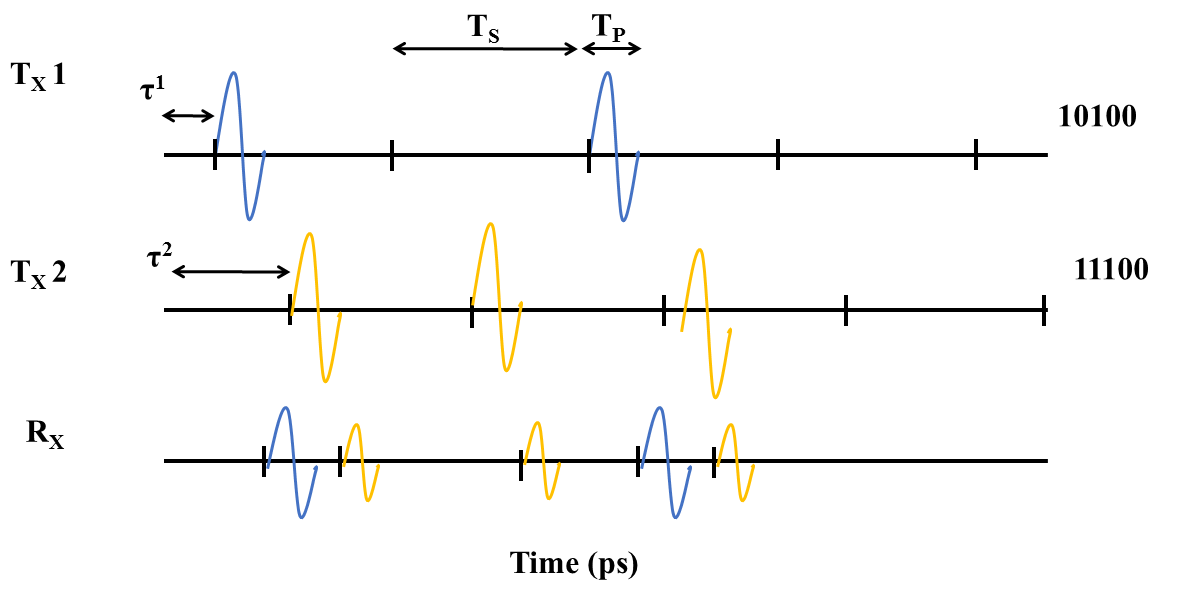}
\caption{An example of Rate Division Time-Spread channel access mechanism used for pulse based communication in Terhertz Nano communication networks\protect\cite{phlame2012}.}
\label{fig:rdtsook}
\end{figure*}

\saim{A scheduling mechanism for distributed networks is presented in~\cite{drih-mac2015} (as DRIH-MAC) using edge colors. Its centralized version is presented in~\cite{RIH-MAC2014} using probabilistic method. The edge coloring problem is considerably challenging in ad-hoc based networks due to the absence of a centralized coordinator. In DRIH-MAC~\cite{drih-mac2015}, the medium access control relies on the receiver-initiated and distributed scheduling for nano nodes in which each pair of nano nodes within a communication range will have an edge with different color. The main objective is to determine a minimum number of colors required to color the edges of a graph i.e., two edges incident on a common node do not have the same color, where each color represents a timeslot in which a nano node can communicate with one of its neighbors. At most $(\delta + 1)$ timeslots are needed at least to reach an agreement/disagreement on color with all neighbors through RTR packet assuming no RTR packet failure. These works are shown to be efficient, however, the limited memory capacity of nano nodes is not considered in these works.}  

\saim{Due to limited battery capacity, frequent transmission can not occur using a nanodevice. While designing a channel access mechanism, energy harvesting must be considered to achieve optimal network performance. A two-state MAC is proposed in~\cite{2stateMAC2018} in which two states are used for a node a harvesting only and harvesting with transmission. Another work using harvesting in sleeping and transmission modes is presented in~\cite{GMAC2016} for grid-based nanonetworks.  }

\saim{For accommodating bursty traffic in a distributed environment, an adaptive pulse interval scheduling (APIS) scheme is presented in~\cite{APIS2017}. In this scheme, the arrival pattern of pulses transmitted is scheduled by nano sinks based on the access bandwidth. It has two scheduling steps such as transmission shifting and interleaving which are based on information collected from short channel sensing. When nano sinks start transmitting pulses, they are first shifted in sequence within interval of $I_S$. After which multi-user transmissions are interleaved by separating pulses with the interval that evenly shares the bandwidth among nano sinks and in response the pulses arrive at the gateway in an ideal pattern. }

\mage{The main issues while designing a scheduled access mechanism for nanonetworks are to consider the energy harvesting and consumption, limited computational capacity, Terahertz band features and optimal performance with node density. The works in~\cite{MDP2014,GMAC2016,phlame2012,2stateMAC2018,TCN2015} considers energy problem, however, they lack in considering other aspects of memory, collisions, and Terahertz band features. In~\cite{phlame2012}, channel parameters, and coding scheme aware protocol is presented but it does not consider the high node density, balance in energy harvesting and consumption and limited computational capacity of nanodevices. The work is required for distributed nanonetworks while considering the unique aspects of Terahertz band. In~\cite{EEWNSN2017}, multihop protocol is discussed but energy efficiency is not considered. The nano nodes require an efficient mechanism to balance between energy harvesting and consumption. The work in~\cite{wangw2019}, is shown to be better in performance than~\cite{phlame2012,RIH-MAC2014,EEWNSN2017}, and considers self slot allocation with energy for both centralized and distributed environment, but band features are not considered and the trade-off between energy harvesting and consumption is not discussed. In~\cite{esaware2013}, the access scheme is provided while considering this trade-off with fair throughput and optimal lifetime, where nodes are aware of energy and spectrum information. In this work, the high node density is also considered for performance evaluation. }

\textit{Frequency and Time Division Multiple Access (FTDMA) based: } In~\cite{dynamic_CH2016}, dynamic frequency selection strategy (DYNAMIC-FH) is presented which uses FTDMA. An FTDMA is initially considered and for a higher number of nano nodes multi-frequency is proposed with timeslots scheduling for a different number of users. Each node is assigned with different timeslots to avoid collisions in case of higher packet sizes (like multimedia traffic). The main objective of the frequency selection strategies is to minimize energy consumption and increase channel capacity. In~\cite{TSN2014}, also a Markov Decision Process-based frequency hopping scheme (TSN) is proposed in which entire band is divided into $K$ frequency sub-channels where the aim is to determine for each timeslot the subchannels be used. \\


\subsection{Macro Scale Networks}\label{subsec-ch:macro}

Figure~\ref{fig:ch_access_class}, shows the classification of Terahertz band channel access mechanisms for Terahertz macroscale networks. The summary and details of each category are discussed below.

\subsubsection{Random channel access}\label{subsubsec-ch:macro-random}

The examples of random mechanisms are ALOHA and CSMA techniques. Ideally, for the random mechanism, a node should sense a medium before accessing it. Since there is a large bandwidth available, the chances of collision occurrence are less. Therefore, the random mechanism is also being used followed by message confirmation strategies. However, the idea of collision and interference cannot be ignored completely as there could be many users accessing the same medium and might be transferring a large volume of data which could potentially generate collisions among the two nodes (cf. Section~\ref{sec-req:maclayer}). The collisions avoidance schemes and recovery from collision schemes are very essential. The delay, on the other hand, can be minimized due to random access, however, further research is required for the collisions and delay parameters trade-off. 

\textit{CSMA based: } \saim{In carrier sensing based channel access schemes, control packets are exchanged before data transmission to access channel. The high message overhead and directional antennas can create problems. To reduce the message overhead, in~\cite{LLsynch2015}, a one-way handshake based channel access scheme is proposed.} A node in transmission mode listens for messages from other nodes until one is received. As the directional antennas are used, the antenna facing problem can occur, however, it is assumed that the nodes know each other's position. To address the antenna facing problem, the work is extended in~\cite{OPTRS2017} by focusing on the use of highly directional antennas to overcome high path loss. In~\cite{OPTRS2017}, the relaying distance is studied that maximizes the throughput by considering the cross-layer effects between the channel, the antenna and the communication layers. This work also focuses on control message exchange to establish nodes association and follows the random channel access, as in~\cite{LLsynch2015}. Although, these schemes are shown as workable, but are not considering the high node density, message overhead and unique Terahertz band features discussed in Section~\ref{sec-req:features}.

\textit{CSMA with multiple radios or hybrid system: } \saim{The limited power of transceiver and high path loss limits the transmission distance as discussed in Section~\ref{sec-req:features} and can also increase the channel access problem. Therefore, requires directional antennas with beamforming both at the transmission and reception. The beams need to aligned to establish a link and to access a channel before transmitting data packets (cf. Section~\ref{sec-req:phylayer}). The beam alignment takes time when antenna sweeps to discover the neighbors. Therefore, in some works like~\cite{TAB-MAC2016,MRA-MAC2017}, multiple radios are used to divide initial access and data transmissions. These works can increase the message overhead, increase antenna switching delay at higher radio costs. The channel is accessed by sending an request/clear-to-send (RTS/CTS) packets including node positions. However, if nodes are mobile, it will cause repeated discovery phase. In~\cite{MRA-MAC2017}. instead of sending a clear to send packet to the transmitter, the receiver estimates the angle of arrival and sends TTS packet to the transmitter. The transmitter can then switch and adjust its directional antenna and starts pointing towards the receiver antenna to start the data transmission. Although the message overhead is shown to be reduced, the uncertainty of packet loss during user association phase is not considered\blue{. T}he difference is shown in Figure~\ref{fig:tabmra} in which one scheme uses 4 transmissions until antenna direction alignment and other uses two transmissions. The high path and absorption loss are also not discussed in these works, which can cause a packet loss. }

\mage{Works in~\cite{LLsynch2015,OPTRS2017} uses random channel access with a one-way handshake to reduce the control message overhead. However, the high node density can cause collisions problems which are not considered in these works. Further, synchronization can be a problem when using directional antennas in these works. To solve the synchronization issues, the works in~\cite{TAB-MAC2016,MRA-MAC2017}, use multi-band antennas with lower frequency bands and Terahertz bands. Although synchronization problem can be solved, the antenna switching and alignment can increase the delay. }

\begin{figure}[htbp]
\centering
\includegraphics[width=3.2in,height=2.2in]{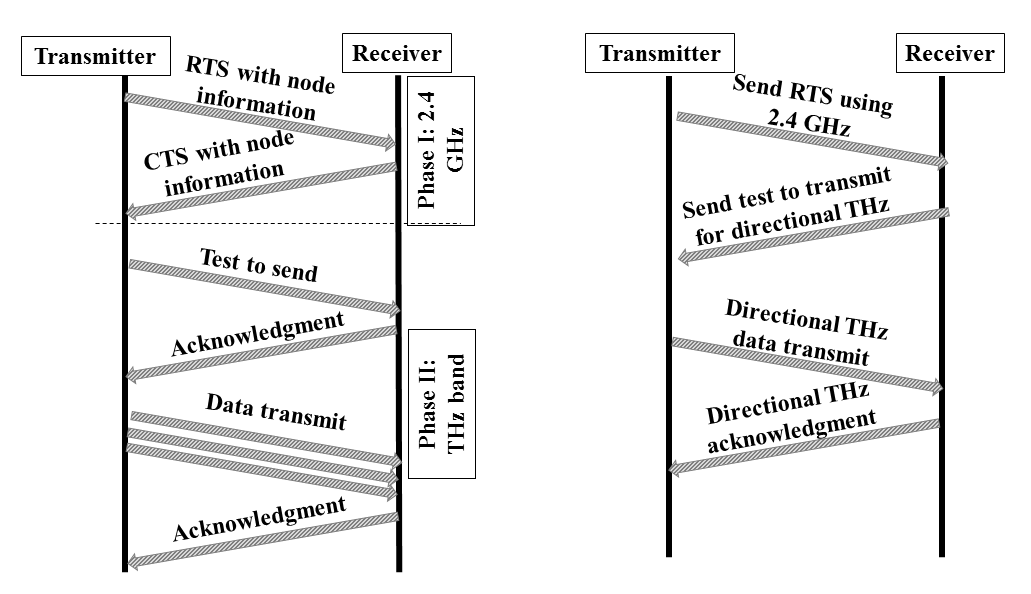}
\caption{Random channel access, node association with antenna direction alignment difference of TAB-MAC~\protect\cite{TAB-MAC2016} and MRA-MAC~\protect\cite{MRA-MAC2017} protocols.}
\label{fig:tabmra}
\end{figure}

\subsubsection{Scheduled channel access}\label{subsubsec-ch:macro-scheduled}

\saim{In distributed networks assigning the schedules to Terahertz nodes is an NP-hard problem~\cite{SDNC2017}. Until the device technology is more mature to allow non-LOS communication, scheduled channel access can enhance the network performance in the presence of the above-mentioned constraints. In scheduled access, each node is assigned with a particular timeslot. Some of the variations of scheduled access are discussed below like FTDMA and TDMA based channel access mechanisms in Terahertz. }

\textit{FTDMA: } \saim{An FTDMA based technique is used in~\cite{ISCT2018} in which the available frequency is further divided into sub-bands and assigned frequency slot numbers. In FTDMA, frequency is divided into different timeslots. The frequency used by any user in a particular time could be represented by a sequence $S^k$. To avoid jamming different sequences or transmission strategies are adopted by each user. For $n$ users transmitting at the same time, the sequences used by each user must be orthogonal. The performance is shown to be improved for security and throughput, however, the unique features of Terahertz bands are not considered like path loss and noise. }

\textit{TDMA: } \saim{A TDMA based channel access scheme is used in~\cite{TRPLE2014} to avoid fighting for access. In~\cite{TRPLE2014}, a fully directional MAC protocol for the Terahertz network is presented which relies on pulse-level beam-switching and energy control. A MAC frame structure is presented which is composed of a POLL period, a Downlink (DL) period and an Uplink (UL) period. In the POLL period, the AP learns the traffic demands of the users and schedules the DL/UL transmissions. In DL and UL, each different user is assigned a separate timeslot to access the channel. }

\textit{TDMA with Multiple radios: } \saim{In Terahertz band, directional narrow beams are required to enhance the transmission distance, as discussed in Section~\ref{sec-req:phylayer}, which can increase the delay to establish initial access, handovers and beam tracking. The TDMA based approaches suits to assign schedules for beam alignment and channel access which also requires synchronization among nodes~\cite{SDNC2017,B5G2018}. A 2.4 GHz band and mmWave with non-LOS are being used to achieve initial synchronization and beam alignment, and a Terahertz band can be used to transfer data. In this way, next time when a node enters into a communication range or requires data, beams can be aligned in advance to perform seamless communication. }

A TDMA based channel access scheme is used in~\cite{SDNC2017}, in which an SDN based controller (SDNC) is used to switch between mmWave and Terahertz band for vehicular communication for high bandwidth data transfer operation. An optimal procedure at the SDN controller for scheduling multiple vehicles for accessing a given small cell tower is also given using a time division approach. The objective is to maximize the bits exchange between the cell tower and vehicle where the condition is to at least schedule one car in each timeslot while considering the distance. 

For link switching, it is proposed that the Terahertz band should be switched whenever the link between the vehicles and the cell tower is less than $d_{th}$, and to mmWave otherwise. Another work discussing the hybrid usage by switching between the mmWave, $\mu$W band, and Terahertz band is given in~\cite{B5G2018}. In this work, a higher capacity link like the Terahertz band is used for data transfer and mmWave for ACK transfer. For error recovery stop and wait is followed with data transmission using Terahertz band and ACKs using the mmWave band. However, this alternate band usage can introduce excessive overhead, higher delay for receiving an ACK, and also introduces the beamforming overhead as the communication must be directional for the Terahertz band. 

A memory assisted angular division multiplexing MAC protocol (MA-ADM) is proposed in~\cite{MAADM2017} for centralized architectures in which an access point is responsible for the coordination and scheduling the transmissions to achieve fairness and efficiency. It also uses Omni-directional antennas to overcome beam alignment and discovery problems and uses directional antennas for data transmission. Memory-Guided message transmission is used by the AP, in which during the network association phase, a node establishes the connection with the AP using an angular slot and register it in the memory. The AP switches the narrow beam by checking the memory towards the registered angular slots for data transmissions to avoid the empty scanning of the unregistered angular slots. The initial use of omnidirectional antenna can limit the service range which can affect the connection of nodes with AP, which is not considered. In addition, the switching delay and cost are also analyzed. To maintain the fairness in which transmission completion is verified by the reception of an ACK message or repeated failure occurrence the service discovery phase is triggered to update the guided transmissions. The scheduling although is considered for data transmission but not focused in detail.

\mage{In macro-scale network\blue{,} synchronization with directional narrow beams is a challenge. The scheduled access can provide contention-free network but increases the delay. A synchronization with narrow beams directional antennas can be used but beam alignment requires more focus to reduce the switching and transmission delay. Although in works like~\cite{TRPLE2014}, TDMA approach is followed for channel access, efficient synchronization is still required. To solve the synchronization problem, multiple bands are used in~\cite{SDNC2017,B5G2018} but overhead is high and specific Terahertz band features (cf. Section~\ref{sec-req:features}) are not considered. A memory assisted approach can be useful, as proposed in~\cite{MAADM2017}, for using the beam direction from the memory. It is not considering the Terahertz band features. }

\subsubsection{Hybrid channel access mechanism}\label{subsubsec-ch:macro-hybrid}

The random access mechanisms can increase the delay and can cause collisions when node density is high but can enhance throughput. However, the scheduled access schemes can reduce the impact of collisions but can enhance the throughput performance. Therefore, hybrid mechanisms are required to overcome the limitations of these schemes. A hybrid channel access mechanism is proposed in~\cite{IHTLD-MAC2017,MAC-TUDWN2013,MAC-TC2013}. In hybrid mechanisms CSMA and TDMA are used in which the channel time is divided into multiple superframes. Each superframe consists beacon period, channel time allocation period (CTAP) and channel access period (CAP). The CSMA/CA is used to compete for the channel in CAP period in which the device which wants to transmit data need to send a channel time request command to PNC. The PNC broadcast slot assignment information in the net beacon frame according to request frame received. The devices can synchronize themselves based on the synchronization information and obtain CTAP slot allocation information, which is made of channel time allocation. The devices access channel in TDMA mode where each device transmits data in its allocated slot. An on-demand retransmission mechanism is also proposed to decrease the message overhead with reserved slots mechanism based on channel condition. This work is an extension of~\cite{HLMAC2016} which also uses a hybrid system and describes a high throughput and low delay Terahertz MAC protocol. The throughput is also shown to be improved by updating the timeslot request numbers with a reduction in latency efficiency.  

\mage{The hybrid channel access mechanisms can improve the limitations of both random and scheduled channel access schemes. The schemes in~\cite{IHTLD-MAC2017,MAC-TUDWN2013,MAC-TC2013} can improve the performance, the poor network conditions are not considered. The new scheme should consider the Terahertz band features (cf. Section~\ref{sec-req:features}) and design considerations (cf. Section~\ref{sec-req:design-issues}).}



\subsection{Summary and discussion}

The traditional channel access schemes are based on continuous signals which cannot be used for nanonetworks due to the size and energy constraint. Instead, short pulses (100 fs) can be generated using simple devices (\blue{G}raphene antenna) and transmitted at the nanoscale. Therefore, novel channel access mechanisms are required for nanoscale networks. Mostly, scheduling based channel access mechanisms are used to avoid frequent messaging to contend for channel access and energy consumption. The new mechanisms should consider the high network density and limited energy availability and in some cases nodes mobilities. Using short pulses can reduce the collision probability and therefore for high-density network random channel access mechanisms can be used by increasing the duration between two transmissions. But, in their MAC the collision avoidance mechanism should be considered to avoid any possible collision and to reduce the message overhead. By using isotropic antenna, a node at nanoscale can communicate with more nodes in its communication range. At the macro scale, the directional antennas are preferred due to high path loss and coverage enhancement but require antenna direction alignment and beam management. To solve the problem of antenna alignment different mechanisms are proposed in the existing literature which includes searching the space and adding an additional tuning phase~\cite{ieee802153d2017}. But, these mechanisms can increase the link establishment delay and energy. It can cause a hidden node or deafness problem (cf. \blue{Section}~\ref{sec-req:maclayer}), in which a node remains unaware of the existence of the nearby node due to limited coverage or antenna misalignment.

Pulse based communication is mainly used in nanonetworks with channel access schedules using a nano controller. In pulse-based communication, short Terahertz pulses can be generated using specific devices to reach higher throughput. Whereas, time division technique for Terahertz MAC is mostly related to the fact that each node has equal share for channel access and to avoid possible collision. However, the TDMA based approaches require efficient and optimal scheduling schemes while considering antenna direction, beam alignment, multi-user interference, and Terahertz band features. To avoid tight synchronization requirement, asynchronous MAC protocols requires further research with energy efficiency. The channel access for macro scale relies more on beam steering. Therefore, an alignment phase should be considered at the MAC layer. The antenna pattern for nanoscale is assumed as isotropic and directional for macro-scale network for which the gain requirements are also high. Further, the high interference for nanoscale due to isotropic properties of antenna requires efficient algorithm while considering the channel characteristics, antenna, gain and transmission power. New realistic measurement studies are also required for modeling of channel behavior for different indoor and outdoor applications at macro scale.

\section{Transmitter and receiver initiated Terahertz MAC protocols}\label{sec:communication}

\begin {figure*}
\centering
\begin{adjustbox}{width=\linewidth}
\scriptsize
\tikzstyle{block} = [rectangle, draw, text width=4cm, text centered, rounded corners, minimum height=5em,fill=blue!20]
\tikzstyle{line} = [draw,thick, -latex']
\tikzstyle{cloud} = [draw, ellipse, text width=4.5cm, text centered]
\tikzstyle{edge from parent}=[->,thick,draw]
\begin{tikzpicture}[auto,edge from parent fork down]
\tikzstyle{level 1}=[sibling distance=60mm,level distance=18ex]
\tikzstyle{level 2}=[sibling distance=25mm,level distance=15ex]
\tikzstyle{level 3}=[sibling distance=30mm,level distance=34ex]
\tikzstyle{level 4}=[sibling distance=35mm,level distance=34ex]
\node [block,text width=10cm,minimum height=4em,,fill=black!30] (cst) {\textbf{Transmitter and Receiver Initiated THz MAC Protocols \\ (Sect.\ref{sec:communication})} }
{
child{node [block,text width=5cm,fill=green!20] (opm) {\textbf{Transmitter Initiated MAC protocols \\ (Sect.\ref{subsec-comm:tx})  } }
  		child{node [block,text width=2.8cm,fill=orange!20,xshift=-0.1cm, yshift=-2.2cm] (pmt) {\textbf{Nanoscale networks \\ (Sect.\ref{subsubsec-comm:tx-nano})  }       \\ 
        PHLAME \cite{phlame2012}\\
        ES-AWARE \cite{esaware2013}\\
		TCN \cite{TCN2015}\\ 
		LL-Modelling \cite{LLModelling2011} \\
		DMDS \cite{DMDS2016}\\ 
		2-state MAC \cite{2stateMAC2018} \\
		CSMA-MAC \cite{slottedCSMAMAC2018}\\ 
		G-MAC \cite{GMAC2016} \\
		APIS \cite{APIS2017}\\ 
		ERR-CONTROL \cite{joint_error2016} \\
		MGDI \cite{MGDI2016}\\ 
		PS-OPT \cite{pkt_size2016} \\
		NS-MAC \cite{NS-MAC2016}\\ 
		DYNAMIC-FH \cite{dynamic_CH2016} \\
		EESR-MAC \cite{EESR-MAC2012}\\ 
		RBMP \cite{RBMP2017}\\
		SSA-MAC \cite{wangw2019} \\
  	    }}
        child{node [block,text width=2.8cm,fill=orange!20,xshift=0.5cm, yshift=-1.1cm] (pmt) {\textbf{Macro scale networks \\ (Sect.\ref{subsubsec-comm:tx-macro}) }   \\
        IHTLD-MAC \cite{IHTLD-MAC2017} \\
        ATLR \cite{ATLR2018} \\
  		TRPLE \cite{TRPLE2014} \\
  		SDNC \cite{SDNC2017} \\
  		B5G \cite{B5G2018} \\
  		MA-ADM \cite{MAADM2017 } \\
		TAB-MAC \cite{TAB-MAC2016} \\
		MRA-MAC \cite{MRA-MAC2017} \\
		MAC-YUGI \cite{MAC-Yugi2018} \\   
        }        }
        }
child{node [block,text width=5cm,fill=green!20] (opm) {\textbf{Receiver Initiated MAC Protocols \\ (Sect.\ref{subsec-comm:rx}) }}
		child{node [block,text width=2.8cm,fill=orange!20,xshift=0.2cm, yshift = -0.4cm] (pmt) {\textbf{Nanoscale networks \\ (Sect.\ref{subsubsec-comm:rx-nano}) }   \\
		DRIH-MAC \cite{drih-mac2015} \\
		MDP \cite{MDP2014} \\		
		RIH-MAC \cite{RIH-MAC2014} \\
		LLsynch \cite{LLsynch2015} \\	
		}}
		child{node [block,text width=2.8cm,fill=orange!20,xshift=0.85cm, yshift = -0.1cm] (pmt) {\textbf{Macro scale networks \\ (Sect.\ref{subsubsec-comm:rx-nano}) }  \\
		OPT-RS \cite{OPTRS2017} \\
		LLsynch \cite{LLsynch2015} \\	
		} } 
}
};
\end{tikzpicture}
\end{adjustbox}
\caption {Classifications of Terahertz MAC protocols based on Transmitter and Receiver initiated communications.}
\label{fig:communications}
\end{figure*}
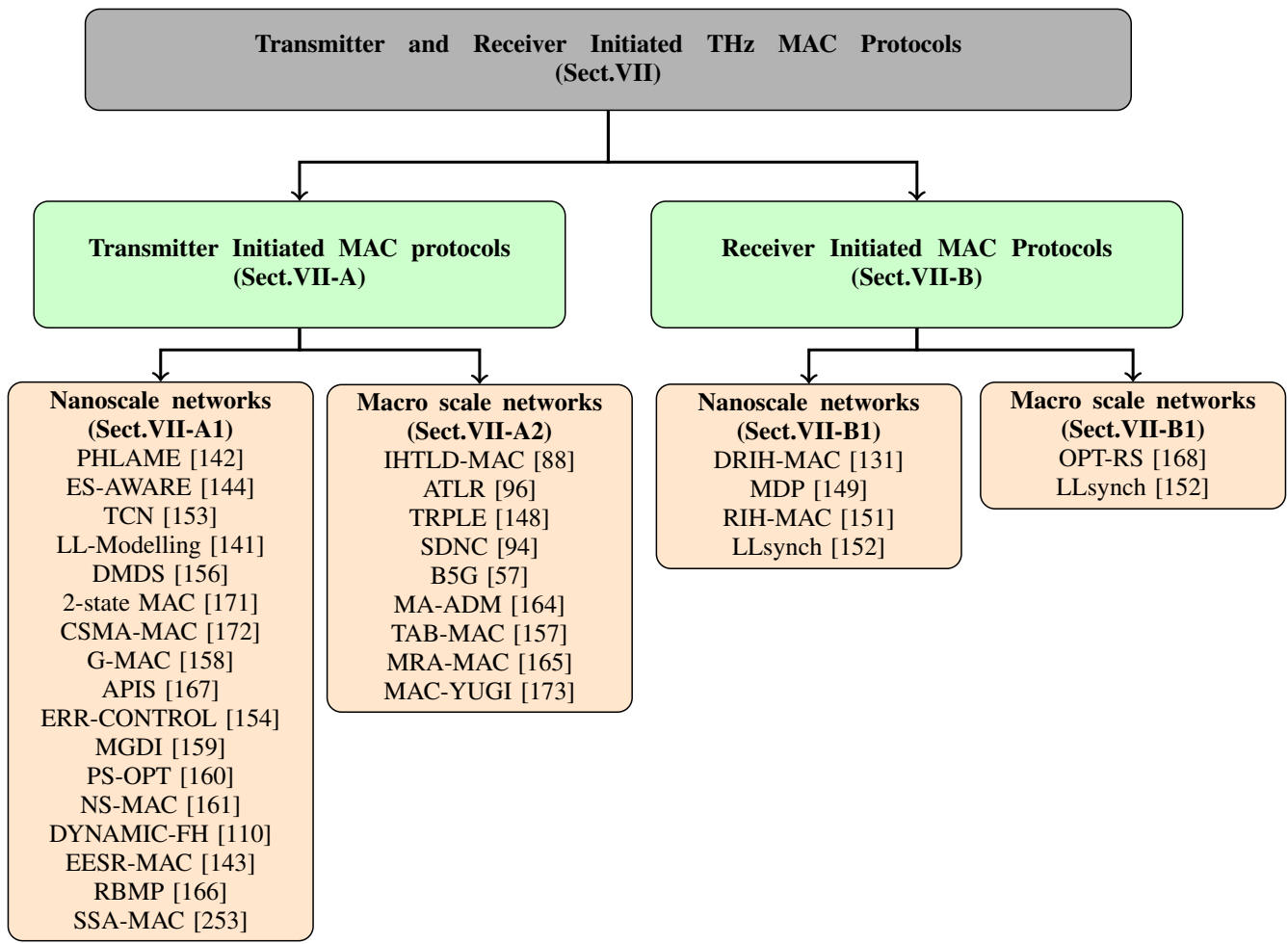


\begin{figure}%
    \centering
    \subfloat[Tx initiated communication]{{\includegraphics[width=2.4in,height=1.3in]{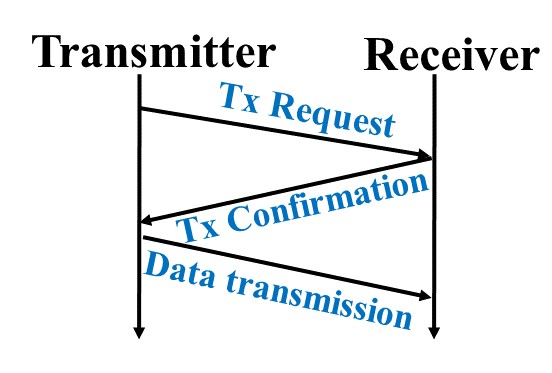} }\label{fig:TIHS}}\\
    \qquad
    \subfloat[Rx initiated communication]{{\includegraphics[width=2.4in,height=1.3in]{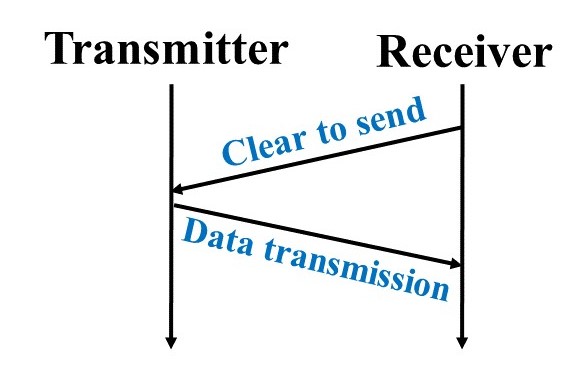} }\label{fig:RIHS}}%
    \caption{Message transmission flow for handshake in Terahertz MAC protocols for link establishment. a) Transmitter initiated handshake mechanism which requires confirmation from receiver of its sent packet before starting a data transmission, b) Receiver initiated handshake mechanism in which receiver initiated communication when it required some information or have enough energy to receive a message, mostly used in nano communication networks.}%
    \label{fig:txrx-handshake}%
\end{figure}

\saim{Establishing a link is an essential part to obtain before starting a communication that can be used to achieve synchronization and exchange information to establish a network like neighbor information, physical parameters and beam alignment (cf. Section~\ref{sec-req:maclayer}).  The handshake mechanisms should be carefully designed to reduce link establishment delay while considering energy efficiency. Two kinds of handshaking mechanisms are generally followed in Terahertz communication which are receiver-initiated and transmitter initiated communication to establish links among different nodes. In particular, the receiver-initiated MAC aimed at reducing the number of transmission in resource-constrained nano and macro-scale networks. Whereas the transmitter-initiated communication focuses on the performance efficiency of the network in a traditional way. Typically, the directional antennas are used for transmitter initiated communication in TCNs due to its narrow beam requirement and distance-dependent bandwidth~\cite{phlame2012}. Other than directional antenna usage, some proposals use multiple antennas to establish initial coordination between multiple nodes. The solutions so far on receiver and transmitter initiated coordination protocols are mentioned below and are shown in Figure~\ref{fig:communications}.}

\subsection{Transmitter initiated MAC protocols}\label{subsec-comm:tx}

In a traditional way, the transmitter is mainly responsible for link establishment, data transmission and synchronization of nodes parameters like scheduling times and channel information. Most of the Terahertz MAC protocols are following the transmitter-initiated communication due to its simplicity and distributed nature. However, the distance-dependent behavior of Terahertz band due to absorption and path loss; directional antenna usage; high bandwidth and throughput support, increases the challenges. These challenges include the antenna facing problem introduced when the transmitter and receiver remain unaware initially about the position and antenna direction; the hidden node problem; and reliability of communication i.e., the packet is lost due to path loss or collision. A general example of transmitter initiated communication is shown in Figure~\ref{fig:txrx-handshake}, in which a transmitter or a node which has information can send a packet, a receiver which receives that packet can trigger an ACK or confirmation to send and after that, a sender can send the Data packet. The transmitter and receiver can agree on the common parameters like physical and MAC layer parameters to reduce the complexity of communication and to reduce the delay.

\subsubsection{Nanoscale networks}\label{subsubsec-comm:tx-nano}

In nanonetworks, mostly a nano controller is used to forward and collect data to/from nanodevices in a centralized network. Transmitter-initiated communication is used mostly to allow the nodes to transmit when they have data to send. The node which has data to send will initiate the communication and perform handshake process.  

A transmitter initiated communication scheme is proposed in~\cite{phlame2012} for nanonetworks, which is built on top of RD TS-OOK and takes benefits of a low weight channel coding scheme. The main aim is to negotiate between the transmitter and receiver the communication parameters and channel coding scheme to minimize the interference and maximize the probability of efficient decoding of the received information. The communication is established by sending a request by a node that carries information as Transmission Request (TR) and a node that receives it will agree to the communication parameters and generate an ACK and sends a Transmission Confirmation (TC) message. The TR contains the synchronization trailer, the transmission ID, packet ID, transmitting Data Symbol rate and Error Detecting Code. \blue{Although, it offers benefits in terms of delay and
throughput, it also has few limitations including the handshake process overhead which limits the Terahertz communication performance and limited computational power of nanodevices which requires optimal communication parameters.} \blue{I}n~\cite{esaware2013}\blue{, an} energy and spectrum aware MAC protocol is proposed, in which the computational load is shifted towards the nano controllers. The works in which energy harvesting is also considered and which uses Tx initiated communication include~\cite{MGDI2016,EESR-MAC2012,2stateMAC2018}. Other works on Tx initiated communication are mentioned in Figure~\ref{fig:communications}.

\mage{The Transmitter initiated communication is used mainly in papers in which distributed architecture is followed and where the goal is to achieve maximum throughput~\cite{DMDS2016,APIS2017} and not the energy efficiency. However, Tx initiated communication can increase the control message overhead. As nanonetworks are limited in computational capacity, handshake mechanisms are required which can balance the energy harvesting and consumption with minimum delay for link establishment. The works which consider energy as well are~\cite{esaware2013,MGDI2016,EESR-MAC2012,2stateMAC2018}. A Tx initiated communication is also followed in~\cite{2stateMAC2018} which presents an energy model and also considers the high node density.}

\subsubsection{Macro scale networks}\label{subsubsec-comm:tx-macro}

For a macro-scale network with directional antennas, the steerable narrow beam is essential to overcome the high path loss at the Terahertz band and to extend the communication range. \saim{Tx initiated communication is used in many macro-scale applications, where main goal is to increases network throughput performance.} A Terahertz based communication model and autonomous relaying mechanism for Vehicular communication are presented in~\cite{ATLR2018}, by considering the channel capacity, autonomous relaying and network establishment. An antenna switching mechanism is proposed in~\cite{SDNC2017,B5G2018} for vehicular and small cell deployment using Terahertz band in which lower frequency bands are used to achieve synchronization. 

The approaches in~\cite{TAB-MAC2016,MAADM2017,MAC-Yugi2018,MRA-MAC2017} uses multiple antennas to separate out the signaling and data transmissions parts. \saim{The 2.4 GHz band is proposed to use for signaling and antenna alignment using an omnidirectional antenna to overcome the antenna facing problem. The initial access and control information part are performed using IEEE 802.11 RTS/CTS mechanism and for data transmission directional antennas are used which also uses Tx initiated communication to transmit data between the Terahertz nodes}. Besides the high path loss, the directional antennas in these works are shown to be beneficial in reaching the distance beyond 1 meter. In~\cite{TRPLE2014}, a pulse level beam switching is used with energy control but Terahertz band features are not considered.

A CSMA/CA-based channel access mechanism is used in~\cite{IHTLD-MAC2017} with on-demand retransmissions mechanism for a TPAN which considers poor link conditions. The beacon frames are used by the piconet coordinator to provide information like channel access, channels slot assignment, and synchronization information. It is shown that network throughput decreases when the channel conditions are poor and proposed MAC protocol shows better performance in comparison with IEEE 802.15.3c and ES-MAC~\cite{esaware2013}.

\blue{The transmitter-initiated protocols are widely used as they incur less complexity and favors the distributed nature, however, they incurs several challenges due to the use of the directional antenna. On one hand, these directional antennas increase the transmission distance and on the other hand, they introduce the antenna facing the problem.} In outdoor scenarios where mobility is involved\blue{,} frequent beam switching occurs which requires a novel mechanism to minimize the synchronization and antenna alignment schemes. The WiFi technology is proposed to minimize the control messages overhead and link establishment in works like~\cite{TAB-MAC2016,MAADM2017,MAC-Yugi2018,MRA-MAC2017} and for antenna alignment which overcomes the facing problem. However, it introduces the high antenna switching overhead and requires efficient scheduling mechanism for seamless control information and data dissemination transmission.


\subsection{Receiver initiated MAC protocols}\label{subsec-comm:rx}

A general example of the receiver-initiated communication is shown in Figure~\ref{fig:txrx-handshake}, in which the receiver announces its existence and readiness to receive a packet from the sender. The Rx initiated communication is mainly used in nano and macro scale networks to save energy and reduce excess message overhead. Different existing solutions for both types of networks are discussed below.

\subsubsection{Nanoscale networks}\label{subsubsec-comm:rx-nano}

In nanoscale networks, which are prone to energy utilization, the excess of transmissions or message exchange means more utilization of energy. In nanonetworks, the amount of energy stored is just enough to hardly transmit a packet~\cite{MDP2014,joint-energy2012}. The transmission remains unsuccessful when the receiver receives the request from the sender but does not acquire enough energy to send an ACK or data packet. Therefore, in receiver-initiated protocols, the receiver takes initiative and announce its status of energy first to all senders by sending a Request-to-Receive (RTR) packet. \saim{The energy efficiency and harvesting are also discussed in Section~\ref{sec-req:maclayer}.} 

The receiver initiated communication is used both in centralized as well as distributed networks~\cite{qxia2019,LLsynch2015,RIH-MAC2014,drih-mac2015,MDP2014}. In centralized nanonetworks, a nano controller is mainly responsible for performing major processing and decision making, as they are energy enrich devices. Since the receivers are usually assumed to generate their own energy resources, they harvest just enough energy to transmit a packet. Therefore, in solutions like~\cite{RIH-MAC2014,drih-mac2015,MDP2014}, the receiver starts the communication by informing the senders, when it is ready to receive a packet and can exchange the information like coding schemes, error rates, and scheduling. Therefore, limited energy resource is one of the main reason, receiver-initiated communications are preferred in centralized networks. The problems occur when the receiver remains busy \blue{in} energy harvesting phase, and senders start sending the packets, which can result in the loss of a packet. Therefore, scheduling becomes an essential part of such kind of schemes. In~\cite{RIH-MAC2014}, a receiver-initiated communication model is presented for centralized topology, in which a receiver announces an RTR packet to nearby nano nodes and then the nano nodes send a DATA packet or ACK in response in a random-access manner with probability $p$ to establish a handshake between the nodes. A distributed scheme is presented in~\cite{drih-mac2015} in which scheduling and harvesting mechanisms are proposed to work together to enhance the energy utilization of nanonetworks. The proposed scheme uses the receiver-initiated approach to achieve the handshake and schedules.

\saim{The unsuccessful transmission-reception because of the node harvesting phase can increase dalay and also cause a hidden node problem.} Therefore, new schemes are required to avoid the hidden node problem in a distributed environment using a receiver-initiated approach. A MAC protocol is discussed in~\cite{MDP2014} in which optimal energy consumption and allocation problem are presented which aims to maximize the data rate. It is also shown that the amount of energy harvested is not enough for transmitting one packet and therefore it can take multiple timeslots to transmit a single packet when the harvesting rate is lower than the energy consumption rate.
	
\mage{In nanonetworks, receiver-initiated communication provides the flexibility to nodes in deciding which when to receive and transmit. The message overhead in these networks can be reduced by following a one-way handshake as provided in~\cite{qxia2019,LLsynch2015} in which energy harvesting is used. Whereas in~\cite{MDP2014} the trade-off between energy harvesting and consumption is also discussed.}

\subsubsection{Macro scale networks}\label{subsubsec-comm:rx-macro}

In these networks, high path loss at Terahertz frequencies affects the achievable distance between the Terahertz devices (cf. Section~\ref{sec-req:features}). It also requires tight synchronization between a transmitter and receiver to overcome the deafness problem~\cite{LLsynch2015,ssingh2011}. A receiver-initiated MAC protocol using directional antennas is discussed in~\cite{qxia2019,LLsynch2015} which uses a sliding window flow control mechanism with a one-way handshake that increases the channel utilization. High speed turning directional antennas are used to periodically sweep the space. The main objective is to prevent unnecessary transmission when the receiver is not available due to antenna facing the problem. In this scheme, a node with sufficient resources broadcasts its current status by using a CTS message by using a dynamically turning narrow beam while sweeping its entire surrounding space. The CTS frame contains the information of receivers' sliding window size. On the other side, the transmitter checks for a CTS frame from the intended receiver and then points its direction for the required period towards the receiver. The initial neighbor discovery of the neighbor nodes is not considered in~\cite{LLsynch2015}. Further, due to the bit error rate and path losses \saim{(cf. Section~\ref{sec-req:phylayer})}, packet reception guarantee is not considered. It is also possible that multiple transmitters might point out at same receiver at the same time, which can result in possible collisions. A MAC protocol focused on cross-layer analysis for relaying strategies is discussed in~\cite{OPTRS2017} with distance dependant throughput maximization analysis while considering the antenna, physical, link, and network layer effects. In particular, receiver-initiated communication is shown to be better than the transmitter-initiated communication in~\cite{qxia2019,LLsynch2015}.

\subsection{Summary and discussion}

The nanonetworks are considered as networks with limited energy and resources in which mainly the schemes are preferred in which energy consumption can be minimized. Other than the traditional way, some receiver-initiated mechanism are proposed in the literature, in which a receiver initiates the communication establishment by announcing its status for the sufficient energy resources to receive a packet. Whereas, in transmitter initiated communication, the node which have information to send can initiate to establish the communication. At nanoscale networks, mostly the networks are centralized and therefore nano controllers are used mostly to manage the control and data transmissions and so form a centralized network. In this kind of network, scheduling the transmission is the main requirement as they use pulse based communication instead of carrier sensing based communication. For higher network density, the collisions cannot be ignored and so require an efficient mechanism to avoid collisions and minimize packet loss.

The receiver initiated communication is used also in the macro scale networks. Although it minimizes the control messages overhead, it increases the complexity and is not preferable in the distributed scenarios. In a distributed environment, where two nearby nodes performing the same operation such as energy harvesting can increase the delay and can cause a hidden node problem. Therefore, efficient synchronization and scheduling mechanism are required for these scenarios.

\section{Challenges and Future Research Directions}\label{sec:issues-challenges}

Designing an efficient Terahertz MAC protocol needs to address different challenges. In this section, these challenges are presented with future research directions.

\subsection{Terahertz communication network topologies}

\subsubsection{Macro scale network}
\saim{The challenges and future research directions for macro scale network topology design are discussed below.}

\saim{\textit{Static and mobile scenarios:} } \saim{In fixed point-to-point connectivity, it is fairly easy to maintain stable links. However, when the nodes are mobile, the network topology changes frequently, in which frequent neighbor discovery and link establishment will be required. To move further, point-to-multipoint connectivity requires more attention in terms of MAC protocols design. A Data Centre environment is a good example, in which establishing point-to-point and multipoint among the inter and intra rack communication to effectively replace wired connections is still a challenge.}

\saim{The Terahertz signal is highly sensitive to the outdoor environment, for example, weather fluctuation, and the presence of blockers between two nodes can affect the communication link. MAC layer for such a system should include fast link re-establishment mechanism and alignment operation in case of sudden miss-alignment between the two nodes by giving alternative paths for transmission.} 

\saim{\textit{Blockage and beam steering:} } \saim{Antenna arrays and reflectors can be used to avoid the chances of blockage and to find a good propagation path. It is critical to find a good beam pointing mechanism to support user mobility with fast and accurate beam tracking, both in LOS and non-LOS conditions~\cite{bonjour2016}. A mechanically steerable antenna is demonstrated in for 300 GHz band~\cite{srey2015}. Overall, a faster electronic beam steering is more practical and required. }

\saim{Unlike lower frequency band such as microwave, Terahertz is very sensitive to blockers. Efficient techniques are required to reduce the effects of blocking in the system. The path diversity can be one of possible solution using beam diversity and MIMO system. The challenge is to track presence of blockers and to keep the link available when direct LOS is absent. The integration of Terahertz intelligent surface is worth considering which enables user tracking and add path diversity to the propagation scenario, users can be reached even in presence of severe blocking scenario.}

\saim{\textit{Data Centre Geometry:}} \saim{The high capacity Terahertz links can help in re-designing the Data Centre geometry by moving the racks closer in a cylindrical architecture. Future work demands the flexible Data Centre architecture to support scalability and energy-efficient design while considering the Terahertz features and limitations. It will reduce the deployment delay and cost for future data centers. Full beam scanning can be used for initial access to serve 360-degree search space. Antenna sectors can also be used for 90-degree search space and communication at the cost of additional hardware. }


\saim{\textit{Coverage: }} \saim{The transmission distance achieved so far for Terahertz communication is still limited to few meters~\cite{Petrov2018}. Extending this range to reach to 10 m or more for an indoor scenario using low power transmission devices and efficient communication protocols is an active field of research for B5G networks. As a future topic, the Terahertz wireless and fiber connectivity for backhaul and fronthaul high data rate communication are also gaining attention (cf. \blue{Section} \ref{sec:applications}). Directional antennas with narrow beams are being encouraged in Terahertz networks to extend the transmission distance. This directional antenna usage can limit the interference and losses (cf. Section \ref{sec-req:features}), they must use the optimal schedules and tight synchronization. The phased array and MIMO techniques can be used to further extend the transmission distance.}

Terahertz communication is characterized by a low coverage zone as transmitted signal experiences high path attenuation. A high gain antenna can be deployed to increase network range, however\blue{,} the communication range is still low compared to lower frequency bands. In order to extend coverage zone\blue{,} it is possible to add additional functionalities to nodes at the coverage area zone to play the role of coverage extension and coordination with nodes out of the coverage zone. Coverage extension is challenging as more interferences occur and additional time is required for out of range nodes discovery is required\blue{. S}ome related work on coverage extension can be found in~\cite{Petrov2018}. From MAC layer point of view, nodes at the edge should coordinate to discover and synchronize with out-of-range nodes. \blue{A second possibility to further extend the coverage area at macro scale is the deployment of intelligent surfaces, in which when a node is near the intelligent reflector in the inner zone can send messages to the surface in order to be reflected out of the inner zone~\cite{wwithaya2018}.} The MAC layer should be aware of new devices required for coverage extension. Edge nodes are selected by the nodes controller for coverage extension if their received power is lower than a threshold and also if they are located near an intelligent surface. This concept is shown in Figure~\ref{fig:coverage}.

\begin{figure}
\centering
\includegraphics[width=3in,height=1.6in]{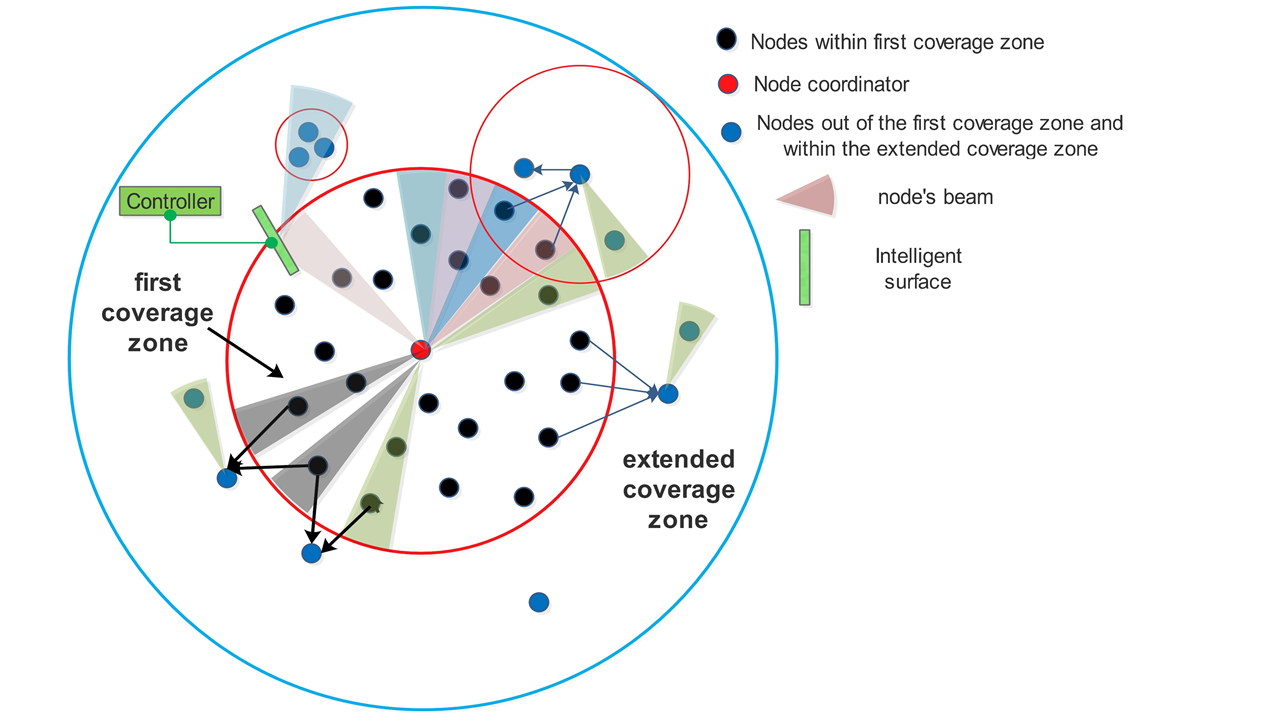}
\caption{Coverage extension in Terahertz networks.}
\label{fig:coverage}
\end{figure}

\subsubsection{Nanoscale network}

\saim{The challenges and future research directions for nanoscale network topology design are discussed below.}

\saim{\textit{Node density: }} In the nano-communication network, thousands of nodes can exist in very small regions which are generally limited in capabilities. These networks require efficient MAC layer techniques to meet network requirements such as high throughput, reliability, and low energy consumption. Models related to node density are used in literature, for example interferers are linearly distributed around the receiver~\cite{zhosain2019}. The density distribution is used to evaluate the interference in the system and assess the limitation of the network\blue{. Moreover,} non-homogeneous density of nodes can affect the MAC choice, then MAC protocols should be able to track the fast change in node density in each part of the network.

With limited energy storage capacity, establishing links and data transmission will require more energy to facilitate larger number of nodes. The nodes mobility in a distributed environment can also change the topology of the nodes frequently which needs to be handled using new MAC layer protocols. In different applications, like agricultural or in-body health monitoring, the nodes scalability is required (cf. \blue{Section} \ref{subsec-app:nano}). Managing node discovery and link stability are still challenges while dealing with limited energy resources at nanoscale. Models to quantify the resources like time, storage and amount of communications are required to reduce the computational complexity in centralized as well as distributed networks.

\saim{\textit{Energy harvesting and transmission trade-off: }} The limited capacity can allow only an energy harvesting or transmission at one time which requires to be scheduled efficiently. When a node \blue{is} busy in harvesting\blue{,} it can not transmit and when it transmits\blue{,} it can not generate energy for next transmissions\blue{,} which can affect the overall latency and throughput of a nanonetwork. Therefore, efficient schedules are required to improve the overall throughput and latency while managing the harvesting and transmissions.

\saim{\textit{Computational complexity: }} \saim{Nano devices are small devices with limited computational capacity and power management and so require simple communication protocols. More functionalities can be provided by optimizing the nano device's behavior. Therefore, to manage the nanodevices and processing the complex computations externally, hierarchical architecture is required with decisions to be taken at devices with higher computational capacity and powerful communication layers~\cite{Galal2018}. Gathering the data in a distributed network architecture to perform complex operations is a challenging task. Therefore, new mechanisms are required for synchronization and coordination for distributed networks. Further, the nanonetworks can also be combined with current network technologies like software-defined networks, IoTs and virtual networks to solve the complex problems at those technologies and higher layers~\cite{Abbasi2016}. }

\saim{\textit{High capacity controller: }} The nanonetwork topology can be dynamic at times, due to dynamic channel conditions which can affect the reliable transmissions. In response, the nanodevices in a network might not share the same topology information. \blue{Further, due to limited memory, storage, and computational processing capabilities, the nanodevice also faces difficulty in storing the routing tables and heavy algorithms.} To solve this, a controller with software-defined functionality can be utilized to do the extra computations, which can also change and reconfigure the behavior of nanonetwork.

\subsection{Terahertz channel access mechanisms}

\subsubsection{Macro scale network}

In this section, challenges and future research directions are mentioned for efficient Terahertz channel access mechanisms.

\saim{\textit{Hybrid protocols:} } \saim{The scheduled channel access techniques can reduce the collisions, whereas, random access techniques can enhance the throughput and latency requirements. The future research should be on hybrid mechanisms to improve the fairness and throughput among the users using combination of both techniques. }

\saim{\textit{Adaptive band selection: }} \saim{The Terahertz transmission bands can be affected by high absorption and path loss. Multiple transmission windows are available in the Terahertz band with different attenuation values. Efficient mechanisms are required to efficiently utilize the distance and bandwidth dependent Terahertz band allocation. }

\saim{\textit{Multiple band usage:}} \saim{To solve the synchronization and initial access, different bands like microwave, mmWave, and Terahertz are being used together. The microwave band can provide wider coverage which can be used for initial access and antenna alignment. However, when to switch among these bands requires new methods to be defined to improve the switching overhead and delay.}

\saim{\textit{Multiple antennas and beam usage:}} \saim{MIMO antennas can be used to mitigate the Terahertz channel effects (cf. Section \ref{sec-req:features}) and can improve the path diversity, and antenna gain which can improve the overall coverage. } 

\saim{\textit{High order modulation: }} New waveforms, and modulation and coding schemes for the Terahertz system are also required to improve the data rate and quality of service for beyond 5G communications~\cite{chanbicen2016}. The use of OOK mapped with ultra-short pulses at one hand can reduce the transceiver complexity, however, it also introduces challenges for antenna arrays for ultra-broadband design. The complexity of generating and processing high order modulated signals at transmitter and receiver side using Terahertz device is a challenge. \blue{For good channel conditions, throughput can be improved by using these high order modulation techniques like 64 QAM~\cite{idan2020}}. Further, the decoding and delay complexity at receiver also needs to be focused.

\saim{\textit{Beam management: }} Fast beam scanning and steering mechanisms are required to obtain channel state information to improve the channel usage decisions at macro-scale networks. The random channel access can solve the problem of antenna alignment, however, the range becomes short due to high path loss. An efficient TDMA based scheduling mechanism is required for both centralized and distributed networks to avoid multi-user interference.

\subsubsection{Nanoscale network}

In nanonetworks, due to shorter distance and low path loss\blue{,} huge bandwidth can be used which can result in very high data rate, short transmission time with low collision probability~\cite{Abbasi2016}. The MAC protocol role is very important in regulating the link behavior, arranging the channel access and coordinating the transmission in a distributed environment.

\saim{\textit{Random access and scheduling: }} \saim{At the nanoscale, huge bandwidth availability and pulse-based communication, reduces the collision probability. The random access techniques can take benefit out of it but require efficient scheduling of transmission with a larger duration between packet transmissions. However, for high node density scenario, this needs to be carefully designed due to energy limitations~\cite{vpetrovsinr2015}.}

\textit{Node density and error control: } Although the actual Terahertz nanoscale network is characterized by a basic MAC layer, it is possible to implement efficient error control protocols to reduce packet retransmissions and packet waiting time. Optimized mechanisms are required overall to run communication tasks at the nanoscale, due to energy limitations and to support a large number of nanodevices. Efficient scheduling mechanisms for nanonetworks are required overall to balance the energy harvesting and channel access for data transmissions. They need to be further optimized, to enable fast communication among a larger number of nanodevices. Further, additional functionalities such as scheduling transmissions are required between inter-cluster and TDMA based intracluster communication to avoid collisions and to increase throughput~\cite{mehfuz2015}.

\subsection{Terahertz Receiver and Transmitter initiated communication}

\subsubsection{Macro scale network}

\blue{In this section, challenges and future research directions are mentioned for Transmitter and Receiver initiated communication. }

\saim{\textit{Message overhead:} } \blue{T}he transmission of control messages in excess can cause the message overhead problem which can easily occur in directional Terahertz communication. Efficient techniques are required to reduce the message overhead. Different windows or bands can be used to utilize the channel resource efficiently. Although energy is not a high constraint like in nanonetworks, fast link establishment techniques, especially for distributed Terahertz networks and networks with high mobility require novel solutions, for example, small cells and vehicular communications. The control packets transmission to establish a link can cause high message overhead. New handshaking mechanism is required to establish link with reduced message overhead and delay.

\saim{\textit{Memory assisted communication:} } \saim{A memory assisted communication can enhance the throughput and latency in a Terahertz network~\cite{MAADM2017}. Nodes position, neighbor tables, beam directions, and weights can be used in memory assisted communication to reduce the alignment and training time for antenna. Relays can be selected by coordinating the neighbor information and transmissions can be scheduled in advance while tracking the node's availability.}

\textit{Antenna direction and communication: } At macro scale, communication is linked with directional antennas. To initiate communication, nodes should have enhanced MAC functionalities such as simultaneous sensing and beam steering capabilities to track channel and node position jointly. Additional technology can assist in transmission initiation, such as RADAR and LIDAR for mobile nodes, and lower band technologies. The radar-based sensing is gaining attention to solve the deafness problem in the vehicle to vehicle communication to efficiently align the antenna direction and communication establishment~\cite{vpetrovv2v2019}. A preamble injection is shown useful in low SNR scenarios to avoid deafness problem in~\cite{pkumari2018}.

\subsubsection{Nanoscale network}
For nanoscale networks, nodes with higher energy harvesting capabilities can take the responsibility of link establishment and transmission initiation, however\blue{,} designing sophisticated algorithms for rapid link initiation still remains an issue. Research is required for distributed MAC protocols for Nanosensor networks to reduce message overhead with energy control mechanism. Receiver initiated transmission suits the low power nanodevices link establishment process due to reduced message overhead, it should consider energy as well as data size to be transferred. A combination of distributed communication in presence of a controller can help in managing the transmission schedules among nano devices~\cite{idemirkol2006}.

\subsection{General challenges and future research directions}

The Terahertz technology is capable of delivering high data rate traffic with an acceptable quality of services such as low delay, and minimized energy consumption. However, many challenges still exist and require further research and attention. Some of them are discussed below related to Terahertz MAC protocols in general. 

\subsubsection{Interference management}

Although interference can be reduced by using highly directional antenna, it is not considered by many research works~\cite{HAN201967}. It can be considered for large network requiring high data rate connection per nodes, such as top of the rack data center network, where nodes should transmit all the time with high data rate and low delay. Interference for dense indoor scenarios should be deeply studied and interference model needs to be established. The MAC layer should be aware of interference in the channel to elaborate further the rapid scheduling and fast channel access and switching based on channel interference information. The new and dynamic channel selection mechanism is also required while considering the Terahertz band's unique characteristics and band-specific interference and achievable distance. 

The interference management module can track channel status and decides on the transmission time slot as well as the carrier to be used and physical parameters to be set, such as modulation and coding scheme and annulating side lobes in some directions. The additional procedure can be also implemented such as adoption, at the design stage, of a specific frequency plan for network and setting a sub-band spacing strategy for each application. At operational mode, each node can use a fixed frequency pattern at the deployment stage or adopts the frequency hopping strategy to keep an acceptable SINR and to overcome the molecular absorption of the propagation medium~\cite{TSN2014}, as the noise generated by molecule depends on frequency. Using frequency hopping scheme is promising as it tracks the channel switching, however, the designing of the frequency hopping algorithm is a challenge. The second challenge is to explore the number of frequencies the MAC layer can manage to improve the throughput.

\subsubsection{Antenna design}

The link quality depends on the physical layer and channel\blue{. F}or Terahertz communication antenna technology improvement is considered as a key factor for link budget enhancement. The MAC layer should monitor antenna by fast switching beams to serve all users in a fair way. MAC can also select antenna carrier frequency and polarization. To monitor efficiently a Terahertz communication, MAC layer should interface with the antenna system, for example controlling the steering angles, beam width and switching time. Because a good command of antenna system can increase data throughput and reduce delays due to misalignment errors. Antennas properties should be optimized, the MAC layer can monitor Terahertz antenna via frequency switching, beamforming, and diversity in order to meet network requirements\blue{.} 

\textit{Polarization capabilities: } Using two polarizations (horizontal and vertical) with sophisticated algorithms of cross-polarization cancellation is promising to boost the Terahertz system performance toward higher throughput and lower total system signal to interference ratio~\cite{mkolano2018}. The main challenges with the dual-polarization approach from the MAC point of view are to balance the traffic between the two polarizations and to mitigate errors. Moreover, channel impulse response and then received power depends on polarization~\cite{spriebejacob2011}\blue{. O}ne challenge is how to exploit efficiently Terahertz wave polarization to increase data throughput by balancing the data flow simultaneously between horizontal and vertical polarization.

\textit{Wideband and multi-band antenna: } The design of multi wideband antennas are required to increase MAC efficiency and meet system requirements in term of high throughput, as more bandwidth will be available to transmit more data rate~\cite{msrabbani2015}. Using a multiband antenna can also reduce system latency by deploying separate bands for data transmission and control message exchange.

\textit{High antenna gain: } To mitigate channel impairment and extend the communication range, antenna gain should be maximized, Horn, logarithmic antenna and phased array are promising for designing high antenna gain Terahertz communication. For high antenna gain, it is possible to increase node reachability, but more care should be addressed to antenna side lobes as they can generate more interferences.

\textit{Spatial diversity: } To mitigate channel impairment and increase channel capacity, multiple antennas along with phased array, exploiting Terahertz propagation diversity, can be deployed for Terahertz links, such as MIMO and ultra massive MIMO. Using MIMO increase spectral and energy efficiency for the link, however, it requires efficient signal processing performance to encode and decode MIMO signals and exploit diversity. From MAC point of view, deployment of ultra massive MIMO will affect resource allocation techniques~\cite{chan2018mimo}.

\textit{Fast switching capability: } To increase data throughput and reduce latency per link, the beam switching time should be minimized. Switching can occur at pulse, symbol or frame level.

\textit{Adaptive beamforming: } \blue{Directional antenna is considered an alternative solution to mitigate channel impairment and increase link budget. Nevertheless, antenna pattern can take different shapes, and it should be optimized for a network use case. By using adaptive weighting of antenna elements monitored at the physical layer and MAC, it is possible to reduce effect of interference by annulling lobes in some direction to avoid interferences.} A MAC module can be implemented to control antenna beamforming and adapt the antenna technology to the network topology. In~\cite{plu2017}, a log-periodic toothed antenna is optimized for beamforming and beam steering for the Terahertz band. A concept of intelligent communication by using ultra massive MIMO is proposed in~\cite{snie2019}, to increase the communication distance and data rates at Terahertz band frequencies.

\subsubsection{Synchronization}
Synchronization adds accuracy to network operations and coordination and reduces frame collisions among nodes, as a result, it contributes to QoS enhancement. Moreover, it is responsible for more computation complexity and requires additional time slots before data transmission starts. Nodes memory and time for link establishment are the main cost to pay, in order to deploy the synchronous network. \blue{At the MAC level, the challenge is
to design algorithms for nodes and frame synchronization, such that nodes become aware of the transmission time.} Another challenge is to efficiently allocate the radio resources for synchronization procedures such as frequency, time and power, which can increase the delay. To reduce the delay\blue{,} transceivers with more capabilities such as memory and processing can be used.

\subsubsection{Transceiver design}
An efficient transceiver is required to deal with different MAC functionalities ranging from framing, synchronization, error control, scheduling and buffering. Authors in~\cite{DLLC2017}, demonstrate that it is possible to optimize transceiver architecture to bear high data rate reaching $100 Gbps$ using parallelism and optimized memory usage along with frame length and error control techniques. Using efficient processing technique at the transceiver and sufficient memory size, it is possible to implement MAC protocols dealing with fast channel access, efficient scheduling technique and multi traffic communication\blue{.}

\subsubsection{Link establishment, neighbor discovery and deafness problem}

Before any communication starts, an establishment phase should be initiated. Link establishment is the duty of the MAC layer when a node needs to transmit to another node. This phase starts with setting up all the required parameters such as physical layer parameters, timers, synchronization procedure\blue{. T}hen after receiving the \blue{acknowledgement} from the receiver, a new transmission can begin. The challenge is how fast we can establish any Terahertz connection, and how to increase the success probability of link establishment phase\blue{?} 

Deafness can complicate the neighbor discovery due to transceiver misalignment and prevents the control messages to be exchanged in a timely manner (cf. \blue{Section} \ref{subsubsec-req:design-mac-nbrdisc}). To avoid the antenna \blue{miss-alignment} and link establishment, the mmWave standard IEEE 802.15.3c and IEEE 802.11ad use beam training, in which one node operates in a directional mode and other node search space in a sequential directional mode. After a complete search, sector-level training occurs to perfectly align the beams. For neighbor discovery the challenges are to discover all the nodes with minimum delay, and techniques to search the space for beam alignment in a short time when two nodes are not initially aware of their beam directions. 


Particularly, for Terahertz communication the neighbor discovery is challenging due to unique band characteristics and antenna directionality, and for nanoscale networks due to limited energy resources. Neighbor discovery is required to synchronize nodes within a network and rapidly consider new nodes in the network. As a future research direction, optimization of discovery time by correctly choosing reference time for antenna alignment is an important challenge with timely information exchange. The discovery can be enhanced by using multibeam with fast switching and coordination among distributed nodes~\cite{yzhao2017}. \saim{A neighbor discovery protocol with directional antennas with side lobes information and full antenna radiation pattern to better detection is proposed in~\cite{qxia2019}. However, multipath effects and LOS blockage were not considered. }

\subsubsection{LOS blockage}

A blockage is a situation when an object crosses the main link between two nodes transmitting to each other, it can also be generated from frequency shifting and reflected signal from surrounding objects. Due to high data rate, a small or temporary blockage can result in very high data loss. Therefore, it is important to propose novel anti-blockage mechanisms to avoid blockage situations and to achieve seamless coverage. At MAC layer, it is important to identify blocked channels to avoid false detection and correction and to distinguish between deafness and unblocked error. In Terahertz band, due to small wavelength (0.3 mm)\blue{,} the directional links can be easily attenuated by LOS obstacles. In mobility scenarios, these obstacles can occur more frequently, and therefore can degrade the Terahertz link performance. Only increasing the transmission power cannot penetrate the obstacles, therefore an alternative unblocked channel is required to steer around. Reflectors can be used to avoid permanent blockage, but new mechanisms are required with beam steering and management functionalities to avoid link blockage. Modeling the blockage phenomena for each use case is required, and MAC layer should be aware of it. The main challenge is to detect blockage and tackle this issue at MAC layer. One alternative is to differ its transmission till the link is cleared or to select an alternative path with new parameters to avoid transmission interruption. \saim{To mitigate temporary blockage of LOS wireless link, an NLOS wireless link with reflection and scattering over multiple rough surfaces is analyzed in~\cite{jma2019}. The use of the NLOS links can broaden the access point options to improve the link performance. However, using complex modulation schemes are yet to be analyzed for its feasible working under different environments.}

\subsubsection{Design of relaying and multihop protocols}

The short-range communication like indoor and Data Centre scenarios, require new and efficient relaying techniques to increase the reachability of nodes. Nodes relaying or forwarding capability can be implemented at the MAC layer. It is activated when the signal from one transmitter needs to be regenerated by intermediary nodes to reach its final destination. However, to activate this, each node must have a complete view of the neighbors which can be exchanged among the nodes as a neighbor table. Due to antenna directionality and tight beam requirement of Terahertz communication, beam switching techniques can be used where antenna can take 0.1 ns to switch antenna beam direction~\cite{jornetcabellos2015}, which can increase the overall delay in forwarding the packets. The relaying protocol using directional antenna must be designed to reduce this delay and to overcome channel impairment problem. 

In Terahertz band only short transmission range is achievable until now. Due to which the signal needs to regenerated by an intermediate node to reach the destination. Designing strategies with relaying capabilities by considering the unique band features and environment is a challenging requirement. Work on node relying on Terahertz band was performed at nanoscale communication~\cite{zrong2017}, where two modes were considered: amplify and forward, and decode and forward, to strengthen the direct path by maximum ratio combining.  

The relay node can be selected from the existing neighbors or can be placed especially in a network. Each mechanism needs to address different challenges including link quality, location, and number of relays. In a distributed environment, where nodes communication range is shorter, multihop communication should be enabled. Multihop protocol design can be challenging when high throughput and low latency are the requirements. Therefore, new multihop strategies are required to fulfill Terahertz communication requirements by considering limited capabilities and behavior of communication layers in case of nanoscale networks. For macro-scale networks, the path loss, molecular noise, and antenna directionality should be considered. Reflectors can also be used for communicating and reaching to nodes with LOS blockage.

\subsubsection{Coordination}
Designing efficient MAC protocols for vehicular and satellite communication with relaying and coordination among the nodes is an open challenge for MAC design. The node should decide the next relay node to strengthen the communication among different nodes using coordination mechanisms. The nodes should be capable of deciding which node it should coordinate.

\subsubsection{Cross layer design}
MAC layer should adapt between traffic flows coming from the upper layer and the channel fluctuations along with the physical layer procedures. For instance, the selection of frame size per transmission period should be adaptively chosen based on packet arrival from the networking layer as well as considering measurements from the physical layer. Scheduling transmissions can be optimized based on measurements gathered from the physical layer. MAC decisions such as band selection and path switching, in case of blockage are also affected by the physical layer and channel status. The transceiver memory should be optimized to support \blue{traffic} with different QoS profiles.

\subsubsection{Scheduling}
In the Terahertz network, scheduling algorithms can enhance the overall quality of service by using radio resources for a given policy such as maximizing throughput, minimizing the total interference in the network or reducing system delay.
 The scheduling module will be interfaced with the medium access module as well as physical layer, where knowledge about the channel conditions and traffic requirements will govern scheduler decisions. Exchanging information related to buffer, channel quality and requirement of each traffic flow should be considered by the scheduler and also schedules of other nodes.

\subsubsection{Framing and Error handling}
Selection of frame size, frames, and multi-frame structure and error control strategies, such as CRC insertion and frame retransmission, can enhance Terahertz link in terms of frame error rate as well as leads to an increase of throughput. Adaptive frame size and control overhead are fundamental to maintain a communication link and reduce errors among transmitted frames. Increasing the packet size can cause a higher number of channel errors which requires more robust error detection and correction schemes. The longer packets can also introduce the buffering problems. Therefore, the optimal packet size and analysis of the trade-off between the size and performance, and flow control policies to avoid congestions and buffer overflow, requires further research.

\subsubsection{Mobility management}
Mobility can easily affect the quality of the established links due to narrow beams. Therefore, frequent re-establishment of links is required to maintain the links and communication over it. Two mobility models are mentioned in~\cite{Park2010}, linear and circular motion. In different Terahertz scenarios, different mobility models need to be studied like in V2X scenarios and small cells, where LOS blockage can also occur frequently. It is important to track the best beam in case of frequent link breakage and beam alignment requirement. 

In V2X networks\blue{,} nodes change their position with variable speed. To keep the connectivity with the network, a management module should be implemented at the MAC layer monitoring node location and tracking its speed. For such an application, a mobility model needs to be proposed for each scenario\blue{.} The mobility management module will decide on the handover and how to make the link robust all the time until the end of the transmission without interruption. It is possible to decide to change a new node as receiver or to transmit by relaying. An update mechanism should be set to sort all neighbor nodes based on their availability. One more challenge is how to sample channel conditions and how fast the MAC can decide for the handover.

\section{Conclusion}\label{sec:conclusion}

In this paper, a comprehensive survey is presented for Terahertz MAC protocols. The Terahertz band has a great potential to support the future ultra-high bandwidth and low latency requirement for beyond 5G networks. Especially, the existing unused frequency operating at disruptive bandwidths of 70 GHz can be a key enabler for beyond 5G networks. In this regard, the key features, design issues for Terahertz MAC and decisions which should be taken at MAC layer to enhance the performance of the network, are highlighted and discussed. Different Terahertz applications for macro and nanoscale are also discussed with their scenario-specific challenges. The survey has identified numerous gaps and limitations of existing Terahertz MAC protocols for enhancing further research in this domain. To highlight the limitations, the existing literature on Terahertz MAC protocols is also classified based on topology, scale, channel access mechanism and transmitter/initiated communication. To push further the research in this domain, challenges and future research directions are also presented with a cross-layer approach.

%



\section*{Acknowledgment}
This project has received funding from the European Union's Horizon 2020 research and innovation programme under grant agreement No 761579 (TERAPOD).

\ifCLASSOPTIONcaptionsoff
  \newpage
\fi

\bibliographystyle{ieeetran}












\end{document}